\documentclass[prx,aps,twocolumn,superscriptaddress,10pt]{revtex4-2}
\usepackage[T1]{fontenc}
\usepackage[utf8]{inputenc}
\usepackage{color}
\usepackage{amsmath}
\usepackage{amsfonts}
\usepackage{hyperref}
\usepackage{amsthm}
\usepackage{bm}
\usepackage{float}
\usepackage{textcomp}
\usepackage{upgreek}
\usepackage{textgreek}
\usepackage{setspace}
\usepackage{stackengine}
\usepackage{multirow}
\usepackage{enumitem}
\usepackage[percent]{overpic}
\usepackage{soul}
\hypersetup{
	colorlinks = true,
	allcolors= blue,
}
\usepackage{tikz,pgfplots}
\usepackage{blkarray}
\usetikzlibrary{patterns}
\usetikzlibrary{fadings}
\usetikzlibrary{decorations.pathmorphing,patterns}
\tikzset{snake it/.style={decorate, decoration=snake}}
\usetikzlibrary{arrows, decorations.markings}
\pgfplotsset{compat=1.14}
\tikzset{
vecArrow/.style={
  thick,
  decoration={markings,mark=at position
   1 with {\arrow[scale=2,thin]{open triangle 60}}},
  double distance=1.4pt, shorten >= 10.5pt,
  preaction = {decorate},
  postaction = {draw,line width=1.4pt, white,shorten >= 4.5pt}
  },
innerWhite/.style={
  semithick,
  white,
  line width=1.4pt,
  shorten >= 4.5pt
  }
}
\definecolor{orange}{rgb}{1,0.5,0}
\definecolor{darkgreen}{rgb}{0,0.4,0.1}

\newcommand{\WidthFigure}{\columnwidth}
\newcommand{\WidthFigureParticular}{0.49\columnwidth}

\newcommand{\nametheory}{Wigner}
\newcommand{\RedNmTh}{LWTE}

\makeatletter
\newcommand{\doublehat}[1]{% 
\begingroup%
  \let\macc@kerna\z@%
  \let\macc@kernb\z@%
  \let\macc@nucleus\@empty%
  \hat{\raisebox{.3ex}{\vphantom{\ensuremath{#1}}}\smash{\hat{#1}}}%
\endgroup%
}
\makeatother

\makeatletter
\newcommand{\doublehatSub}[1]{% 
\begingroup%
  \let\macc@kerna\z@%
  \let\macc@kernb\z@%
  \hat{\raisebox{-.07ex}{\vphantom{\ensuremath{#1}}}\smash{\hat{#1}}}%
\endgroup%
}
\makeatother

\makeatletter
\DeclareFontFamily{OMX}{MnSymbolE}{}
\DeclareSymbolFont{MnLargeSymbols}{OMX}{MnSymbolE}{m}{n}
\SetSymbolFont{MnLargeSymbols}{bold}{OMX}{MnSymbolE}{b}{n}
\DeclareFontShape{OMX}{MnSymbolE}{m}{n}{
    <-6>  MnSymbolE5
   <6-7>  MnSymbolE6
   <7-8>  MnSymbolE7
   <8-9>  MnSymbolE8
   <9-10> MnSymbolE9
  <10-12> MnSymbolE10
  <12->   MnSymbolE12
}{}
\DeclareFontShape{OMX}{MnSymbolE}{b}{n}{
    <-6>  MnSymbolE-Bold5
   <6-7>  MnSymbolE-Bold6
   <7-8>  MnSymbolE-Bold7
   <8-9>  MnSymbolE-Bold8
   <9-10> MnSymbolE-Bold9
  <10-12> MnSymbolE-Bold10
  <12->   MnSymbolE-Bold12
}{}

\let\llangle\@undefined
\let\rrangle\@undefined
\DeclareMathDelimiter{\llangle}{\mathopen}%
                     {MnLargeSymbols}{'164}{MnLargeSymbols}{'164}
\DeclareMathDelimiter{\rrangle}{\mathclose}%
                     {MnLargeSymbols}{'171}{MnLargeSymbols}{'171}
\makeatother

\DeclareMathAlphabet{\mathsfit}{\encodingdefault}{\sfdefault}{m}{sl}
\SetMathAlphabet{\mathsfit}{bold}{\encodingdefault}{\sfdefault}{bx}{sl}
\newcommand{\tens}[1]{\bm{\mathsfit{#1}}}
\newcommand{\tenscomp}[1]{\mathsfit{#1}}

\makeatletter
\usepackage{comment}
\let\wfs@comment@comment\comment
\let\comment\@undefined

\usepackage[final]{changes}

\let\wfs@changes@comment\comment
\let\comment\@undefined

\newcommand\comment{%
    \ifthenelse{\equal{\@currenvir}{comment}}
    {\wfs@comment@comment}
    {\wfs@changes@comment}%
}
\makeatother

\definecolor{dgreen}{rgb}{0,0.45,0}

\makeatletter
\@namedef{Changes@AuthorColor}{red}
\colorlet{Changes@Color}{red}
\setaddedmarkup{\textcolor{dgreen}{#1}}

\makeatother

\begin{document}
%PREPRINT_COMMAND
%\setstretch{2}
%\setcitestyle{super}

%s\title{Wave-particle duality in heat conduction from Wigner's phase-space formalism}
\title{Wigner formulation of thermal transport in solids} %and wave-particle duality
%: wave-particle duality for heat

\author{Michele Simoncelli}
\altaffiliation{Present address:  
Theory of Condensed Matter Group of the Cavendish Laboratory and Gonville \& Caius College, University of Cambridge (UK)}
\email{ms2855@cam.ac.uk}
\affiliation{Theory and Simulation of Materials (THEOS) and National Centre for Computational Design and Discovery of Novel Materials (MARVEL), {\'E}cole Polytechnique F{\'e}d{\'e}rale de Lausanne, Lausanne, Switzerland.}
\author{Nicola Marzari}
\affiliation{Theory and Simulation of Materials (THEOS) and National Centre for Computational Design and Discovery of Novel Materials (MARVEL), {\'E}cole Polytechnique F{\'e}d{\'e}rale de Lausanne, Lausanne, Switzerland.}
\author{Francesco Mauri}
\affiliation{Dipartimento di Fisica, Universit{\`a} di Roma La Sapienza, Piazzale Aldo Moro 5, I-00185 Roma, Italy}


\begin{abstract}
Two different heat-transport mechanisms are discussed in solids: 
in crystals, heat carriers propagate and scatter \replaced{particle-like}{like particles} as described  by  Peierls' formulation of \added{the} Boltzmann transport equation for phonon wavepackets.
In glasses, instead, carriers behave wave-like, diffusing via a Zener-like tunneling between quasi-degenerate vibrational eigenstates, as described by the Allen-Feldman equation.
\replaced{
Recently, it has been shown that these two conduction mechanisms emerge from a Wigner transport equation, which unifies and extends the Peierls-Boltzmann and Allen-Feldman formulations, allowing to describe also complex crystals where particle-like and wave-like conduction mechanisms coexist.  
}{Recently, it has been shown that these two conduction mechanisms emerge as limiting cases from a unified transport equation, which describes on an equal footing solids ranging from crystals to glasses;
moreover, in materials with intermediate characteristics
the two conduction mechanisms coexist, and it is crucial to account for both.}
Here, we discuss the theoretical foundations of such transport equation as is derived from the Wigner phase-space formulation of quantum mechanics, elucidating how the interplay between disorder, anharmonicity, and the quantum Bose-Einstein statistics of atomic vibrations determines thermal conductivity.
This Wigner formulation argues for a preferential phase convention for the 
dynamical matrix in the reciprocal Bloch representation and related off-diagonal velocity operator's elements; such convention is the only one yielding 
a conductivity which is invariant with respect to the non-unique choice of the crystal's unit cell and is size-consistent. 
We rationalize the conditions determining the crossover from particle-like to wave-like heat conduction, showing that phonons below the Ioffe-Regel limit (\textit{i.e.} with a mean free path shorter than the
interatomic spacing) contribute to heat transport due to their wave-like capability to interfere and tunnel.
%Finally, we show that the present approach allows to accurately predict the ultralow and glass-like thermal conductivity of complex crystals used for thermal barrier coatings and thermoelectrics, and relate 
%We showcase these findings with first-principles calculations in crystals with ultralow or glass-like thermal conductivity (used for thermal barrier coatings or thermoelectric energy conversion), showing that in these materials a significant number of phonons 
Finally, we show that the present approach overcomes the failures of the Peierls-Boltzmann formulation for \replaced{crystals}{materials} with ultralow or glass-like thermal conductivity, with case studies of materials for thermal barrier coatings and thermoelectric energy conversion.
\end{abstract}

\maketitle

\section{Introduction} % (fold)
\label{sec:intro}
In 1929 Peierls   \cite{peierls1929kinetischen} formulated the phonon Boltzmann transport equation (BTE) to explain heat conduction in crystalline solids, envisioning that in crystals the microscopic heat carriers are phonon wavepackets that diffuse and scatter as if they were particles of a gas. 
Recent computational advances have allowed to compute the parameters entering in the linearized form of the BTE (LBTE) from first principles   \cite{RevModPhys.73.515,paulatto2013anharmonic,li2014shengbte,hiphive,phono3py}, and to solve it either approximately, in the so-called single mode approximation   \cite{PhysRevLett.106.045901}, or exactly, using iterative   \cite{omini1995iterative,broido2007intrinsic,carrete2017almabte}, variational   \cite{fugallo2013ab}, or exact diagonalization   \cite{PhysRevLett.110.265506,phono3py,cepellotti2016thermal,PhysRevX.10.011019} methods.
Several studies     \cite{broido2007intrinsic,PhysRevLett.106.045901,Esfarjani_PRB_11,Chen_Science_12,Nanoscale_Thermal_14,mcgaughey2019phonon} have highlighted the accuracy of the LBTE in ``simple crystals'', \added{\textit{i.e.} crystals}  characterized by phonon interband spacings much larger than the linewidths.
Notably, in absence of disorder 
\replaced{the LBTE predicts the relation between the temperature and the thermal conductivity ($\kappa$) to follow a universal asymptotic decay $\kappa(T){\sim}T^{-m}$ for $T$ larger than the Debye temperature, 
where $m{=}1$ when anharmonic three-phonon interactions are the dominant source of thermal resistance \cite{ziman1960electrons}. Even faster decays ($m{>}1$) are possible when higher-order phonon scattering processes become relevant \cite{PhysRevB.96.161201}. This is in marked contrast to the much milder decrease in conductivity, or even the increase, observed in disordered or glassy materials as a function of temperature \cite{li2015ultralow,
lee2017ultralow,lory2017direct}.}{the LBTE predicts the temperature ($T$)-thermal conductivity ($\kappa$) relation to follow a universal asymptotic decay $\kappa(T){\sim}T^{-1}$ for $T$ larger than the Debye temperature;  this is at variance with the much milder decays or temperature-independent trend observed in disordered or glassy materials.}
%\added{ and when anharmonic three-phonon scattering processes are the dominant source of phonon scattering, and when the anharmonic interactions at the lowest third order are the dominant source of phonon scattering, } the LBTE predicts the temperature ($T$)-thermal conductivity ($\kappa$) relation to follow a universal asymptotic decay $\kappa(T){\sim}T^{-1}$ for $T$ larger than the Debye temperature \cite{ziman1960electrons}; 
An attempt to explain the microscopic heat-conduction mechanisms in glasses was made by Kittel in 1949 \cite{Kittel_thermal_cond_glasses}, who introduced a phenomenological model where atomic vibrations have a constant mean free path that is determined by the disorder length scale; however, this model lacks rigorous validation and has aroused some controversy   \cite{Freeman_Anderson,allen1989thermal,allen1993thermal,allen1999diffusons}, since it describes nonperiodic glasses through Peierls’ theory for periodic crystals.
A key step forward was made by Allen and Feldman in 1989   \cite{allen1989thermal}: they envisioned that 
in disordered systems heat diffuses in a wave-like fashion, specifically through a Zener-like tunneling between quasi-degenerate vibrational eigenstates, with a tunneling strength dependent on the off-diagonal elements of the velocity operator \cite{hardy1963energy,physrevb.29.2884}.
Notably, this formulation reproduces the temperature-conductivity curves measured in several glasses \cite{allen1989thermal,allen1993thermal,allen1993thermal,Feldman1995,allen1999diffusons}, which increase up to saturation in the high-temperature limit.

Recent work has shown that the LBTE fails to describe materials with ultralow thermal conductivity   \cite{donadio2009atomistic,li2015ultralow,lee2017ultralow,chen2015twisting,Weathers_PRB_2017,lory2017direct,Mukhopadhyay1455}, leading to the speculation that 
\replaced{wave-like transport mechanisms might emerge and coexist with particle-like conductivity}{the particle- and wave-like transport mechanisms discussed above might coexist} \cite{donadio2009atomistic,chen2015twisting,Mukhopadhyay1455,zhu2019mixed}. 
\deleted{In addition, the approach of Allen and Feldman for glasses 
does not account for anharmonicity; thus, it is supposed to be accurate only in the low-temperature (harmonic) regime.} 
%However, it has been shown that in complex amorphous composites with strong anharmonicity the harmonic theory is not accurate even at low temperatures \cite{McGaughey2009predicting}. 
\replaced{Conversely}{Last}, particle-like propagation mechanisms have been suggested also for glasses (propagons, albeit without a formal justification) in order to rationalize experimental results \cite{allen1999diffusons}.
At variance with these formulations, the \nametheory{} transport equation we recently introduced \cite{simoncelli2019unified} 
naturally encompasses the emergence and coexistence of particle-like and wave-like conduction mechanisms, providing a unified approach to heat-transport phenomena in solids, including crystals (where particle-like propagation dominates), glasses (where wave-like tunneling dominates), and all intermediate cases.

Here, we discuss the theoretical foundations for the derivation of the Wigner thermal transport equation 
from the phase-space formulation of quantum mechanics.  
We introduce a general theoretical framework, grounded on Wigner's formalism~\cite{wigner1932,moyal1949quantum}, that allows to describe  transport phenomena in solids in a very general and convenient way.
We employ this to describe thermal transport, showing how this formulation leads to a well defined microscopic energy field and  related microscopic heat flux, and we use the latter to derive the thermal conductivity from the solution of the Wigner transport equation.
We highlight a subtle aspect of this framework, leading to a well defined ``Wallace'' \cite{wallace1972thermodynamics} phase convention that provides a formulation that is size-consistent and invariant with respect to any possible choice of a crystal's unit cell.
We elucidate the physics underlying the crossover from particle-like to wave-like heat-conduction mechanisms and provide a quantitative criterion to distinguish the regimes where particle-like or wave-like conduction mechanisms are dominant.
Specifically, we show that a ``Wigner limit in time'' naturally emerges from this formulation as the time scale determining when particle-like and wave-like conduction mechanisms coexist and are equally important. We discuss how such a time scale is related to the gaps between the energy levels of atomic vibrations, or of the phonon bands in a crystal. 
We rely on these findings to introduce a regime diagram for thermal transport which allows to determine the theoretical framework needed to describe transport in a solid just from the knowledge of its vibrations' frequencies and lifetimes.
Moreover, we demonstrate that atomic vibrations with a lifetime comparable to the Wigner limit in time have a mean free path of the order of the typical interactomic spacings, a length scale which is known in the literature as the ``Ioffe-Regel limit in space''~\cite{iofferegel1960,allen1999diffusons}, and can be used to assess phenomenologically the validity of particle-like transport formulations (\textit{i.e.} kinetic theories in which microscopic carriers propagate particle-like as in the Peierls-Boltzmann transport equation)  \cite{Kittel_thermal_cond_glasses,allen1999diffusons,Mukhopadhyay1455,luo2020vibrational}.

The paper is organized as follows. 
In Sec.~\ref{sec:preliminaries} we briefly summarize the key quantities needed to model thermal transport in dielectric crystals using the well known semiclassical Peierls-Boltzmann formulation.
In Sec.~\ref{sec:density_matrix_formalism_for_atomic_vibrations} we introduce the quantities needed to describe thermal transport beyond the semiclassical Peierls-Boltzmann approach, 
presenting a second-quantization formalism for atomic vibrations 
in real space.
We use this formalism to describe a system driven out of equilibrium by a temperature gradient, where the one-body density matrix is space-dependent and undergoes a Markovian irreversible evolution.
In Sec.~\ref{sec:Wigner_thermal_transport_equation} we discuss the 
connection between the density matrix and Wigner phase-space formalisms.
We show that the Wigner formalism is particularly convenient to describe transport phenomena in solids, and we employ it to derive the \nametheory{} transport equation, which 
generalizes the Peierls-Boltzmann equation accounting not only for the particle-like propagation of phonon wavepackets but also for their wave-like tunneling between different bands. 
In Sec.~\ref{sec:solution_of_the_wigner_equation_and_generalized_conductivity_formula} we use the Wigner formalism to derive an expression for the microscopic harmonic energy field and the related microscopic harmonic heat flux. We discuss thoroughly the linearized form of the Wigner transport equation (\RedNmTh), showing how its solution determines the microscopic heat flux and a general \RedNmTh{} thermal conductivity expression that accounts for both particle- and wave-like conduction mechanisms. 
We show that particle-like conduction mechanisms dominate in  ``simple crystals'' having phonon interband spacings much larger than the linewidths, implying that in this regime the \RedNmTh{} conductivity becomes equivalent to the Peierls-Boltzmann conductivity for weakly anharmonic crystals~\cite{peierls1929kinetischen,peierls1955quantum}.
In contrast,  \replaced{we show analytically that in the limiting case of a disordered harmonic solid, wave-like mechanisms are dominant and the}{we show that wave-like mechanisms are dominant in harmonic glasses, where the} \RedNmTh{} conductivity becomes equivalent to the Allen-Feldman formula for the conductivity of glasses~\cite{allen1989thermal}.
Most importantly, we show that the \RedNmTh{} conductivity covers in all generality all intermediate cases, including ``complex crystals''  having phonon interband spacings comparable or smaller than the linewidths and ultralow or glass-like conductivity.
In Sec.~\ref{sec:size_consistency_in_silicon_supercell} 
we show that a phase convention (discussed e.g. in the book of Wallace \cite{wallace1972thermodynamics}) 
needs to be employed in the derivation of the Wigner transport equation (in particular, in the computation of the dynamical matrix in the reciprocal Bloch representation and of the related off-diagonal velocity operator's elements).
We report numerical calculations demonstrating that only with such convention the thermal conductivity is size-consistent and invariant with respect to the multiple choices for a crystal's unit cell.
In Sec.~\ref{sec:numerical_results} we show that the present approach allows to predict the ultralow thermal conductivity of complex crystals used for thermal barrier coatings and thermoelectrics, with applications to the zirconate La$_2$Zr$_2$O$_7$ and the perovskite CsPbBr$_3$ as materials representative of these classes.
In Sec.~\ref{sec:Particle_wave_crossover_phonons} we discuss how the \RedNmTh{} predicts the coexistence of particle-like and wave-like heat-conduction mechanisms, with a crossover between phonons that mainly propagate particle-like and phonons that mainly %interfere and 
tunnel wave-like. We show how phonons at the center of such crossover have a lifetime approximately equal to the Wigner limit in time, and a mean free path approximately equal to the Ioffe-Regel limit in space (\textit{i.e.} the  typical interatomic spacing).
{Finally, Sec.~\ref{sec:software_implementation} discusses the
implementation in computer programs of the general \RedNmTh{} thermal conductivity formula, providing numerical recipes
apt to modify  LBTE solvers~\cite{alamode,li2014shengbte,carrete2017almabte,phono3py,fugallo2013ab,phonts} with minimal effort.}

\section{Preliminaries} % (fold)
\label{sec:preliminaries}
We start by considering a 3-dimensional bulk crystal, \textit{i.e.} an infinite lattice with a basis having primitive vectors $\bm{a}_\alpha$  ($\alpha{=}1,{\dots},3$ is a Cartesian index), Bravais lattice vectors $\bm{R}=n_\alpha\bm{a}_\alpha$ ($n_\alpha\in \mathbb{Z}\; \forall \alpha$), a primitive cell of volume $\mathcal{V}$ containing $N_{\rm at}$ atoms at positions $\bm{\tau}_b\in \mathcal{V}$ ($b{=}1,{\dots},N_{\rm at}$ is an atomic label) \cite{ziman1960electrons,wallace1972thermodynamics,srivastava1990physics}. 
Here with ``primitive cell'' we mean \added{as usual} a minimal-cardinality set of atoms whose periodic repetition allows to describe the crystal; \replaced{with}{in the following we will use the term} ``unit cell'' \deleted{to denote} 
a set of atoms, not necessarily of minimal cardinality, whose periodic repetition allows to describe the crystal. \deleted{(e.g. a unit cell can be defined as a $2{\times}3{\times}2$ supercell of a primitive cell). Even if we are focusing first on crystals, it is worth mentioning that the assumptions above can be used also as a starting point to describe amorphous solids---these will be discussed later as limiting cases of disordered but periodic crystals in the limit of infinitely large primitive cells.}
We want to describe \added{here the ionic contribution to} thermal transport, \deleted{in non-metallic solids \textit{i.e.}} focusing on the evolution of the atomic vibrational energy in the presence of a temperature gradient. The quantum operator describing this \deleted{atomic vibrational energy} is the nuclear Hamiltonian, taken here in the Born-Oppenheimer approximation
\begin{equation}
  \hat{H}=\sum_{\bm{R},b,\alpha}{\frac{\hat{p}^2(\bm{R})_{b\alpha}}{2 M_{b}}} + \hat{V}\big(\{\hat{u}(\bm{R})_{b\alpha}\}\big);
  \label{eq:Born_Oppenheimer}
\end{equation}
where $\hat{p}(\bm{R})_{b\alpha}$ and $\hat{u}(\bm{R})_{b\alpha}$ are the \replaced{momentum and displacement-from-equilibrium operators}{associated with the momentum and with the displacement from equilibrium} \footnote{Formally, the deviation-from-equilibrium operator is $\hat{u}(\bm{R})_{\hspace*{-0.3mm}b\alpha}{=}\hat{{x}}(\bm{R})_{b\alpha}{-}{{x}}^0_{\bm{R}b\alpha}\hat{\bm{1}}$, where $\hat{{x}}(\bm{R})_{b\alpha}$ is the $\alpha$ component of the position operator of the nuclei $b$ in unit cell $\bm{R}$, and ${x}^0_{\bm{R}b\alpha}$ the corresponding constant equilibrium position.} for atom $b$  along the Cartesian direction $\alpha$ in the primitive cell labeled by the Bravais vector $\bm{R}$. $M_{b}$ is the mass of the atom $b$ and $\hat{V}(\{\hat{u}(\bm{R})_{b\alpha}\})$ is the lattice-periodic 
interatomic potential, which depends on all the displacement operators. The usual canonical commutation relations are satisfied:
\begin{equation}
  [\hat{u}(\bm{R})_{b\alpha},\hat{p}(\bm{R}')_{b'\hspace*{-0.5mm}\alpha'}]{=}i\hbar \delta_{\bm{R},\bm{R}'}\delta_{b,b'}\delta_{\alpha,\alpha'}.
  \label{eq:canonical_comm}
\end{equation}
In a solid, atoms oscillate around their equilibrium positions, allowing to approximate the potential with a $n$-th order Taylor expansion of the displacement operators $\{\hat{u}(\bm{R})_{b\alpha}\}$. 
For now it is sufficient to consider explicitly the leading (second-order, or harmonic) term in such expansion:
\begin{equation}
\begin{split}
  {V}\big(\{{u}(\bm{R})_{b\alpha}\} \big){=}&\frac{1}{2}\!\!\!\sum_{\substack{\scriptscriptstyle{\bm{R},b,\alpha}\\ \scriptscriptstyle{\bm{R'}\!,b'\!,\alpha'}} } \!\!
\frac{\partial^2 {V} }{\partial {u}(\bm{R})_{b\alpha} \partial {u}(\bm{R}')_{b'\hspace*{-0.5mm}\alpha'}\! }\bigg\lvert_{\!\rm eq}\!\!
{u}(\bm{R})_{b\alpha}
{u}(\bm{R}')_{b'\hspace*{-0.5mm}\alpha'}\\
  &{+}\mathcal{O}(
  {u}(\bm{R})_{\hspace*{-0.3mm}b\alpha}
 {u}(\bm{R}')_{\hspace*{-0.3mm}b'\hspace*{-0.5mm}\alpha'} 
 {u}(\bm{R}'\!')_{\hspace*{-0.3mm}b'\!'\hspace*{-0.5mm}\alpha'\!'}
 ),
  \label{eq:expansion_pot}
  \raisetag{5mm}
\end{split}
\end{equation}
where the zeroth-order term can be taken as reference for the energy (\textit{i.e.} equal to zero) without loss of generality, the first-order term  %the first-order derivative 
vanishes when evaluated at equilibrium \cite{ziman1960electrons}, and $\mathcal{O}$ represents the higher-order terms in the displacements.
We consider the atomic displacements from equilibrium to be small and below Lindemann's threshold \cite{Lindemann,hofmann2015solid}, as it is the case in solids.
Using perturbation theory one can decompose the Hamiltonian~(\ref{eq:Born_Oppenheimer}) as the sum of a leading Hamiltonian and a perturbation, $\hat{H}{=}\hat{H}^{\rm har}{+}\hat{H}^{\rm per}$, where $\hat{H}^{\rm har}$ is the leading Hamiltonian given by Eq.~(\ref{eq:Born_Oppenheimer}) limiting $\hat{V}$ to the harmonic term, 
and $\hat{H}^{\rm per}$ is the perturbation in the potential due to anharmonic third- or higher-order terms in Eq.~(\ref{eq:expansion_pot}).
It is worth mentioning that also the presence of isotopic mass disorder can be treated as a kinetic-energy perturbation and taken into account in $\hat{H}^{\rm per}$ \cite{tamura1983isotope,PhysRevLett.106.045901}.
We first focus on the description of the leading Hamiltonian $\hat{H}^{\rm har}$; the perturbation introducing transitions between the eigenstates of $\hat{H}^{\rm har}$ will be discussed later.
It is useful to recast the harmonic coefficients in Eq.~(\ref{eq:expansion_pot}) in terms of the tensor of mass-renormalized interatomic force constants commonly employed in the literature \cite{wallace1972thermodynamics,revmodphys_maradudin68},
\begin{equation}
\tenscomp{G}_{\bm{R}b\alpha,\bm{R'}b'\hspace*{-0.5mm}\alpha'}{=}\frac{1}{\sqrt{M_bM_{b'}}}\frac{\partial^2 {V} }{\partial {u}(\bm{R})_{b\alpha} \partial {u}(\bm{R}')_{b'\hspace*{-0.5mm}\alpha'} }\Big\lvert_{\rm eq},
\label{eq:matrix_G}
\end{equation}
which is symmetric and translation invariant:
\begin{align}
  &\tenscomp{G}_{\bm{R}b\alpha,\bm{R'}b'\hspace*{-0.5mm}\alpha'}{=}\tenscomp{G}_{\bm{R'}b'\hspace*{-0.5mm}\alpha',\bm{R}b\alpha},
  \label{eq:symmetry}\\
  &\tenscomp{G}_{\bm{R}b\alpha,\bm{R'}b'\hspace*{-0.5mm}\alpha'}{=}\tenscomp{G}_{(\bm{R-R'})b\alpha,\bm{0}b'\hspace*{-0.5mm}\alpha'}{=}\tenscomp{G}_{\bm{0}b\alpha,(\bm{R'-R})b'\hspace*{-0.5mm}\alpha'}.
  \label{eq:translation_inv}
\end{align}
{In fact, as extensively discussed in textbooks \cite{wallace1972thermodynamics,srivastava1990physics}, at the leading harmonic order atomic vibrations in crystals can be  \replaced{decomposed in a linear combination}{as a superimposition} of the normal modes of vibration (phonons), and the properties of  these normal modes can be obtained computing the Fourier transform of the mass-renormalized interatomic force constants~(\ref{eq:matrix_G}), hereafter referred to as ``dynamical matrix in the reciprocal Bloch representation''
\begin{equation}
  \tenscomp{D}(\bm{q})_{b\alpha,b'\!\alpha'}{=}\sum_{\bm{R}}\tenscomp{G}_{\bm{R}b\alpha,\bm{0}b'\!\alpha'}e^{-i\bm{q}\cdot(\bm{R}{+}\bm{\tau}_b{-}\bm{\tau}_{b'})},
  \label{eq:dynamical}
\end{equation}
where $\bm{q}$ is a wavevector that can be restricted to the first Brillouin zone of the crystal $\mathfrak{B}$, and \deleted{we note that} the sum in Eq.~(\ref{eq:dynamical}) runs over a single lattice vector due to \deleted{the} translation invariance~(\ref{eq:translation_inv}) \cite{wallace1972thermodynamics}.
\deleted{Specifically} The 
eigenvalues $\omega^2(\bm{q})_s$, where $s$ is a vibrational-mode index running from 1 to $3\; N_{\rm at}$, and eigenvectors $\mathcal{E}(\bm{q})_{s,b\alpha}$ of the dynamical matrix~(\ref{eq:dynamical})
\begin{equation}
  \sum_{b'\alpha'}\tenscomp{D}(\bm{q})_{b\alpha,b'\!\alpha'}\mathcal{E}(\bm{q})_{s,b'\!\alpha'}=\omega^2(\bm{q})_s\mathcal{E}(\bm{q})_{s,b\alpha},
  \label{eq:diagonalization_dynamical_matrix}
\end{equation}
provide the vibrational frequencies $\omega(\bm{q})_s$ and displacement patterns of the normal modes (we recall that $\mathcal{E}(\bm{q})_{s,b\alpha}$ describes how atom $b$ moves along the Cartesian direction $\alpha$ when the phonon with wavevector $\bm{q}$ and mode $s$ is excited). 
In Eq.~(\ref{eq:dynamical}) the dynamical matrix is computed using the ``smooth phase convention'' employed e.g. in the book of Wallace~\cite{wallace1972thermodynamics}, where the phases depend on the atomic positions in real space ($\bm{R}{+}\bm{\tau}_b{-}\bm{\tau}_{b'}$).
It is worth mentioning that a different ``step-like'' phase convention is \added{often} employed in the literature to compute the dynamical matrix, where the phases depend only on the Bravais lattice vector $\bm{R}$ and thus vary discontinuously (\replaced{as explained in the following, see also}{see e.g.} the book of Ziman~\cite{ziman1960electrons}),
\begin{equation}
  \underline{\tenscomp{D}(\bm{q})}_{b\alpha,b'\!\alpha'}{=}\sum_{\bm{R}}\tenscomp{G}_{\bm{R}b\alpha,\bm{0}b'\!\alpha'}e^{-i\bm{q}\cdot\bm{R}}.
  \label{eq:dynamical_Ziman}
\end{equation}
Hereafter we will refer to the matrix $\tenscomp{D}(\bm{q})_{b\alpha,b'\!\alpha'}$ (Eq.~(\ref{eq:dynamical})) as the ``smooth dynamical matrix'', and to the other matrix $\underline{\tenscomp{D}(\bm{q})}_{b\alpha,b'\!\alpha'}$ (Eq.~(\ref{eq:dynamical_Ziman})) as the ``step-like dynamical matrix''. 
\replaced{It can be shown analytically}{It is easy to show}  that these two dynamical matrices are related by a unitary transformation $\tenscomp{U}(\bm{q})_{b\alpha,b'\alpha'}{=}e^{+i\bm{q}\cdot\bm{\tau}_{b}}\delta_{\alpha,\alpha'}\delta_{b,b'}$,
\begin{equation}
\begin{split}
\underline{\tenscomp{D}(\bm{q})}_{b\alpha,b'\!\alpha'}{=}
\sum_{b'\!'\!\alpha'\!',\bar{b}\bar{\alpha}}
    \tenscomp{U}(\bm{q})_{b\alpha,b'\!'\!\alpha'\!'}    {\tenscomp{D}(\bm{q})}_{b'\!'\!\alpha'\!',\bar{b}\bar{\alpha}} \tenscomp{U}^\dagger(\bm{q})_{\bar{b}\bar{\alpha},b'\alpha'};
%{=}e^{+i\bm{q}\cdot\bm{\tau}_b} {\tenscomp{D}(\bm{q})}_{b\alpha,b'\!\alpha'}e^{-i\bm{q}\cdot\bm{\tau}_{b'}};
\label{eq:relation_dynamical_matrices}
\end{split}
\raisetag{13mm}
\end{equation}
as such, they have the same eigenvalues $\omega^2(\bm{q})_s$ and their eigenvectors are related by the unitary transformation 
\begin{equation}
	\underline{\mathcal{E}(\bm{q})}_{s,b\alpha}{=}
\sum_{b'\alpha'} \tenscomp{U}(\bm{q})_{b\alpha,b'\alpha'} \mathcal{E}(\bm{q})_{s,b'\alpha'}.
  %{=}\mathcal{E}(\bm{q})_{s,b\alpha}e^{i\bm{q}\cdot\bm{\tau}_b}.
	\label{eq:eigenvectors_conventions_relation}
\end{equation}
%}
While both these conventions allow to describe the equilibrium vibrational properties of materials ({e.g.} the phonon spectrum), we will show later that the smooth phase convention (``Wallace'') has to be used to describe the out-of-equilibrium propagation of vibrations within the Wigner framework discussed here.

The derivative with respect to wavevector of the smooth dynamical matrix~(\ref{eq:dynamical}) is related to the propagation (group) velocity $\tens{v}^\beta(\bm{q})_s$ of the phonon wave packet centered at wavevector $\bm{q}$ and having mode $s$, which 
in absence of degeneracies ($\omega(\bm{q})_s{\neq}\omega(\bm{q})_{s'}$ for $s{\neq}s'$) 
is~\cite{fugallo2013ab,wang2008quantum}
\begin{equation}
\begin{split}
  \tenscomp{v}^\beta(\bm{q})_{s}{=}\frac{1}{2\omega(\bm{q})_s}
  \!\!\!\!\sum_{\;\;{b\alpha,b'\!\alpha'}\!\!}\!\!\!\!
  \mathcal{E}^\star(\bm{q})_{s,b\alpha} \frac{\!\partial \tenscomp{D}(\bm{q})_{b\alpha,b'\!\alpha'\!\!}}{\partial q^\beta}\mathcal{E}(\bm{q})_{s,b'\!\alpha'\!}.
  \end{split}
  \raisetag{3.5mm}
  \label{eq:group_velocity}
\end{equation}
{In the presence of degeneracies  ($\omega(\bm{q})_s{=}\omega(\bm{q})_{s'}$ for $s{\neq}s'$), the eigenvectors appearing  in Eq.~(\ref{eq:group_velocity}) \replaced{should}{must} be chosen so that the matrix 
$\tenscomp{m}^\beta(\bm{q})_{s,s'}{=}\tfrac{1}{2\omega(\bm{q})_s}
  \sum_{{b\alpha,b'\!\alpha'}}\!
  \mathcal{E}^\star(\bm{q})_{s,b\alpha} \tfrac{\!\partial \tenscomp{D}(\bm{q})_{b\alpha,b'\!\alpha'\!\!}}{\partial q^\beta}\mathcal{E}(\bm{q})_{s',b'\!\alpha'\!}$ is diagonal in the indexes $s,s'$ corresponding to degenerate frequencies \cite{fugallo2013ab}.
We will show later that in the Wigner framework quantities related to the non-degenerate off-diagonal elements of the matrix $\tenscomp{m}^\beta(\bm{q})_{s,s'}$ become relevant. 
In this regard, it is possible to highlight already now that the smooth phase convention appears to be more suitable for the computation of these off-diagonal elements. For example, the smooth dynamical matrix~(\ref{eq:dynamical}) 
\replaced{is not changed if}{is invariant with respect to} using two primitive cells that differ \replaced{just by}{because of} one atom rigidly shifted by a Bravais-lattice vector $\bm{R'}$.
This can be verified \deleted{straightforwardly} by replacing $\bm{\tau}_{b'}\to \bm{\tau}_{b'}{+}\bm{R'}$  in Eq.~(\ref{eq:dynamical}), and \deleted{noting that this produces no changes, since one can exploit} \added{exploiting} translation invariance. \deleted{and the change of variable $\bm{R}{+}\bm{R'}\to\bm{R'\!'}$ to rewrite the resulting expression exactly as Eq.~(\ref{eq:dynamical}).}
As a consequence, also the eigenvectors of the smooth dynamical matrix~(\ref{eq:diagonalization_dynamical_matrix}), and thus the off-diagonal elements of the matrix $\tenscomp{m}^\beta(\bm{q})_{s,s'}$, do not vary. 
In contrast, the step-like dynamical matrix~(\ref{eq:dynamical_Ziman}) and its eigenvectors~(\ref{eq:eigenvectors_conventions_relation}) do not satisfy this invariance; \deleted{property. This can be verified by} applying the same \deleted{single-atom} shift \deleted{discussed above} to Eq.~(\ref{eq:dynamical_Ziman}) \replaced{leads to a different result.}{, and noting that now the exponential does not account for this shift and thus the resulting expression cannot be rewritten in the same form of the original one.}

The quantities described up to now \replaced{enter into}{parametrize} the well known semiclassical model for thermal transport in crystals developed by Peierls \cite{peierls1929kinetischen,peierls1955quantum}, in which the dynamics of out-of-equilibrium and space-dependent populations of phonon wavepackets ($\tenscomp{n}({\bm{r}},{\bm{q}},t)_{s}$, here $\bm{r}$ is a continuous Cartesian coordinate) is determined by the Boltzmann transport equation:
\begin{equation}
\begin{split}
  &\frac{\partial  }{\partial  t} \tenscomp{n}({\bm{r}},{\bm{q}},t)_{s}
+{\tens{v}}(\bm{q})_{s}{\cdot}{\nabla}_{\bm{r}} \tenscomp{n}(\bm{r},\bm{q},t)_{s}=\frac{\partial \tenscomp{n}({\bm{r}},{\bm{q}},t)_{s} }{\partial  t} \bigg|_{\hat{H}^{\rm per}}\!\!\!\!\!,\hspace*{10mm}
\end{split}
\raisetag{8mm}
\label{eq:BTE_std}
\end{equation}
where \deleted{the dot represents the scalar product between vectors, and} $\frac{\partial \tenscomp{n}({\bm{r}},{\bm{q}},t)_{s} }{\partial  t} \big|_{\hat{H}^{\rm per}}$ is the phonon scattering  superoperator that originates from $\hat{H}^{\rm per}$ \cite{fugallo2013ab,ziman1960electrons} and that will be discussed later. 
The Peierls-Boltzmann equation~(\ref{eq:BTE_std}) is derived under the assumption that phonon wavepackets drift akin to the particles of a classical gas in the presence of a temperature gradient \cite{peierls1955quantum}.
Mathematically, this is apparent from the left-hand side of Eq.~(\ref{eq:BTE_std}), which describes phonons' drift and has a mathematical form analogous to the drift term of the Boltzmann equation for the particles of a classical gas \cite{cercignani1988boltzmann}.
In the next sections we discuss a formalism that allows to generalize Peierls' model beyond this particle-like behavior, and we discuss the conditions under which the Peierls-Boltzmann equation~(\ref{eq:BTE_std}) fails and this generalization is relevant.

\section{Quantum description of atomic vibrations }
\label{sec:density_matrix_formalism_for_atomic_vibrations}
In this section we discuss a second-quantization formalism for atomic vibrations in real space that is suitable to define the one-body density matrix of an out-of-equilibrium system, {where a temperature gradient 
enforces a space-dependent atomic vibrational energy.}

\subsection{Second quantization for atomic vibrations }
\label{sub:bosonic_operators_in_direct_and_reciprocal_space}

The leading Hamiltonian $\hat{H}^{\rm har}$ is quadratic in the momentum and displacement operators, and thus
represents a many-body system of  harmonic oscillators.
With the goal of exploiting the second-quantization formalism to simplify the description of this many-body problem, 
and tracking the space-dependent atomic vibrations that characterize the regime where a temperature gradient drives thermal transport (\textit{i.e.} the regime in which atoms vibrate more in the warm region and less in the cold region),
we introduce the class of bosonic operators in real space $\hat{a}(\bm{R})_{b\alpha}$ %\cite{simoncelli2019unified}
\begin{equation}
  \begin{split}
\hat{a}(\bm{R})_{b\alpha}{=}\frac{1}{\!\sqrt{2 \hbar}}\!\sum_{\bm{R'}\!,b'\!,\alpha'}\!\Big(&\!\sqrt[4]{\!\tenscomp{G}^{{-}1}\!}_{\!\bm{R}b\alpha,\bm{R'}b'\!\alpha'\!}\frac{{\hat{p}}(\!\bm{R'})_{b'\!\alpha'\!}}{\sqrt{M_{b'}}}
 \\&{-}i \sqrt{\!M_{b'}}{\hat{u}}(\!\bm{R'})_{b'\!\alpha'\!} \sqrt[4]{\!\tenscomp{G}}_{\bm{R'}b'\!\alpha'\!,\bm{R}b\alpha}\!\Big),
  \end{split}
\raisetag{13mm}
  \label{eq:def_op_a_real}
\end{equation}
that are constructed so that $\hat{a}(\bm{R})_{b\alpha}$ annihilates a vibration 
centered around atom $b$ in the unit cell $\bm{R}$ and along the Cartesian direction $\alpha$.
In Eq.~(\ref{eq:def_op_a_real}), the matrices $\sqrt[4]{\tenscomp{G}}_{\bm{R'}b'\!\alpha'\!,\bm{R}b\alpha}$ and $\sqrt[4]{\tenscomp{G}^{-1}}_{\bm{R}b\alpha,\bm{R'}b'\!\alpha'}$ are the fourth roots of the 
mass-rescaled harmonic interatomic force-constant matrix~(\ref{eq:matrix_G}) and its inverse; they satisfy the relations 
$\sum_{\!\bm{R'\!'}\!,b'\!'\!,\alpha'\!'}\!\sqrt[4]{\!\tenscomp{G}^{-1}}_{\bm{R}b\alpha,\bm{R'\!'}b'\!'\!\alpha'\!'}\sqrt[4]{\!\tenscomp{G}}_{\bm{R'\!'}b'\!'\!\alpha'\!',\bm{R'}b'\!\alpha'\!}{=}\delta_{\bm{R},\bm{R'}}\delta_{b,b'}\delta_{\alpha,\alpha'}\!$, and are related to the  matrix~(\ref{eq:matrix_G}) via
$\sum_{\bm{R'\!'}\!,b'\!'\!,\alpha'\!'}\sqrt[4]{\tenscomp{G}}_{\bm{R}b\alpha,\bm{R'\!'}b'\!'\!\alpha'\!'}\sqrt[4]{\tenscomp{G}}_{\bm{R'\!'}b'\!'\!\alpha'\!',\bm{R'}b'\!\alpha'\!}{=}\sqrt{\tenscomp{G}}_{\bm{R}b\alpha,\bm{R'\!}b'\!\alpha'\!}$ and 
$\sum_{\bm{R'\!'}\!,b'\!'\!,\alpha'\!'}\sqrt{\tenscomp{G}}_{\bm{R}b\alpha,\bm{R'\!'}b'\!'\!\alpha'\!'}\sqrt{\tenscomp{G}}_{\bm{R'\!'}b'\!'\!\alpha'\!',\bm{R'}b'\!\alpha'\!}{=}{\tenscomp{G}}_{\bm{R}b\alpha,\bm{R'\!}b'\!\alpha'\!}$.
These annihilation-of-localized-vibrations operators~(\ref{eq:def_op_a_real}) are labeled by $\bm{R},b,\alpha$, and  satisfy the bosonic commutation relations together with their adjoints (the creation-of-localized-vibration operators $\hat{a}^\dagger(\bm{R}')_{b'\alpha'}$) \footnote{This can be proved using Eq.~(\ref{eq:def_op_a_real}) and the canonical commutation relation~(\ref{eq:canonical_comm}).}
\begin{equation}
  [\hat{a}(\bm{R})_{b\alpha},\hat{a}^\dagger(\bm{R}')_{b'\alpha'}]{=}\delta_{\bm{R},\bm{R'}}\delta_{b,b'}\delta_{\alpha,\alpha'} .
  \label{eq:commutation_rel_direct}
\end{equation}
To better understand the physical action of the localized bosonic operators in~(\ref{eq:def_op_a_real}) and (\ref{eq:commutation_rel_direct}), we proceed as follows: first, we create a vibrational excitation on the ground state $\left|0\right>$ using the creation operator at position $\bm{R}{+}\bm{\tau}_{b}$, $\left|\psi_{\bm{R}b\alpha}\right>{=}\hat{a}^\dagger(\bm{R})_{b\alpha}\left|0\right>$;
then, we look at how the atom at the generic position $\bm{R'}{+}\bm{\tau}_{b'}$ vibrates in such an excited state, computing the expectation value of the squared atomic displacement operator $\hat{u}^2(\bm{R'})_{b'\alpha'}$.
Specifically, from Eq.~(\ref{eq:def_op_a_real}) it follows that the displacement-from-equilibrium operator for an atom can be written as:
\begin{equation}
\begin{split}
   \hat{u}(\hspace*{-0.3mm}\bm{R'}\hspace*{-0.3mm})_{b'\!\alpha'}{=}i\!\!\!\sum_{\bm{R'\!'}\!,b'\!'\!,\alpha'\!'}\!\!\sqrt{\!\!\frac{\hbar}{\!2M_{b'}\!\!}}\!
    \sqrt[4]{\!\tenscomp{G}^{-\!1}\!}_{\!\bm{R'}\!b'\!\alpha'\!\!,\bm{R'\!'}\!b'\!'\!\alpha'\!'}\!\!\left[\!\hat{a}(\hspace*{-0.3mm}\bm{R'\!'}\hspace*{-0.3mm})_{b'\!'\!\alpha'\!'} {-}\hat{a}^\dagger\!(\hspace*{-0.3mm}\bm{R'\!'}\hspace*{-0.3mm})_{b'\!'\!\alpha'\!'}\! \right]\!.
    \label{eq:displacement}
 \end{split}
 \raisetag{4mm}   
 \end{equation} 
 \begin{figure*}
  \centering
  \includegraphics[width=\textwidth]{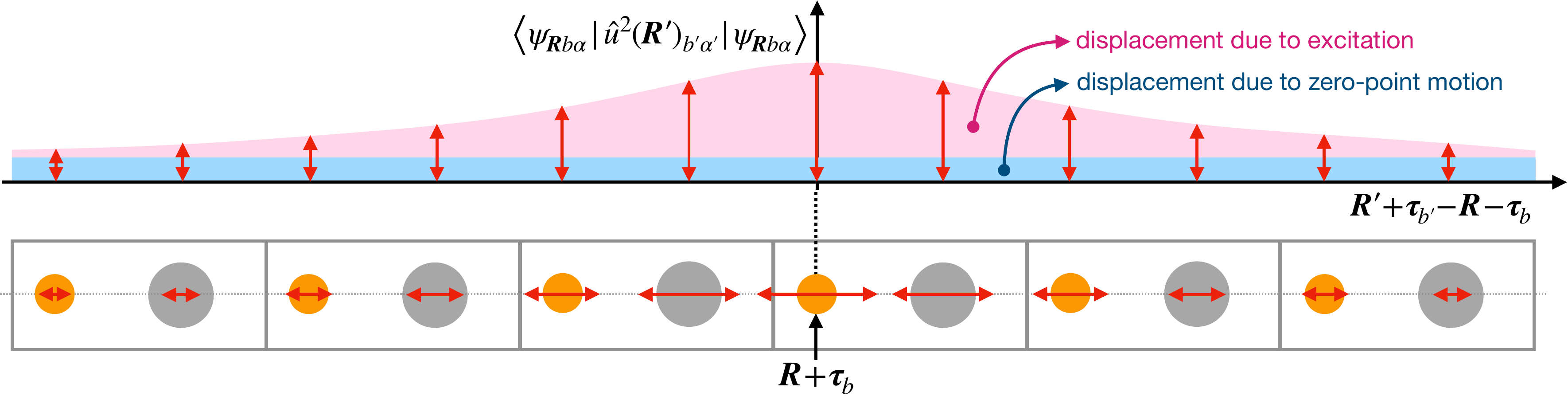}
  \vspace*{-5mm}
  \caption{
$\hat{a}^\dagger(\bm{R})_{b\alpha}$ \textbf{applied to the ground state $\left|0\right>$ creates an atomic vibration centered on the atom at position $\bm{R}{+}\bm{\tau}_b$.} 
The gray rectangles represent primitive cells of a lattice with basis, orange and gray circles symbolize atoms.
  The horizontal red lines show the mean square displacements of the atoms. The vertical lines have the same length of the horizontal lines below them, they show that in the excited state $\left|\psi_{\bm{R}b\alpha}\right>{=}\hat{a}^\dagger(\bm{R})_{b\alpha}\left|0\right>$ (where $\left|0\right>$ is the ground state), atoms at position $\bm{R'}{+}\bm{\tau}_{b'}$ have a mean square displacement (MSD) that becomes smaller  as the  distance $|\bm{R}{+}\bm{\tau}_{b}{-}\bm{R'}{-}\bm{\tau}_{b'}|$ increases.
  The envelope of the atomic MSD (vertical red lines) is decomposed into pink and blue, to emphasize that
the total MSD in the excited state is the result of the sum of two terms: a space-dependent MSD due to the excitation created by $\hat{a}^\dagger(\bm{R})_{b\alpha}$ (pink), and a space-independent MSD due to the zero-point motion (blue rectangle).}
  \label{fig:displ_real}
\end{figure*}
Then, the expectation value of the squared atomic displacement on the excited state is:
\begin{equation}
\begin{split}
    &\left<0\right|\hat{a}(\bm{R})_{b\alpha}|\hat{u}^2(\bm{R'})_{b'\alpha'}\big|\hat{a}^\dagger(\bm{R})_{b\alpha}\left|0\right>
    \\
    &=\frac{\hbar}{M_{b'}}\left(\sqrt[4]{\tenscomp{G}^{-1}}_{\bm{R'}b'\alpha',\bm{R}b\alpha}\right)^2 
    +\left<0\right|\hat{u}^2(\bm{R'})_{b'\alpha'}\left|0\right>,
\end{split}
  \label{eq:ref_excitation}
\end{equation}
where $\left<0\right|\hat{u}^2(\bm{R'})_{b'\alpha'}\left|0\right>= \frac{\hbar}{2M_{b'}}\sqrt{\tenscomp{G}^{-1}}_{\bm{0}b'\alpha',\bm{0}b'\alpha'}$ is the mean square displacement due to the zero-point motion.
{The force constant matrix $\tenscomp{G}_{\bm{R}b\alpha,\bm{R'}\!b'\!\alpha'}$ goes to zero for $|\bm{R}{+}\bm{\tau}_b{-}\bm{R'}{-}\bm{\tau}_{b'}|{\to}\infty$ \cite{RevModPhys.73.515}, and this implies that also $\sqrt[4]{\tenscomp{G}^{-1}}_{\bm{R'}\!b'\!\alpha'\!,\bm{R}b\alpha}$ goes to zero in the same limit \cite{SimoncelliPhD}.} 
Therefore, Eq.~(\ref{eq:ref_excitation}) shows that the bosonic operator $\hat{a}^\dagger(\bm{R})_{b\alpha}$ 
creates atomic vibrations along direction $\alpha$ and centered around the position $\bm{R}{+}\bm{\tau}_b$. In other words, after the action of $\hat{a}^\dagger(\bm{R})_{b\alpha}$ on the ground state, atoms close to $\bm{R}{+}\bm{\tau}_b$ have a larger mean square displacement than atoms far from $\bm{R}{+}\bm{\tau}_b$ (Fig.~\ref{fig:displ_real}). 
We note that an analogous calculation can be done for the expectation value of the square momentum operator; such a calculation is reported in Appendix~\ref{sub:app_momentum} and shows that considerations analogous to those for the mean square displacement reported here hold also for the mean square momentum.
We have thus shown how these bosonic operators~(\ref{eq:def_op_a_real}) can be used to describe space-dependent atomic vibrations.% in real space.

The localized bosonic operators~(\ref{eq:def_op_a_real}) allow also to write the leading harmonic Hamiltonian in second-quantized form as
\begin{equation}
\begin{split}
  \hat{H}^{\mathrm{har}}{=} \hbar\!\!\sum\limits_{\substack{\scriptscriptstyle{\bm{R},b,\alpha}\\\scriptscriptstyle{ \bm{R'}\!,b'\!,\alpha'} }}\!\! 
   \sqrt{\!\tenscomp{G}}_{\!\bm{R}b\alpha,\bm{R'}b'\hspace*{-0.5mm}\alpha'}
    \hat{a}^{\dagger}\!(\bm{R})_{b\alpha}\hat{a}(\bm{R'})_{b'\hspace*{-0.5mm}\alpha'}
{+}E_0,
%{+}\tfrac{\delta_{\bm{R},\bm{R'}}\delta_{b,b'}\delta_{\alpha,\alpha'}}{2}.
\end{split}
\raisetag{5mm}
\label{eq:H_real_space}
\end{equation}
where the additive constant $E_0{=}\tfrac{\hbar}{2}\sum_{\bm{R}b\alpha}\!\!
 \sqrt{\tenscomp{G}}_{\!\bm{R}b\alpha,\bm{R}b\alpha}$ represents the zero-point energy, which can be taken as reference energy and thus will be omitted in the following.
We see from Eq.~(\ref{eq:H_real_space}) that the harmonic vibrational energy is a one-body operator; therefore, to compute its expectation value the knowledge of the complex many-body density operator $\hat{\rho}$ is not needed, but it is sufficient to know the much simpler one-body density matrix 
\begin{equation}
  \varrho(\bm{R},\bm{R'},t)_{b\alpha,b'\hspace*{-0.5mm}\alpha'}={\rm Tr} \big(\hat{\rho}(t)\hat{a}^{\dagger}(\bm{R'})_{b'\hspace*{-0.5mm}\alpha'}\hat{a}(\bm{R})_{b\alpha}\big),
  \label{eq:one_body_def}
\end{equation}
{where ${\rm Tr}$ denotes the trace operation over the Fock space.}
In fact, it is \replaced{possible to show}{easy to verify} that 
\begin{equation}
\begin{split}
    {\rm Tr}\big(\hat\rho(t)\hat{H}^{\mathrm{har}} \big){=}
\hbar\!\sum\limits_{\substack{\scriptscriptstyle{\bm{R},b,\alpha}\\\scriptscriptstyle{ \bm{R'}\!,b'\!,\alpha'} }}\!\!
   \sqrt{\tenscomp{G}}_{\bm{R}b\alpha,\bm{R'}\!b'\!\alpha'\!}
   \varrho(\bm{R'}\!,\!\bm{R},t)_{b'\alpha',b\alpha}. %{+}E_0.\hspace*{6mm}
\end{split}
   \raisetag{5mm}
   \label{eq:expectation_value}
\end{equation}

Up to now we have managed to reduce the complexity of the many-body problem by introducing the localized bosonic operators in real space~(\ref{eq:def_op_a_real}) 
and the one-body density matrix~(\ref{eq:one_body_def}). 
Now we show that the formulation can be further simplified exploiting the invariance under translation of the $\tenscomp{G}$ tensor in~(\ref{eq:matrix_G}).
We start by defining the Fourier transform $\hat{a}(\bm{q})_{b\alpha}$ of the localized bosonic operator $\hat{a}(\bm{R})_{b\alpha}$
\begin{equation}
\begin{split}
  &\hat{a}(\bm{q})_{b\alpha}{=}{\textstyle\sum_{\bm{R}}}\hat{a}(\bm{R})_{b\alpha}e^{- i \bm{q}\cdot (\bm{R}+\bm{\tau}_b)}\\
  &{=}\frac{1}{\sqrt{2 \hbar}}\sum_{b'\alpha'}\hspace*{-0.8mm} \Big[\hspace*{-1mm}\sqrt[4]{\hspace*{-0.5mm}\tenscomp{D}^{\hspace*{-0.4mm}{-}1}\hspace*{-0.6mm}(\bm{q})}_{\hspace*{-0.5mm}b\alpha,b'\hspace*{-0.5mm}\alpha'}\hspace*{-0.5mm}{\hat{P}}(\bm{q})_{\hspace*{-0.4mm}b'\hspace*{-0.5mm}\alpha'} {-}i \hspace*{-0.5mm}\sqrt[4]{\hspace*{-0.5mm}\tenscomp{D}(\bm{q})}_{\hspace*{-0.5mm}b\alpha\hspace*{-0.2mm},\hspace*{-0.2mm}b'\hspace*{-0.5mm}\alpha'}\hspace*{-0.5mm}{\hat{U} }^\dagger\hspace*{-0.5mm}(\bm{q})_{\hspace*{-0.4mm}b'\hspace*{-0.5mm}\alpha'}\hspace*{-1mm}\Big],
\end{split}
\raisetag{14mm}
  \label{eq:bosonic_reciprocal}
\end{equation}
where ${\hat{U} }(\bm{q})_{b\alpha}{=}{\sqrt{M_b}}\sum_{\bm{R}} \hat{u}(\bm{R})_{b\alpha}e^{{+}i\bm{q}\cdot (\bm{R}+\bm{\tau}_b)}$ and ${\hat{P}}(\bm{q})_{b\alpha}{=}(M_b)^{-1/2}\sum_{\bm{R}} \hat{p}(\bm{R})_{b\alpha}e^{-i\bm{q}\cdot(\bm{R}+\bm{\tau}_b)}$ are the Fourier transforms of the displacement and momentum operators \footnote{the Fourier transforms of $\hat{u}(\bm{R})_{b\alpha}$ and $\hat{p}(\bm{R})_{b\alpha}$ have phases with opposite sign because these operators are canonically-conjugate \cite{ziman1960electrons}.}, respectively, 
$\sqrt[4]{\hspace*{-0.5mm}\tenscomp{D}(\bm{q})}_{\hspace*{-0.5mm}b\alpha\hspace*{-0.2mm},\hspace*{-0.2mm}b'\hspace*{-0.5mm}\alpha'}{=}
\sum_{\bm{R}}\sqrt[4]{\hspace*{-0.5mm}\tenscomp{G}}_{\bm{R}b\alpha\hspace*{-0.2mm},\bm{0}\hspace*{-0.2mm}b'\hspace*{-0.5mm}\alpha'}
e^{-i\bm{q}\cdot(\bm{R}{+}\bm{\tau}_b{-}\bm{\tau}_{b'})}$ and
$\!\sqrt[4]{\hspace*{-0.5mm}\tenscomp{D}^{-1}\!(\bm{q})}_{\hspace*{-0.5mm}b\alpha\hspace*{-0.2mm},\hspace*{-0.2mm}b'\hspace*{-0.5mm}\alpha'}{=}\!
\sum_{\bm{R}}\!\!\sqrt[4]{\hspace*{-0.5mm}\tenscomp{G}^{-1}}_{\!\bm{R}b\alpha\hspace*{-0.2mm},\bm{0}\hspace*{-0.2mm}b'\hspace*{-0.5mm}\alpha'\!}
e^{-i\bm{q}\cdot(\bm{R}+\bm{\tau}_b-\bm{\tau}_{b'}\!)}$ 
are the Fourier transforms of the fourth root of the mass-rescaled harmonic interatomic force-constant matrix and of its inverse. 
It is \replaced{possible}{easy} to verify that, analogously to the matrix $\sqrt[4]{\hspace*{-0.5mm}\tenscomp{G}}$, 
also the fourth root of the smooth dynamical matrix in reciprocal Bloch representation 
satisfies $\sqrt{\hspace*{-0.5mm}\tenscomp{D}(\bm{q})}_{b\alpha,b'\!\alpha'\hspace*{-0.2mm}}{=}\sum_{b'\!'\!,\alpha'\!'}\sqrt[4]{\hspace*{-0.5mm}\tenscomp{D}(\bm{q})}_{\hspace*{-0.5mm}b\alpha\hspace*{-0.2mm},\hspace*{-0.2mm}b'\!'\hspace*{-0.5mm}\alpha'\!'} \sqrt[4]{\hspace*{-0.5mm}\tenscomp{D}(\bm{q})}_{b'\!'\hspace*{-0.5mm}\alpha'\!'\!,b'\!\alpha'\hspace*{-0.2mm}}$, 
and an analogous calculation allows to obtain the smooth dynamical matrix~(\ref{eq:dynamical}) from its square root. Analogous considerations hold for the roots of the inverse smooth dynamical matrices in the reciprocal Bloch representation. 

Combining Eq.~(\ref{eq:commutation_rel_direct}) and Eq.~(\ref{eq:bosonic_reciprocal}), it is \replaced{possible}{easy} to show that also the  operators in reciprocal space~(\ref{eq:bosonic_reciprocal}) satisfy the bosonic commutation relation 
\begin{equation}
\begin{split}
[\hat{a}(\bm{q})_{b\alpha},\hat{a}^\dagger(\bm{q'})_{b'\alpha'}]
&{=}N_c \delta_{\bm{q},\bm{q'}}\delta_{b,b'}\delta_{\alpha,\alpha'}\\
&\stackrel{\lim\limits_{N_c\to\infty}}{=}\tfrac{(2\pi)^3}{\mathcal{V}}\delta({\bm{q}{-}\bm{q'}})\delta_{b,b'}\delta_{\alpha,\alpha'},
\end{split}
\label{eq:commutator_bosonic_reciprocal}
\end{equation}
where we have highlighted that the Dirac delta is obtained from the Kronecker delta in the bulk limit, \textit{i.e.} when the crystal is made up of $N_c{\to}\infty$  Bravais-lattice sites \cite{Mahan_book,marder2010condensed,callaway1991quantum}. 

The key role of the bosonic operators in reciprocal space~(\ref{eq:bosonic_reciprocal}) is their capability to \replaced{simplify}{reduce} to 
\deleted{a simple} block-diagonal form the representation of translation-invariant quantities such as the harmonic Hamiltonian~(\ref{eq:H_real_space}): 
\begin{equation}
\begin{split}
\hat{H}^{\mathrm{har}}{=} \frac{\mathcal{V}}{\!(2\pi)^3\!\!}\!
  \int_{\mathfrak{B}} 
  \sum_{\substack{\scriptscriptstyle{ b,\alpha}\\
  \scriptscriptstyle{b'\!,\alpha'}}}\!
   \hbar\sqrt{\tenscomp{D}(\bm{q})}_{\!b\alpha,b'\!\alpha'}
  \hat{a}^\dagger\!(\bm{q})_{b\alpha}
  \hat{a}(\bm{q})_{b'\!\alpha'} d^3\!q.
\label{eq:H_reciprocal_space}
\end{split}
\raisetag{4mm}
\end{equation}
{So, the bosonic operators $\hat{a}(\bm{R})_{b\alpha}$, $\hat{a}^{\dagger}(\bm{R})_{b\alpha}$, localized in real space, allow to discuss local vibrational excitations, and their Fourier transforms $\hat{a}(\bm{q})_{b\alpha}$, $\hat{a}^{\dagger}(\bm{q})_{b\alpha}$ allow to describe translation-invariant quantities in block-diagonal form in reciprocal space.}

To compute the expectation value of the harmonic vibrational energy from Eq.~(\ref{eq:H_reciprocal_space}),
it is sufficient to know the one-body density matrix in reciprocal space, 
\begin{equation}
  \varrho(\bm{q},\bm{q'},t)_{b\alpha,b'\hspace*{-0.5mm}\alpha'}{=}{\rm Tr} \big(\hat{\rho}(t)\hat{a}^{\dagger}(\bm{q'})_{b'\hspace*{-0.5mm}\alpha'}\hat{a}(\bm{q})_{b\alpha}\big).
  \label{eq:one_body_rec}
\end{equation}
To describe a weakly non-homogeneous out-of-equilibrium system, in which a small temperature gradient is present in real space, it is convenient to work in reciprocal space with {the average and displacement coordinates (\textit{i.e.} given a generic couple of  wavevectors $\bm{q}_1$ and $\bm{q}_2$, they are written in terms of the displacement $\bm{q'\!'}{=}\bm{q}_1{-}\bm{q}_2$ from their average $\bm{q}{=}\tfrac{\bm{q_1}{+}\bm{q_2}}{2}$, with 
$\bm{q}_1{=}\bm{q}{+}\tfrac{\bm{q'\!'}}{2}$ and $\bm{q}_2{=}\bm{q}{-}\tfrac{\bm{q'\!'}}{2}$):}
\begin{align}
&\varrho\big(\bm{q}{+}\tfrac{\bm{q'\!'}}{2},\bm{q}{-}\tfrac{\bm{q'\!'}}{2},t\big)_{b\alpha,b'\hspace*{-0.5mm}\alpha'}{=}\!\!
  \sum_{\bm{R},\bm{R'}}\!\!  \varrho(\bm{R},\bm{R'}\!\!,t)_{b\alpha,b'\hspace*{-0.5mm}\alpha'}\label{eq:one_body_T1}\\
&\hspace*{1.5cm}\times e^{- i \bm{q}{\cdot} (\bm{R}{+}\bm{\tau}_b{-}\bm{R'}\!{-}\bm{\tau}_{b'})}
e^{-i \bm{q'\!'}\cdot \tfrac{\bm{R}{+}\bm{\tau}_b+\bm{R'}\!+\bm{\tau}_{b'}}{2}}.
\nonumber
\end{align}
In fact, when the system is homogeneous, e.g. at equilibrium, the one-body density matrix in real space is invariant under translation,
$\varrho(\bm{R},\bm{R'}\!\!,t)_{b\alpha,b'\hspace*{-0.5mm}\alpha'}{=}\varrho(\bm{R}{-}\bm{R'}\!\!,t)_{b\alpha,b'\hspace*{-0.5mm}\alpha'}$; consequently, its Fourier transform~(\ref{eq:one_body_T1}) yields a one-body density matrix diagonal in reciprocal space $\varrho\big(\bm{q}{+}\tfrac{\bm{q'\!'}}{2},\bm{q}{-}\tfrac{\bm{q'\!'}}{2},t\big)_{b\alpha,b'\hspace*{-0.5mm}\alpha'}{=}\varrho(\bm{q},\bm{q})_{b\alpha,b'\hspace*{-0.5mm}\alpha'}\frac{(2\pi)^3}{\mathcal{V}}\delta(\bm{q'\!'})$ \footnote{given a one-body density matrix $\varrho(\bm{q}{+}\tfrac{\bm{q''}}{2},\bm{q}{-}\tfrac{\bm{q''}}{2})_{b\alpha,b'\alpha'}$, we say that it is diagonal in its two arguments ($\bm{q}{+}\tfrac{\bm{q''}}{2}$ and $\bm{q}{-}\tfrac{\bm{q''}}{2}$) if and only if it is nonzero only when the two arguments are equal (\textit{i.e.} for $\bm{q''}=\bm{0}$).}.
In the weakly non-homogeneous out-of-equilibrium regime of interest here, the one-body density matrix in real space $\varrho(\bm{R},\bm{R'}\!\!,t)_{b\alpha,b'\hspace*{-0.5mm}\alpha'}$ is close to the  translation-invariant form. In the context of Eq.~(\ref{eq:one_body_T1}) this implies that the one-body density matrix in reciprocal space $\varrho\big(\bm{q}{+}\tfrac{\bm{q'\!'}}{2},\bm{q}{-}\tfrac{\bm{q'\!'}}{2},t\big)_{b\alpha,b'\hspace*{-0.5mm}\alpha'}$
is close to the diagonal form, \textit{i.e.} appreciably different from zero
only for $|\bm{q'\!'}|{\to} 0$.
We will discuss later how this property can be exploited to simplify to solvable form the equation describing the system's evolution.

The transformation~(\ref{eq:one_body_T1}) is a reminder that the one-body density matrix in direct and reciprocal space contain the same information. 
Heat transport in solids is driven by a temperature gradient in real space, 
thus the formulation in real space is convenient to keep track of the vibration's locations and relate them to the space-dependent temperature driving transport. Conversely, we have seen that the formulation in reciprocal space simplifies to block-diagonal form the expression of key quantities for the description of thermal transport (e.g. the harmonic Hamiltonian~(\ref{eq:H_reciprocal_space})), and we will show later that using Wigner's phase-space framework it is possible to combine the advantages of these two formulations in real and in reciprocal space.

We note in passing that the standard phonon operators \cite{ziman1960electrons,wallace1972thermodynamics}\\[-5mm]
\begin{equation}
\begin{split}
\hat{a}(\bm{q})_{s}{=}\!\sum_{b,\alpha}\!\mathcal{E}^{\star}\!(\bm{q})_{s,b\alpha}\!\bigg[\!
\frac{1}{\!\sqrt{\!2 \hbar\omega(\bm{q})_s\!\!}}
{\hat{P}}(\bm{q})_{\hspace*{-0.4mm}b\alpha} 
{-}i \frac{\sqrt{\!\omega(\bm{q})_s\!}}{2 \hbar}{\hat{U}}^\dagger\!(\bm{q})_{b\alpha}\!\bigg]\!,\hspace*{9mm}
%\\
%&=\mathcal{E}^{\star}(\bm{q})_{s,b\alpha}\hat{a}(\bm{q})_{b\alpha},
\end{split}
\raisetag{15mm}
  \label{eq:ph_op_mode_basis}
\end{equation}
where $s$ is the phonon-band index, have a real-space representation that \deleted{in general} is not suitable to track where vibrations are centered. In fact, the eigenvectors of the dynamical matrix in the reciprocal Bloch representation $\mathcal{E}(\bm{q})_{s,b\alpha}$ (Eq.~(\ref{eq:diagonalization_dynamical_matrix})) have an undetermined phase, which allows to apply arbitrary unitary transformations to $\hat{a}(\bm{q})_{s}$
%e.g. of type $\mathcal{E}(\bm{q})_{s,b\alpha}\to \mathcal{E}(\bm{q})_{s,b\alpha}e^{+i\bm{q}\cdot\bm{R}_U}$ (where $\bm{R}_{M}$ is an arbitrary Bravais lattice vector)
and thus shift \replaced{by}{of} an arbitrary Bravais lattice vector the corresponding real-space operator
$\hat{a}(\bm{R})_{s}{=}\frac{\mathcal{V}}{(2\pi)^3}\int_{\mathfrak{B}}\hat{a}(\bm{q})_{s}e^{+i\bm{q}\cdot\bm{R}}d^3q$.
Details are discussed in Appendix~\ref{sub:localization_problems_of_the_eigenmodes_basis}, where it is shown 
that the non-uniqueness of the real-space representation of the phonon operators $\hat{a}(\bm{q})_{s}$ mirrors the electronic case, where the phase indeterminacy of the Bloch orbitals is reflected in the non-uniqueness of their transformation into Wannier functions.

\subsection{Evolution equation for the density matrix }
\label{sub:evolution_equation_for_the_density_matrix}
We recall that the goal of the present work is to describe the energy transfer across a solid material driven by a temperature gradient. More precisely, we consider the regime where temperature variations are appreciable over a macroscopic length scale $L$ that is much larger than the interatomic spacing $a$. 
We assume that such a temperature gradient derives from a continuous energy exchange with some heat baths.
This implies that the system \deleted{in focus here} undergoes an irreversible evolution \cite{blum2013density}, which can be modeled as Markovian and described by the master equation \cite{rossi2011theory,breuer2002theory,vasko2006quantum} 
\begin{equation}
\label{eq:lindblad_d_matrix_S}
\frac{{\partial} \hat{\rho}(t)}{\partial t}+\frac{i}{\hbar}\Big[\hat{H}^{\rm har},\hat{\rho}(t)\Big]=\frac{\partial \hat{\rho}(t) }{\partial  t }\bigg|_{\hat{H}^{\rm per}};
\end{equation}
where the scattering  superoperator $\frac{\partial \hat{\rho}(t) }{\partial  t }\big|_{\hat{H}^{\rm per}}$ is determined by the perturbative Hamiltonian $\hat{H}^{\rm per}$ (which accounts e.g. for the presence of anharmonicity and isotopes) and describes transitions between the eigenstates of the leading Hamiltonian $\hat{H}^{\rm har}$.

Recalling Eq.~(\ref{eq:H_reciprocal_space}), we have that $\hat{H}^{\rm har}$ 
is a one-body operator; thus, applying the operator $\hat{a}^{\dagger}(\bm{q}{-}\tfrac{\bm{q'\!'}}{2})_{b'\hspace*{-0.5mm}\alpha'}\hat{a}(\bm{q}{+}\tfrac{\bm{q'\!'}}{2})_{b\alpha}$ to both sides of Eq.~(\ref{eq:lindblad_d_matrix_S}) and taking the trace as in Eq.~(\ref{eq:one_body_rec}), gives us the evolution equation for the one-body density matrix
\begin{equation}
\begin{split}
  &\frac{\partial }{\partial  t}{{\varrho}}\big(\bm{q}{+}\tfrac{\bm{q'\!'}}{2},\bm{q}{-}\tfrac{\bm{q'\!'}}{2},t\big)_{b\alpha,b'\!\alpha'}\\
  &+i\bigg[
  \sum_{b'\!'\!,\alpha'\!'}\sqrt{\!{\tenscomp{D}}\big(\bm{q}{+}\tfrac{\bm{q'\!'}}{2}\big)}_{b\alpha,b'\!'\!\alpha'\!'}{\varrho}\big(\bm{q}{+}\tfrac{\bm{q'\!'}}{2},\bm{q}{-}\tfrac{\bm{q'\!'}}{2},t\big)_{b'\!'\!\alpha'\!',b'\!\alpha'}\\
  &-\sum_{b'\!'\!,\alpha'\!'}{\varrho}\big(\bm{q}{+}\tfrac{\bm{q'\!'}}{2},\bm{q}{-}\tfrac{\bm{q'\!'}}{2},t\big)_{b\alpha,b'\!'\!\alpha'\!'}\sqrt{\!{\tenscomp{D}}\big(\bm{q}{-}\tfrac{\bm{q'\!'}}{2}\big)}_{b'\!'\!\alpha'\!',b'\!\alpha'}\bigg]\\
  &=\frac{\partial {\varrho}\big(\bm{q}{+}\tfrac{\bm{q'\!'}}{2},\bm{q}{-}\tfrac{\bm{q'\!'}}{2},t\big)_{b\alpha,b'\!\alpha'}\!\!}{\partial   t}\bigg|_{\!\hat{H}^{\rm per}}.
\end{split}
\raisetag{8mm}
  \label{eq:evol_density_matrix}
\end{equation}
We also note that Eq.~(\ref{eq:evol_density_matrix}) can be derived starting from real space \cite{SimoncelliPhD}, applying the operator $\hat{a}^{\dagger}(\bm{R'})_{b'\hspace*{-0.5mm}\alpha'}\hat{a}(\bm{R})_{b\alpha}$ to both sides of Eq.~(\ref{eq:lindblad_d_matrix_S}), then taking the trace as in Eq.~(\ref{eq:one_body_def}), and finally performing the Fourier transform~(\ref{eq:one_body_T1}).

We recall that Eq.~(\ref{eq:one_body_T1}) implies that at equilibrium, where the system is homogeneous (translation-invariant), the one-body density matrix is diagonal in reciprocal space, \textit{i.e.} ${\varrho}\big(\bm{q}{+}\tfrac{\bm{q'\!'}}{2},\bm{q}{-}\tfrac{\bm{q'\!'}}{2},t\big)_{b\alpha,b'\!\alpha'}{=}{\varrho}\big(\bm{q},\bm{q},t\big)_{b\alpha,b'\!\alpha'}\frac{(2\pi)^3}{\mathcal{V}}\delta(\bm{q'\!'})$. 
As a consequence, Eq.~(\ref{eq:evol_density_matrix}) yields a non-trivial evolution only out of equilibrium, where a perturbation (temperature gradient) in real space implies that the one-body density matrix  has non-zero off-diagonal elements.
In the next section we will discuss a theoretical framework that makes the description of this non-homogeneous regime particularly \deleted{simple and} intuitive \added{and manageable}.

\section{Phase-space formalism }
\label{sec:Wigner_thermal_transport_equation}
The ideal framework to describe heat transport in solids should keep track of: (i) the equilibrium eigenstates towards which the system relaxes, {for which the quasi-momentum $\hbar \bm{q}$ is a good quantum number; {(ii) the spatial dependence of the perturbation driving transport, whose localization in real space would be conveniently described by an Hilbert-space basis having as quantum number the Bravais lattice vector $\bm{R}$.}   
However, the quasi-momentum and Bravais-lattice operators ($\hbar\hat{\bm{q}}$ and $\hat{\bm{R}}$, respectively) 
are a pair of canonically conjugate operators   \cite{Zak1975_lattice_operator}, and thus their eigenvalues ($\hbar{\bm{q}}$ and ${\bm{R}}$, respectively) cannot be used to label simultaneously a quantum-mechanical representation within the usual Dirac formalism.
The Wigner phase-space formalism   \cite{weyl1927quantenmechanik,wigner1932,groenewold1946principles,moyal1949quantum,imre1967wigner} allows to describe quantum mechanics in terms of distributions having as arguments eigenvalues of canonically conjugate operators, and in this section we discuss the application of such formalism to solids.
We show that the central quantity appearing in this formulation generalizes the semiclassical distribution appearing in the Peierls-Boltzmann equation   \cite{peierls1955quantum}, describing not only intraband propagation of particle-like phonon wavepackets, but also wave-like interband (Zener) tunneling of phonons.
\deleted{This formalism is general and can be used to describe solids with any degree of disorder, ranging from ordered crystals with small primitive cell to disordered glasses (these  treated as the limit of disordered crystals with extremely large primitive cells).}

\subsection{Wigner transform }
\label{sub:weyl_wigner_transform}
As anticipated, the central objects of the Wigner formalism are phase-space distributions having as arguments eigenvalues of non-commuting operators~\cite{hillery1984distribution,moyal1949quantum,imre1967wigner}.
In order to show how such a framework allows to describe conveniently transport in non-homogeneous, out-of-equilibrium systems, it is useful to start from the Wigner transformation that maps the matrix elements of a one-body operator
${O}{\big(}{\bm{q}}{+}\tfrac{\bm{q'\!'}}{2}\hspace*{-1mm},\hspace*{-0.4mm}{\bm{q}}{-}\tfrac{\bm{q'\!'}}{2}\hspace*{-1mm}{\big)}_{\hspace*{-0.7mm}{b}{\alpha}{,}{b'}\hspace*{-0.5mm}{\alpha'}}$ into a ``phase-space'' distribution that depends on a wavevector $\bm{q}$ (belonging to a Brillouin zone $\mathfrak{B}$) and a Bravais lattice vector $\bm{R}$. More generally, we will discuss soon that the spatial dependence of a Wigner distribution  can be \replaced{extended}{promoted} from the discrete Bravais lattice vectors $\bm{R}$ to a continuous Cartesian position $\bm{r}$; \deleted{the formula below is written in terms of the position $\bm{r}$} for the sake of generality \added{we use this definition then for the Wigner transformation $\mathcal{W}_{[{O}]}$ of a one-body operator $O$}:
\begin{equation}
\begin{split}
  \mathcal{W}_{[{O}]}(\!\bm{r},\!\bm{q},t)_{{b}{\alpha}{,}{b'}\hspace*{-0.5mm}{\alpha'}}&{=}
   \frac{\mathcal{V}}{(2\pi)^3\!} 
  \!\!\!\int\limits_{\mathfrak{B}}\!\!\!
{O}{\big(}{\bm{q}}{+}\tfrac{\bm{q'\!'}}{2}\hspace*{-1mm},\hspace*{-0.4mm}{\bm{q}}{-}\tfrac{\bm{q'\!'}}{2}\hspace*{-1mm},t{\big)}_{\hspace*{-0.7mm}{b}{\alpha}{,}{b'}\hspace*{-0.5mm}{\alpha'}}   e^{i\bm{q'\!'}\!\cdot \bm{r}}     d^3q'\!'\!.\;\;\;\;\;
\end{split}
\raisetag{15mm}
  \label{eq:Wigner_transform_Rec}
\end{equation}
When the Wigner transformation~(\ref{eq:Wigner_transform_Rec}) is applied to the one-body density matrix~(\ref{eq:one_body_rec}), it becomes apparent that the Wigner distribution $\tenscomp{w}(\!\bm{r},\!\bm{q},t)_{{b}{\alpha}{,}{b'}\hspace*{-0.5mm}{\alpha'}}{=}\mathcal{W}_{[{\varrho}]}(\!\bm{r},\!\bm{q},t)_{{b}{\alpha}{,}{b'}\hspace*{-0.5mm}{\alpha'}}$ does not depend on space if the system is homogeneous (\textit{i.e.} translation-invariant; as discussed before, Eq.~(\ref{eq:one_body_T1}) implies that for a translation-invariant system the one-body density matrix is nonzero only for $\bm{q'\!'}{=}\bm{0}$).
Conversely, for an out-of-equilibrium, non-homogeneous system the one-body density matrix is non-diagonal in $\bm{q'\!'}$ and thus its Wigner representation depends on space. 

In general, the position $\bm{r}$ appearing in Eq.~(\ref{eq:Wigner_transform_Rec}) can be any (continuous) position in real space. Nevertheless, restricting such a position to the discrete Bravais-lattice vectors is sufficient to obtain a phase-space distribution that contains the same information of the Dirac matrix element of the corresponding operator. 
In fact, knowing the phase-space distribution %(defined by Eq.~(\ref{eq:Wigner_transform_Rec})) 
at the discrete Bravais vectors $(\bm{r}{=}\bm{R})$ allows to determine the matrix element of the corresponding operator 
(proof in Appendix~\ref{sec:properties_of_the_wigner_transform})
\begin{equation}
\begin{split}
{O}{\big(}{\bm{q}}{+}\tfrac{\bm{q'\!'}}{2}\hspace*{-1mm},\hspace*{-0.4mm}{\bm{q}}{-}\tfrac{\bm{q'\!'}}{2}\hspace*{-1mm},t{\big)}_{\hspace*{-0.7mm}{b}{\alpha}{,}{b'}\hspace*{-0.5mm}{\alpha'}}
  =
\sum_{\bm{R}}
\mathcal{W}_{[{O}]}\big(\bm{R},\bm{ q},t\big)_{{b}{\alpha}{,}{b'}\hspace*{-0.5mm}{\alpha'}}
  e^{-i\bm{q'\!'}\cdot \bm{R}}\!.
  \label{eq:Wigner_transform_Rec_inv}
  \raisetag{4mm}
  \end{split}
\end{equation}

We note in passing that the time dependence for  $O$ and $\mathcal{W}_{[O]}$ in Eqs.~(\ref{eq:Wigner_transform_Rec},\ref{eq:Wigner_transform_Rec_inv}) has been reported for generality; in the present paper the only operator (phase-space distribution) that can depend on time is the density matrix (Wigner distribution $\tenscomp{w}(\!\bm{R},\bm{q},t)_{{b}{\alpha}{,}{b'}\hspace*{-0.5mm}{\alpha'}}$).

{
In order to shed light on why the Wigner framework is particularly convenient to describe transport, it is useful to discuss how expectation values are computed in this framework, and compare it the usual Dirac formalism.
We start by recalling that in the Dirac formalism the expectation value of a quantum mechanical observable $\hat A$, on a state represented by the density matrix $\hat \rho(t)$ is determined from the trace operation ${\rm Tr}\big(\hat \rho(t)\hat A \big){=}\big<\hat A\big>$.
In the Wigner framework, instead, expectation values are computed as  phase-space-like integrals of the product between the Wigner representations of the density matrix and that of the observable \cite{imre1967wigner}.
For the one-body operators and phase-space distributions in focus here and \replaced{related via}{connected by} Eq.~(\ref{eq:Wigner_transform_Rec}) 
these two equivalent methods to compute expectation values are summarized in the following equation (proof in Appendix~\ref{sec:properties_of_the_wigner_transform}):}
\begin{equation}
  \begin{split}
 &{\rm Tr}\big(\hat \rho(t)\hat A \big)  \\
&{=}\frac{\mathcal{V}^2}{(2\pi)^6\!\!}
 \sum_{\substack{b,\alpha\\b'\!,\alpha'}}\;   \int\limits_{\!\!\mathfrak{B}\mathfrak{B}}\!\!\!\!\!\!\!\!\!\! \int
   \!{\varrho}\hspace*{-0.3mm}\big(\!{{\bm{q}{+}\tfrac{\bm{q'\!'}\!}{2}}}\!,\!\bm{q}{-}\tfrac{\bm{q'\!'}\!}{2}\!,t\big)_{\!b\alpha,b'\!\alpha'\!}
   {A}\!\big(\!\bm{q}{-}\tfrac{\bm{q'\!'}\!}{2},\!\bm{q}{+}\tfrac{\bm{q'\!'}\!}{2}\!\big)_{\!b'\!\alpha'\!,b\alpha}\!\!
 d^3\!q d^3\!q'\!'\\
 &=\frac{\mathcal{V}}{(2\pi)^3}\!
 \sum_{\substack{\bm{R}\\b,\alpha,b'\!,\alpha'}}
    \int_{\mathfrak{B}}
\!\mathcal{W}_{[\varrho]}(\bm{R},\bm{q},t)_{b\alpha,b'\!\alpha'}
\mathcal{W}_{[A]}(\bm{R},\bm{q})_{b'\!\alpha',b\alpha}
d^3\!q.     
\label{eq:trace_property_main}
\raisetag{32mm}
  \end{split}
\end{equation}
{The third line in Eq.~(\ref{eq:trace_property_main}) contains a sum over all lattice vectors in real space ($\bm{R}$) and an integral over wavevectors in reciprocal space ($\bm{q}$); therefore, carrying out only the integration in reciprocal space yields an expression for the expectation value as a spatial integration of a space-dependent quantity (here with ``spatial integration'' we mean the sum over $\bm{R}$).
This shows that the Wigner framework provides prescriptions to resolve 
in space expectation values~\cite{imre1967wigner,physreve.81.021119}. 
Explicit calculations will be reported later, when we will exploit Eq.~(\ref{eq:trace_property_main}) to define a space-dependent vibrational energy field, compute the related heat flux, and thus determine the thermal conductivity.}

\subsection{\nametheory{} transport equation }
\label{sub:evolution_of_the_wigner_density}
The first key step in obtaining the equation describing the temporal evolution of the Wigner distribution at the Bravais site $\bm{R}$ and wavevector $\bm{q}$, $\tenscomp{w}(\!\bm{R},\bm{q},t)_{{b}{\alpha}{,}{b'}\hspace*{-0.5mm}{\alpha'}}$ is applying the transformation~(\ref{eq:Wigner_transform_Rec}) to the evolution equation for the one-body density matrix~(\ref{eq:evol_density_matrix}). 
Then, we note that we are interested in the close-to-equilibrium regime characterized by weak non-homogeneities in real space, \textit{i.e.} in the regime where temperature varies slowly and causes spatial variations of the one-body density matrix~(\ref{eq:one_body_def}) to be appreciable only at ``mesoscopic scales'' $l$ that are much larger than the lengths $a$ at which atomic positions can be resolved ($l\gg a$).

We start by recalling that as the spatial variations of the one-body density matrix in real space become smaller, the one-body density matrix in reciprocal space becomes closer to being diagonal in the arguments (see Eq.~(\ref{eq:one_body_T1}) and related discussion). This implies that, close to equilibrium, the one-body density matrix ${\varrho}\big(\!\bm{q}{+}\tfrac{\bm{q'\!'}\!}{2},\!\bm{q}{-}\tfrac{\bm{q'\!'}\!}{2}\!\big)_{b\alpha,b'\alpha'}$ is sharply peaked around $\bm{q}$, and significantly different from zero only for $|\bm{q'\!'}|\ll 2\pi/|\bm{a}_i|$ ($\bm{a}_i$ is the $i$-th direct lattice vector).
This property allows us to simplify the evolution equation~(\ref{eq:evol_density_matrix}), performing a Taylor expansion for  $|\bm{q'\!'}|{\to} \bm{0}$ of the coefficients appearing in that equation:
\begin{equation}
\begin{split}
  \sqrt{\!{\tenscomp{D}}\big(\bm{q}{\pm}\tfrac{\bm{q'\!'}}{2}\big)}_{b\alpha,b'\!'\!\alpha'\!'}{\approx}\sqrt{\!{\tenscomp{D}}(\bm{q})}_{b\alpha,b'\!'\!\alpha'\!'}{\pm}{\nabla}_{\bm{q}}\sqrt{\!{\tenscomp{D}}(\bm{q})}_{b\alpha,b'\!'\!\alpha'\!'}{\cdot} \frac{\bm{q'\!'}}{2}.
  \label{eq:Taylor}
  \raisetag{13mm}
  \end{split}
\end{equation}
Inserting such approximation into Eq.~(\ref{eq:evol_density_matrix}), one obtains a simplified evolution equation:
\begin{equation}
\begin{split}
  &\frac{\partial}{\partial  t} {{\varrho}}\big(\bm{q}{+}\tfrac{\bm{q'\!'}}{2},\bm{q}{-}\tfrac{\bm{q'\!'}}{2},t\big)_{b\alpha,b'\!\alpha'}\\
  &+i\Big(\sum_{b'\!'\!,\alpha'\!'\!}\!\sqrt{\!{\tenscomp{D}}(\bm{q})}_{b\alpha,b'\!'\!\alpha'\!'}{\varrho}\big(\bm{q}{+}\tfrac{\bm{q'\!'}}{2},\bm{q}{-}\tfrac{\bm{q'\!'}}{2},t\big)_{b'\!'\!\alpha'\!',b'\!\alpha'}\\
  &\hspace*{6mm}-\sum_{b'\!'\!,\alpha'\!'\!}{\varrho}\big(\bm{q}{+}\tfrac{\bm{q'\!'}}{2},\bm{q}{-}\tfrac{\bm{q'\!'}}{2},t\big)_{b\alpha,b'\!'\!\alpha'\!'}\sqrt{\!{\tenscomp{D}}(\bm{q})}_{b'\!'\!\alpha'\!',b'\!\alpha'}\Big)\\
  &+\frac{1}{2}i\bm{q'\!'}\bigg(\sum_{b'\!'\!,\alpha'\!'\!}\Big(\nabla_{\bm{q}}\sqrt{\!{\tenscomp{D}}(\bm{q})}_{b\alpha,b'\!'\!\alpha'\!'}\Big){\varrho}\big(\bm{q}{+}\tfrac{\bm{q'\!'}}{2},\bm{q}{-}\tfrac{\bm{q'\!'}}{2},t\big)_{b'\!'\!\alpha'\!',b'\!\alpha'}\\
  &\hspace*{10mm}+\sum_{b'\!'\!,\alpha'\!'\!}{\varrho}\big(\bm{q}{+}\tfrac{\bm{q'\!'}}{2},\bm{q}{-}\tfrac{\bm{q'\!'}}{2},t\big)_{b\alpha,b'\!'\!\alpha'\!'}
  \Big(\nabla_{\bm{q}}\sqrt{\!{\tenscomp{D}}(\bm{q})}_{b'\!'\!\alpha'\!',b'\!\alpha'}\Big)\bigg)\\
  &=\frac{\partial {\varrho}\big(\bm{q}{+}\tfrac{\bm{q'\!'}}{2},\bm{q}{-}\tfrac{\bm{q'\!'}}{2},t\big)_{b\alpha,b'\!\alpha'}\!\!}{\partial   t}\bigg|_{\!\hat{H}^{\rm per}}.
\end{split}
\raisetag{5mm}
  \label{eq:evol_density_matrix_approx}
\end{equation}
Multiplying all terms in Eq.~(\ref{eq:evol_density_matrix_approx}) by $\tfrac{\mathcal{V}}{(2\pi)^3}e^{i\bm{q'\!'}{\cdot}\bm{R}}$ and integrating $\bm{q'\!'}$ over the Brillouin zone yields the evolution equation for the Wigner distribution  
\begin{equation}
\begin{split}
&\frac{\partial  }{\partial  t} \tenscomp{w}({\bm{R}},{\bm{q}},t)_{b\alpha,b'\!\alpha'\!}
{+}i\Big[\sqrt{{\tenscomp{D}}(\bm{q})},\tenscomp{w}(\bm{R},\bm{q},t)\Big]_{\!b\alpha,b'\!\alpha'\!}+\\
&{+}\frac{1}{2}\Big\{\!\nabla_{\bm{q}}\! \sqrt{\!\tenscomp{D}(\bm{q})},\nabla_{\bm{R}} \tenscomp{w}(\bm{R}\hspace*{-0.2mm},\bm{q}\hspace*{-0.3mm},t)\!\Big\}_{\!b\alpha,b'\!\alpha'\!}\!\!{=}\frac{\partial \tenscomp{w}(\bm{R}\hspace*{-0.2mm},{\bm{q}}\hspace*{-0.3mm},t)_{\!b\alpha,b'\!\alpha'\!} \!}{\partial   t}\bigg|_{\hat{H}^{\rm per}}\!\!,
\end{split}
\raisetag{42pt}
\label{eq:Wigner_evolution_equation}
\end{equation}
where we have introduced the notation 
$[\sqrt{{\tenscomp{D}}(\bm{q})},\tenscomp{w}(\bm{R},\bm{q},t)]_{\!b\alpha,b'\!\alpha'\!}$ to denote the matrix element of the commutator, \textit{i.e.} the quantity
${\scriptstyle \sum_{b'\!'\!\!,\alpha'\!'}\!\!\big(\!\!\sqrt{{\tenscomp{D}}(\bm{q})}_{\!b\alpha,b'\!'\!\alpha'\!'\!}\tenscomp{w}(\bm{R},\bm{q},t)_{\!b'\!'\!\alpha'\!'\!\!,b'\!\alpha'\!}{-}\tenscomp{w}(\bm{R},\bm{q},t)_{\!b\alpha,b'\!'\!\alpha'\!'\!}\sqrt{{\tenscomp{D}}(\bm{q})}_{\!b'\!'\!\alpha'\!'\!\!,b'\!\!\alpha'\!}}\big)\!$; an analogous notation is employed for the anticommutator $\{,\}$.
{Moreover, the derivative of the Wigner distribution with respect to the position $\bm{R}$ is obtained taking first the derivative with respect to the continuous position $\bm{r}$ (see Sec.~\ref{sub:weyl_wigner_transform}), and then evaluating such derivative at the Bravais lattice vector $\bm{R}$.}

After having performed the first-order Taylor expansion~(\ref{eq:Taylor}), for which localization and differentiation properties are crucial, one can apply any unitary transformation to the simplified evolution equation~(\ref{eq:Wigner_evolution_equation}), leaving the physics unchanged.
Therefore, we can apply to Eq.~(\ref{eq:Wigner_evolution_equation}) the unitary transformation~(\ref{eq:diagonalization_dynamical_matrix}) that diagonalizes the square root of the dynamical matrix in the reciprocal Bloch representation, obtaining distributions that depend on position $\bm{R}$, wavevector $\bm{q}$, and phonon band indexes ($s,s'$), and thus can be compared directly with the distribution appearing in the Peierls-Boltzmann equation.
Specifically, we introduce the quantities
\begin{equation}
\begin{split}
   \hspace*{-3mm}\tenscomp{n}(\bm{R},\bm{q},t)_{s,s'}{=}\!\!\sum_{\substack{b,\alpha,b'\!,\alpha'}}\!\mathcal{E}^\star(\bm{q})_{{s},b\alpha}{\tenscomp{w}(\bm{R},\bm{q},t)}_{b\alpha,b'\hspace*{-0.5mm}\alpha'} \mathcal{E}(\bm{q})_{s'\hspace*{-0.5mm},b'\hspace*{-0.5mm}\alpha'},
   \raisetag{12.5mm}
\end{split}   
   \label{eq:Population_Peierls_Generalized}
\end{equation}
\vspace*{-6mm}   
\begin{equation} 
\begin{split}     
 \hspace*{-3mm}\tenscomp{v}^\beta(\bm{q})_{s,s'}{=}\!\!\sum_{\substack{b,\alpha,b'\!,\alpha'}}\!\!\!\mathcal{E}^\star(\bm{q})_{{s},b\alpha}{{\nabla^{\beta}_{\bm{q}} \sqrt{\tenscomp{D}(\bm{q})}} }_{b\alpha,b'\hspace*{-0.5mm}\alpha'} \mathcal{E}(\bm{q})_{s'\hspace*{-0.5mm},b'\hspace*{-0.5mm}\alpha'},\\[-3mm]
\end{split}    
   \raisetag{10mm} 
\label{eq:vel_op}
\end{equation}
{which generalize the concepts of phonon populations and group velocities beyond the particle-like interpretation provided by the semiclassical Peierls-Boltzmann equation~(\ref{eq:BTE_std}). In fact, 
%\added{we will show later in Sec.~\ref{sec:solution_of_the_wigner_equation_and_generalized_conductivity_formula}  that} 
in the absence of degeneracies (\textit{i.e.} for $s{\neq}s'$, $\omega(\bm{q})_s{\neq}\omega(\bm{q})_{s'}$) their diagonal elements ($s{=}s'$) coincide with the phonon populations ($\tenscomp{n}(\bm{R},\bm{q},t)_{s,s}=\tenscomp{n}(\bm{R},\bm{q},t)_{s}$) and group velocities~(\ref{eq:group_velocity}) ($\tenscomp{v}^\beta(\bm{q})_{s,s}{=}\tenscomp{v}^\beta(\bm{q})_{s}$) appearing in the Peierls-Boltzmann equation~(\ref{eq:BTE_std}),} respectively, while their off-diagonal elements emerge from the wave-like nature of atomic vibrational eigenstates and  are related to the phase coherence between pairs of vibrational eigenstates $s$ and $s'$ \cite{rossi2011theory,PhysRevB.72.125347,RevModPhys.74.895,vasko2006quantum}.
\added{
From now on we will use the textbook nomenclature   \cite{cohen1992atom} and refer to the diagonal elements ($s{=}s'$) of the Wigner distribution~(\ref{eq:Population_Peierls_Generalized}) as ``populations'' and to the off-diagonal ones ($s{\neq}s'$) as ``coherences''.
As discussed in the literature \cite{rossi2011theory,PhysRevB.72.125347,RevModPhys.74.895,vasko2006quantum,blum2013density}, populations have a well defined energy (frequency) and therefore can be interpreted as particle-like excitations with a well defined wavevector $\bm{q}$ and mode index $s$. In contrast, coherences do not have an absolute energy  and cannot be related to a single eigenstate; they describe oscillations between pairs of eigenstates and correspond to an evolution which does not preserve the nature of the excitation (e.g. we will show later that coherences describe couplings between two different vibrational modes $s,s'$ at the same wavevector $\bm{q}$).} 
%\cite{rossi2011theory,PhysRevB.72.125347,RevModPhys.74.895,vasko2006quantum,blum2013density,RevModPhys.74.895} }
%\added{More details on how Eq.~\ref{eq:Population_Peierls_Generalized} and Eq.~\ref{eq:vel_op} generalize the concepts of phonon population and group velocity are discussed later in Sec.~Appendix~\ref{sec:physical_meaning_coherences}}
Moreover, the velocity operator~(\ref{eq:vel_op}) has the following properties: it is Hermitian (since it is the incremental ratio of an Hermitian operator), its different Cartesian components do not commute in general, {$\sum_{s'\!'}(\!\tenscomp{v}^\alpha\!(\bm{q})_{s,s'\!'\!}\tenscomp{v}^\beta\!(\bm{q})_{s'\!'\!,s'}{-}\tenscomp{v}^\beta\!(\bm{q})_{s,s'\!'}\tenscomp{v}^\alpha\!(\bm{q})_{s'\!'\!,s'}\!){\neq}0$ if $\alpha{\neq} \beta$}, and from the time-reversal symmetry of the dynamical matrix in reciprocal Bloch representation \cite{revmodphys_maradudin68} it follows that $\tenscomp{v}^\beta(\bm{q})_{s,s'}{=}{-}\tenscomp{v}^\beta(-\bm{q})_{s',s}$.

\deleted{From now on we will use the textbook nomenclature   \cite{cohen1992atom} and refer to the diagonal elements ($s{=}s'$) of the Wigner distribution~(\ref{eq:Population_Peierls_Generalized}) as ``populations'' and to the off-diagonal ones ($s{\neq}s'$) as ``coherences''.}
{Representing Eq.~(\ref{eq:Wigner_evolution_equation}) in this phonon\deleted{-mode} basis,
and denoting with $\tfrac{\partial \tenscomp{n}(\bm{R},{\bm{q}},t)_{s,s'\hspace*{-0.5mm}} }{\partial   t}   \big|_{\hat{H}^{\rm per}}$ the scattering  superoperator in the basis of the phonon eigenmodes (which will be discussed later), we obtain the \nametheory{} transport equation}\vspace*{-5mm}
\begin{widetext}
\vspace*{-1.5mm}
\begin{equation}
\begin{split}
\frac{\partial  }{\partial  t} \tenscomp{n}({\bm{R}},{\bm{q}},t)_{s,s'}
&+i\Big[{\omega}(\bm{q})_s\tenscomp{n}(\bm{R},\bm{q},t)_{s,s'}-\tenscomp{n}(\bm{R},\bm{q},t)_{s,s'}{\omega}(\bm{q})_{s'}\Big]  \\
&+\frac{1}{2}\Big\{\sum_{s'\!'}{\tens{v}}(\bm{q})_{s,s'\!'}{\cdot}\nabla_{\bm{R}} \tenscomp{n}(\bm{R},\bm{q},t)_{s'\!',s'}+
\sum_{s'\!'}\nabla_{\bm{R}} \tenscomp{n}(\bm{R},\bm{q},t)_{s,s'\!'} {\cdot}{\tens{v}}(\bm{q})_{s'\!',s'}
\Big\}=\frac{\partial\tenscomp{n}(\bm{R},{\bm{q}},t)_{s,s'} }{\partial  t}   \bigg|_{\hat{H}^{\rm per}},
\end{split}
\label{eq:Wigner_evolution_equation_N}
\end{equation}
\vspace*{-1.5mm}
\end{widetext}
where \added{the} terms in \added{the} square bracket are reminiscent of the commutator appearing in the quantum evolution equation~(\ref{eq:evol_density_matrix}); the time derivative and the terms in curly brackets generalize the drift term of the LBTE~(\ref{eq:BTE_std}), since the diagonal elements ($s{=}s'$) of these terms are equal to the product between the phonon group velocities and phonon populations appearing in the LBTE (this will be discussed more in detail later in Sec.~\ref{sec:solution_of_the_wigner_equation_and_generalized_conductivity_formula}).
The Wigner scattering  superoperator in Eq.~(\ref{eq:Wigner_evolution_equation_N}) is derived from the many-body scattering  superoperator of Eq.~(\ref{eq:lindblad_d_matrix_S})  following a textbook approach~\cite{vasko2006quantum}. Here, we account for scattering due to third-order anharmonicity and perturbative mass disorder~\cite{tamura1983isotope}; we neglect energy-renormalization effects \cite{rossi2011theory}, and we follow the standard procedure of treating scattering at linear order in the deviation from equilibrium~\cite{ziman1960electrons}. The result is
\begin{equation}
\begin{split}
    &\frac{{\partial }\tenscomp{n}(\!{\bm{R},\!\bm{q},\!t})_{s{,}s'} \!}{{\partial }  t}   \bigg|_{\hat{H}^{\rm per}}\!\!\!{=}{-}\delta_{s,s'}\Gamma(\bm{q})_{s}\Big[\tenscomp{n}(\!{\bm{R},\!\bm{q},\!t})_{s,s}{-}\overline{\tenscomp{N}}({\bm{q}})_{s}\Big]\\
    &{+}\frac{\delta_{s,s'}}{(2\pi)^3}\!\sum\limits_{{s'\!'}}\!\! \int\limits_{\mathfrak{B}}\!\!\mathsf{\Omega}^{\rm in}(\bm{q},{\bm{q'\!'}})_{s,{s'\!'}}\Big[\tenscomp{n}(\!{\bm{R},\!\bm{q'\!'},\!t})_{{s'\!'},{s'\!'}}{-}\overline{\tenscomp{N}}({\bm{q'\!'}})_{{s'\!'}}\Big] d^3q'\!'\!\\
    &{-}(1{-}\delta_{s,s'})
    \frac{\Gamma(\bm{q})_{s}{+}\Gamma(\bm{q})_{s'}}{2} \tenscomp{n}(\!{\bm{R},\!\bm{q},\!t})_{s,s'}.
\end{split}
\raisetag{4mm}
\label{eq:formula_scattering}
\end{equation}
Here, $\overline{\tenscomp{N}}({\bm{q'\!'}})_{{s'\!'}}{=}\big\{{\exp\big[\hbar\omega(\bm{q'\!'})_{s'\!'}/(k_B \bar{T}) \big]{-}1}\big\}^{-1}$ is the Bose-Einstein distribution at the equilibrium temperature $\bar{T}$, and $\mathsf{\Omega}^{\rm in}(\bm{q},{\bm{q'\!'}})_{s,{s'\!'}}$ is the scattering matrix that describes the repumping of the populations due to the incoming scattered phonons ($\mathsf{\Omega}^{\rm in}(\bm{q},{\bm{q'\!'}})_{s,{s'\!'}}{=}{-}{A^{\rm in}_{\bm{q}s,\bm{q'\!'}\!s'\!'} }/\big\{{\overline{\tenscomp{N}}(\bm{q'\!'})_{s'\!'}[\overline{\tenscomp{N}}({\bm{q'\!'}})_{{s'\!'}}{+}1]}\big\}$, where ${A^{\rm in}_{\bm{q}s,\bm{q'\!'}\!s'\!'}}$ is defined by the anharmonic and mass-disorder contributions in Eq.~(15) of Ref.~\cite{fugallo2013ab}). 
$\hbar\Gamma(\bm{q})_{s}$ is the linewidth of the phonon with wavevector $\bm{q}$ and mode $s$ (related to the phonon lifetime $\tau(\bm{q})_{s}$  via $[\Gamma(\bm{q})_{s}]^{-1}{=}\tau(\bm{q})_{s}$, see e.g. Eq.~(14) of Ref.~\cite{fugallo2013ab} for the definition of the phonon lifetime $\tau(\bm{q})_{s}$), and accounts for all the scattering events that decrease \deleted{(depumping)} the populations or coherences involving phonons with wavevector $\bm{q}$ and mode $s$.
The conservation of energy in anharmonic and mass-disorder scattering \cite{spohn2006phonon} implies the following relation between
 the ``depumping rate'' $\Gamma(\bm{q})_{s}$ and the ``repumping matrix''  $\mathsf{\Omega}^{\rm in}\!(\bm{q},{\bm{q'\!'}})_{s,{s'\!'}}$ 
\begin{equation}
\begin{split}
   \Gamma(\bm{q})_{s}{=}\!\sum\limits_{{s'\!'}}\!\! \int\limits_{\mathfrak{B}}\!\!\!\mathsf{\Omega}^{\rm in}\!(\bm{q},{\bm{q'\!'}})_{s,{s'\!'}}\!\frac{\!\omega({\bm{q'\!'}})_{s'\!'}\overline{\tenscomp{N}}({\bm{q'\!'}})_{{s'\!'}}\![\overline{\tenscomp{N}}({\bm{q'\!'}})_{{s'\!'}}{+}1]\!}{\omega({\bm{q}})_{s}\overline{\tenscomp{N}}({\bm{q}})_{{s}}[\overline{\tenscomp{N}}({\bm{q}})_{{s}}{+}1]}\frac{d^3q'\!'}{\!(2\pi)^3}.
   \label{eq:relation_scatt_in_out} 
\end{split}
\raisetag{2mm}
 \end{equation} 
From Eq.~(\ref{eq:formula_scattering}) it is evident that populations scatter only with populations, and coherences scatter only with coherences. Importantly, Eq.~(\ref{eq:formula_scattering}) shows that scattering does not create but only destroys off-diagonal coherences $s{\neq}s'$ (since the scattering term for coherences in the third line of Eq.~(\ref{eq:formula_scattering}) has negative sign), while it can both create (term with plus sign in the second line of Eq.~(\ref{eq:formula_scattering}))
or destroy (term with minus sign in the first line of Eq.~(\ref{eq:formula_scattering})) populations.
It follows that scattering drives the Wigner distribution towards a distribution 
that is diagonal in the phonon-mode indexes, which coincides with the 
local-equilibrium state towards which the Peierls-Boltzmann transport equation evolves to \cite{spohn2006phonon,lepri2016thermal}. We conclude by noting that the scattering superoperator has been linearized for convenience in the deviation from equilibrium, to report equations containing quantities available in LBTE solvers \cite{phono3py,alamode,paulatto2015first,fugallo2013ab,carrete2017almabte,li2014shengbte}, but  it can be extended to the non-linear order to describe thermal transport in the far-from-equilibrium regime~\cite{ziman1960electrons,vasko2006quantum}. Moreover, the description of scattering employed here relies on the assumption that phonons are well defined excitations, \textit{i.e.} their linewidth is small compared to their energy, $\hbar\gamma(\bm{q})_s{<}\hbar\omega(\bm{q})_s$ \cite{landau1980statistical}. When this assumption breaks down, it is no longer possible to describe scattering in terms of the phonon wavevector $\bm{q}$ and mode $s$ as in Eq.~(\ref{eq:formula_scattering}), and one has to consider phonon spectral functions as discussed in Refs.~\cite{caldarelli2022,dangic2020origin}.

\section{Thermal conductivity }
\label{sec:solution_of_the_wigner_equation_and_generalized_conductivity_formula}

\subsection{Vibrational energy field and flux }
\label{sub:vibrational_energy_field_and_flux}

We \replaced{have seen}{start by recalling} that the Wigner framework provides prescriptions to resolve in space expectations values (Eq.~(\ref{eq:trace_property_main}) and related discussion). This property allows to compute the expectation value of the harmonic vibrational energy of a crystal integrating the space-dependent Wigner energy field.
% determined following the aforementioned prescriptions.
To see this, we first recast the expectation value of the harmonic Hamiltonian~(\ref{eq:expectation_value}) in reciprocal space using the inverse of Eq.~(\ref{eq:one_body_T1}):
\begin{equation}
\begin{split}
{\rm Tr}\big(\hat\rho(t)\hat{H}^{\mathrm{har}} \big){=}&
\frac{\mathcal{V}^2}{(2\pi)^6}
\!\!\sum_{\substack{b,\alpha, b'\!,\alpha'}}\;
\int\!\!\!\!\!\int\limits_{\!\!\!\mathfrak{B}\mathfrak{B}}\!\!
   \hbar\sqrt{\tenscomp{D}(\bm{q})}_{b\alpha,b'\!\alpha'\!}
\frac{(2\pi)^3}{\mathcal{V}}\delta(\bm{q'\!'})\\
 &{\times}  \varrho\big(\bm{q}{-}\tfrac{\bm{q'\!'}}{2},\bm{q}{+}\tfrac{\bm{q'\!'}}{2},t\big)_{b'\!\alpha'\!,b\alpha}d^3q\; d^3q'\!'.
\end{split}
\label{eq:trace_Fourier}
\raisetag{5mm}
\end{equation}
{Combining Eq.~(\ref{eq:trace_Fourier}) with Eq.~(\ref{eq:trace_property_main}) allows to compute the expectation value of the harmonic vibrational energy as spatial integration of Wigner's energy field $E(\bm{R})$ \cite{physreve.81.021119}, ${\rm Tr}(\hat{\rho}\hat{H}^{\rm har}){=}\mathcal{V}\sum_{\bm{R}}E(\bm{R})$, where}
\begin{align}
     E(\bm{R})&=\frac{1}{(2\pi)^3}
     \sum_{\substack{b,\alpha,b'\!,\alpha'}}
    \int_{\mathfrak{B}}
\hbar\sqrt{\tenscomp{D}(\bm{q})}_{b\alpha,b'\!\alpha'}
\tenscomp{w}(\bm{R},\bm{q})_{b'\!\alpha',b\alpha}
d^3\!q\nonumber \\
&=\frac{1}{(2\pi)^3}
\sum_{s,s'}
    \int_{\mathfrak{B}}
\hbar{\omega(\bm{q})}_{s}\delta_{s,s'}
\tenscomp{n}(\bm{R},\bm{q})_{s'\!,s}
d^3\!q.
\label{eq:local_energy_field}
\end{align}
{The first line of Eq.~(\ref{eq:local_energy_field}) shows that the  Wigner representation of the translation-invariant square root mass-renormalized force-constant matrix $\sqrt{\tenscomp{G}}_{\bm{R}b\alpha,\bm{R'}\!b'\!\alpha'}$ does not depend on space and is equal to the square root of the dynamical matrix in reciprocal Bloch representation $\sqrt{\tenscomp{D}(\bm{q})}_{b\alpha,b'\!\alpha'}$ (see Appendix~\ref{sec:properties_of_the_wigner_transform} for details).}
\replaced{The second line of Eq.~(\ref{eq:local_energy_field}) has been obtained recasting the first line in the phonon eigenmodes basis (see Eq.~(\ref{eq:diagonalization_dynamical_matrix}) for the definition of the phonon eigenmodes basis).}{The second line of Eq.~(\ref{eq:local_energy_field}) has been obtained recasting the first line of such an equation in the phonon eigenmodes basis defined in Eq.~(\ref{eq:diagonalization_dynamical_matrix}) and used in Eq.~(\ref{eq:Population_Peierls_Generalized}).}

The time \replaced{derivative}{derive} of the Wigner  energy field $E(\bm{R})$, which is intensive, can be related directly to the gradient of the intensive heat flux $\bm{J}(\bm{R},t)$. \deleted{Specifically,} Using Eq.~(\ref{eq:Wigner_evolution_equation_N}) to evaluate the time derivative of Eq.~(\ref{eq:local_energy_field}) one obtains a continuity equation that relates the time derivative of the energy field to the heat flux $\bm{J}(\bm{R},t)$: $\frac{\partial}{\partial t} {E(\bm{R},t)}{=}{-}\nabla{\cdot} \bm{J}(\bm{R},t){+}\sigma(\bm{R},t)$, where
\begin{equation}
\begin{split}
  \bm{J}(\bm{R},t)
  &{=}\frac{\hbar}{(2\pi)^3}\!\!
\sum_{s,s'}
  \int_{\mathfrak{B}}\!\!\!\!
  \frac{\omega(\bm{q})_{\!s}{+} \omega(\bm{q})_{\!s'}\!}{2} { {\tens{v}}(\bm{q})_{s,{s'} }\tenscomp{n}(\bm{R},\bm{q},t)_{s'\!,s}} d^3q
\end{split}
\raisetag{14mm}
  \label{eq:def_flux}
\end{equation}
is the heat flux that originates from the leading harmonic Hamiltonian $\hat H^{\rm har}$, while $\sigma(\bm{R},t)$ originates from the perturbation $\hat H^{\rm per}$. In the following $\sigma(\bm{R},t)$ will be neglected, since with respect to $\nabla\cdot \bm{J}(\bm{R},t)$ it is of higher order in the perturbative treatment of anharmonicity.

Importantly, the spatial average of the \added{harmonic} Wigner heat-flux~(\ref{eq:def_flux}) differs from the well-known harmonic heat-flux derived by Hardy \cite{hardy1963energy}.
Such a difference is discussed in detail in Ref. \cite{caldarelli2022}, and originates from a difference in the expressions for the harmonic microscopic energy field used as starting point for the computation of the heat flux here (Eq.~(\ref{eq:local_energy_field})) and in Hardy's work \cite{hardy1963energy}.

\subsection{Steady-state solution of the \RedNmTh{} }
\label{sub:steady_state_solution_of_the_wbte}
The thermal conductivity is defined as the tensor $\kappa^{\alpha \beta}$ that relates a temperature gradient to the heat flux generated in response to it: ${J}^\alpha{=}{-}\sum_\beta\kappa^{\alpha \beta}\nabla T^\beta$.
Consequently, one can determine the thermal conductivity as follows:
(i) fix a temperature gradient constant in time;
(ii) solve Eq.~(\ref{eq:Wigner_evolution_equation_N}) with boundary conditions corresponding to the temperature gradient in (i);
(iii) determine the heat flux inserting the solution obtained in (ii) into Eq.~(\ref{eq:def_flux});
(iv) determine the thermal conductivity as a tensor which relates the temperature gradient (i) and the heat flux (iii).
In this section we will determine the thermal conductivity following the protocol above.

We consider a system where a steady and space-dependent temperature $T(\bm{r}){\simeq} \bar{T}{+}\nabla T{\cdot} \bm{r}$ is enforced along a certain direction.
In addition, the temperature gradient 
is assumed to be small and to be related to temperature variations appreciable over a length scale $L$ much larger than the interatomic spacing $a$,  \textit{i.e.} $\nabla T{\simeq}{-}\frac{\delta T}{L}$, where $\delta T {\ll}\bar{T}$.  
These considerations allow us to look for a steady-state solution for Eq.~(\ref{eq:Wigner_evolution_equation_N}) as a perturbation of order $\nabla T$ in the local-equilibrium distribution corresponding to the local temperature $T(\bm{r})$ (see e.g. Ref.~\cite{hinch1991perturbation} for the mathematical details). Therefore, we look for a solution of the form
\begin{equation}
\begin{split}
    \tenscomp{n}(\bm{R},{\bm{q}})_{\!s,s'}{=}\overline{\tenscomp{N}}({\bm{q}})_{\!s}\delta_{s,s'}{+}\bar{\tenscomp{n}}({\bm{R}}, {\bm{q}})_{\!s}\delta_{s,s'}{+}{\tens{n}}^{(1)}\hspace*{-0.5mm}({\bm{q}})_{\!s,s'}{\cdot}\nabla T,\hspace*{7mm}
  \raisetag{4mm}
 \end{split}
  \label{eq:expansion}
\end{equation}
where $\overline{\tenscomp{N}}(\bm{q})_s$ is the equilibrium  Bose-Einstein distribution at temperature $\bar{T}$,  $\bar{\tenscomp{n}}({\bm{R}}, {\bm{q}})_{s}\delta_{s,s'}{=}
\frac{d \bar{\tenscomp{N}}({\bm{q}})_{s}}{d \bar{T}}(T(\bm{R}){-}\bar{T})\delta_{s,s'}$ is the distribution that 
describes the local-equilibrium state corresponding to a space-dependent temperature (such a local-equilibrium distribution has been discussed e.g. in Refs.~\cite{hardy1970phonon,PhysRevB.98.085427}),
and ${\tens{n}}^{(1)}\hspace*{-0.5mm}({\bm{q}})_{s,s'}{\cdot}\nabla T$  is in general  non-diagonal in $s,s'$ and contains the information concerning the deviation of the full solution from the local equilibrium solution (it is assumed to be of the order of the temperature gradient, as in previous work    \cite{fugallo2013ab,cepellotti2016thermal}).
We recall that, even though the phase-space distributions~(\ref{eq:Wigner_transform_Rec}) are defined everywhere in space, only the values of these distributions at the Bravais lattice sites appear in Eq.~(\ref{eq:Wigner_evolution_equation_N}) and Eq.~(\ref{eq:expansion}), since the knowledge of the Wigner distributions at these points is sufficient to fully describe the problem, as explained in Sec.~\ref{sub:weyl_wigner_transform}.

Inserting the expansion~(\ref{eq:expansion}) into the \RedNmTh{}~(\ref{eq:Wigner_evolution_equation_N}) at steady state, and considering only terms linear in the temperature gradient, we obtain a matrix equation that is decoupled in its diagonal and off-diagonal parts.
The (diagonal) equation for the populations is the usual steady-state LBTE \cite{fugallo2013ab}:
\begin{equation}
\begin{split}
{\tens{v}}&(\bm{q})_{s,s} {\cdot}\frac{\partial \overline{\tenscomp{N}}(\bm{q})_{s} }{\partial \bar{T} } \nabla T{=}{-}\Gamma(\bm{q})_s{\tens{n}}^{(1)}(\bm{q})_{{s},{s} }{\cdot} \nabla T \\
&{+}\frac{1}{(2\pi)^3}\!\sum_{{s'\!'}}\!\! \int_{\mathfrak{B}}\!\!\!\!\mathsf{\Omega}^{\rm in}(\bm{q},{\bm{q'\!'}})_{s,{s'\!'}}\big[{\tens{n}}^{(1)}(\bm{q'\!'})_{{s'\!'}\!\!,{s'\!'}\! }{\cdot} \nabla T \big] d^3\!q'\!',
\end{split}
\label{eq:populations_steady_state}
\end{equation}
where on the left-hand side of this equation we have used the property that the local-equilibrium distribution $\bar{\tenscomp{n}}({\bm{R}}, {\bm{q}})_{s}\delta_{s,s'}$ is an eigenvector with zero eigenvalue of the populations' scattering integral (the sum of the first two terms in Eq.~(\ref{eq:formula_scattering}),    
$\mathsf{\Omega}^{\rm tot}(\bm{q},{\bm{q'\!'}})_{s,{s'\!'}}{=}\mathsf{\Omega}^{\rm in}(\bm{q},{\bm{q'\!'}})_{s,{s'\!'}}{-}(2\pi)^3\Gamma(\bm{q})_s\delta(\bm{q}{-}{\bm{q'\!'}})\delta_{s,s'\!'}$)
\cite{hardy1970phonon,spohn2006phonon,PhysRevX.10.011019}; thus, $\sum_{{s'\!'}}
\int_{\mathfrak{B}}\!\mathsf{\Omega}^{\rm tot}(\bm{q},{\bm{q'\!'}})_{s,{s'\!'}}\bar{\tenscomp{n}}(\bm{R},\bm{q'\!'})_{{s'\!'}} d^3q'\!' {=}0$. 
Eq.~(\ref{eq:populations_steady_state}) can be solved exactly using iterative   \cite{carrete2017almabte,phono3py,alamode}, variational   \cite{fugallo2013ab} or exact diagonalization methods   \cite{cepellotti2016thermal,PhysRevLett.110.265506}.
The off-diagonal equation for the coherences ($s{\neq}s'$) is
\begin{equation}
\begin{split}
&\frac{1}{2} \left( \frac{\partial \overline{\tenscomp{N}}(\bm{q})_{s} }{\partial \bar{T}}+\frac{\partial \overline{\tenscomp{N}}(\bm{q})_{s'} }{\partial \bar{T} } \right){\tens{v}}(\bm{q})_{s,s'} \cdot \nabla T=\\
&-\bigg(i\big(\omega(\bm{q})_s-\omega(\bm{q})_{s'}\big)+\frac{\Gamma(\bm{q})_{s}+\Gamma(\bm{q})_{s'}}{2}\bigg) {\tens{n}}^{(1)}({\bm{q}})_{s,s'}{\cdot}\nabla T,
\end{split}
\raisetag{16mm}
\label{eq:coherences_steady_state}
\end{equation}
\replaced{and its solution reads ($s{\neq}s'$):}{and can be solved straightforwardly, yielding ($s\neq s'$):}
\begin{equation}
  \begin{split}
&{\tenscomp{n}}^{(1),\alpha}({\bm{q}})_{s,s'}{=}{-}\frac{\hbar}{k_{B} {T}^2}{\tenscomp{v}^\alpha}(\bm{q})_{s,s'}\\
&\times
\frac{\omega(\bm{q})_{s}\overline{\tenscomp{N}}({\bm{q}})_{s}[\overline{\tenscomp{N}}({\bm{q}})_{s}+1]+\omega(\bm{q})_{s'}\overline{\tenscomp{N}}({\bm{q}})_{s'}[\overline{\tenscomp{N}}({\bm{q}})_{s'}+1]}{2i[\omega(\bm{q})_{s}-\omega(\bm{q})_{s'}]+[\Gamma(\bm{q})_{s}+\Gamma(\bm{q})_{s'}]}.
\end{split}
\raisetag{16mm}
\label{eq:coherences_sol}
\end{equation}
We note that solving exactly the equation for the populations~(\ref{eq:populations_steady_state}) requires accounting for all the interactions between populations (those described by the scattering matrix $\mathsf{\Omega}^{\rm tot}(\bm{q},{\bm{q'\!'}})_{s,{s'\!'}}$); in contrast, the exact evolution equation for the coherences~(\ref{eq:coherences_steady_state}) 
contains only the coupling of a coherence ${\tenscomp{n}}({\bm{R}},{\bm{q}})_{s,s'}$ (with $s{\neq}s'$) to itself.
From a practical point of view, the computational cost for solving exactly the coherences' equation~(\ref{eq:coherences_steady_state}) is negligible compared to that for solving exactly the LBTE (\textit{i.e.} the populations' equation~(\ref{eq:populations_steady_state}))\footnote{Specifically, finding the exact solution of the populations' equation~(\ref{eq:populations_steady_state}) requires applying iterative or variational algorithms to a matrix of size $(N_c{\times}3{\times} N_{\rm at})^2$ (where $N_c$ is the number of q-points used to sample the Brillouin zone in numerical calculations), while solving the equation for coherences requires just knowing the entries of a vector of size $N_c{\times}(3{\times} N_{\rm at})^2$.}.

After having determined the steady-state solution of the \RedNmTh{}~(\ref{eq:expansion}) (\textit{i.e.} the sum of the solutions of Eq.~(\ref{eq:populations_steady_state}) and of Eq.~(\ref{eq:coherences_steady_state})), one can insert it into Eq.~(\ref{eq:def_flux}) and thus evaluate the heat flux. 
At this point it is \replaced{possible}{straightforward} to show that only the deviation-from-equilibrium \deleted{part} ${\tens{n}}^{(1)}\hspace*{-0.5mm}({\bm{q}})_{s,s'}{\cdot}\nabla T$ of the solution~(\ref{eq:expansion}) gives a non-zero contribution to the heat flux, since the Bose-Einstein term ($\overline{\tenscomp{N}}({\bm{q}})_s\delta_{s,s'}$) and the local-equilibrium term ($\bar{\tenscomp{n}}({\bm{R}}, {\bm{q}})_{s}\delta_{s,s'}$) are both even functions of the wavevector; %($f(-\bm{q}){=}f(\bm{q})$), 
\deleted{and} thus, when multiplied by the diagonal elements of the velocity operator --- which have odd parity in the wavevector    --- they yield an odd-parity function whose integral over the symmetric Brillouin zone yields zero \cite{PhysRevX.10.011019}.
Therefore, one obtains a linear relation between heat flux and temperature gradient, $J^\alpha{=}{-}\sum_\beta\kappa^{\alpha\beta}\nabla^\beta T$, with the proportionality tensor being the thermal conductivity:
\begin{widetext}
  \begin{equation}
\begin{split}
\kappa^{\alpha \beta}{=}\kappa^{\alpha \beta}_{\rm P}{+}\frac{1}{(2\pi)^3}\!\int\limits_{\mathfrak{B}}\!\sum_{s\neq s'}\frac{\omega(\bm{q})_{s}{+}\omega(\bm{q})_{s'}}{4}\!
\left[\frac{C(\bm{q})_s}{\omega(\bm{q})_s}{+}\frac{C(\bm{q})_{s'}}{\omega(\bm{q})_{s'}}\right]\!
{\tenscomp{v}^\alpha}(\bm{q})_{s,s'\!}{\tenscomp{v}}^\beta(\bm{q})_{s'\!,s}\frac{\tfrac{1}{2}\big[\Gamma(\bm{q})_{s}{+}\Gamma(\bm{q})_{s'}\big]}{[\omega(\bm{q})_{s'}{-}\omega(\bm{q})_{s}]^2+\tfrac{1}{4}[\Gamma(\bm{q})_{s}{+}\Gamma(\bm{q})_{s'}]^2}d^3q;
  \label{eq:thermal_conductivity_final_sum}
  \end{split}
  \raisetag{15mm}
\end{equation}
\end{widetext}
where $C(\bm{q})_{s}{=}\tfrac{\hbar^{2}\omega^2(\bm{q})_s}{k_B\bar{T}^2}\overline{\tenscomp{N}}(\bm{q})_s\big[\overline{\tenscomp{N}}(\bm{q})_s{+}1\big]$ is the specific heat of a phonon with wavevector $\bm{q}$ and mode $s$, $\kappa^{\alpha \beta}_{\rm P}$ is the ``populations conductivity'' obtained from the exact (\textit{i.e.} iterative   \cite{carrete2017almabte,omini1995iterative}, variational   \cite{fugallo2013ab}, or exact-diagonalization   \cite{PhysRevLett.110.265506,cepellotti2016thermal,phono3py}) solution of the diagonal (population) part of the \RedNmTh{}~(\ref{eq:populations_steady_state}) --- which is exactly the Peierls-Boltzmann equation for phonon wavepackets   \cite{peierls1955quantum} --- and the additional tensor derives from the equation for the coherences~(\ref{eq:coherences_steady_state}), thus it is called here ``coherences conductivity'' ($\kappa^{\alpha \beta}_{\rm C}$).
Eq.~(\ref{eq:thermal_conductivity_final_sum}) can be explicitly calculated by evaluating the integral over the Brillouin zone on a uniform mesh of wavevectors, \textit{i.e.} replacing 
$\frac{1}{(2\pi)^3}\int_{\mathfrak{B}} d^3q\to\frac{1}{\mathcal{V}N_c }\sum_{\bm{q}} $, where $N_c$ is the number of $\bm{q}$-points that sample the Brillouin zone and over which the sum runs, and $\mathcal{V}$ is the volume of the primitive cell \cite{callaway1991quantum}.  

We stress that the \RedNmTh{} conductivity~(\ref{eq:thermal_conductivity_final_sum}) derives from an exact solution of the \RedNmTh{}~(\ref{eq:Wigner_evolution_equation_N}) and thus is also valid in the hydrodynamic regime, where the scattering between phonons that conserve the crystal momentum (``normal processes'') dominate and yield a Peierls-Boltzmann thermal conductivity that is very high ($\kappa_P(300\; K)\gtrsim 10^3$ W/mK) and several order of magnitude larger than the coherences conductivity  \cite{PhysRevX.10.011019}.
It is also worth mentioning that in the case of low-thermal-conductivity solids, where ``Umklapp'' scattering between phonon wavepackets that dissipate the crystal momentum are dominant,
Eq.~(\ref{eq:formula_scattering}) contains depumping (negative sign) terms that are dominant with respect to the repumping (positive sign) terms.
Therefore, one can \replaced{apply}{perform} the \deleted{so-called} single-mode relaxation-time approximation (SMA) \cite{lindsay_first_2016,fugallo2013ab}, which consists in discarding the repumping terms in Eq.~(\ref{eq:formula_scattering}).
In practice, this approximation 
 implies discarding the second line (the repumping term) in the populations' equation~(\ref{eq:populations_steady_state}), and leaves the coherences' equation~(\ref{eq:coherences_steady_state}) unchanged.
It follows that within this approximation nothing changes for the coherences, \replaced{while the populations equation assumes a simpler form that can be solved at a greatly reduced computational cost compared to the full solution.}{and the populations equation can be solved straightforwardly  (and at a computational cost similar to that for solving exactly the equation for the coherences~(\ref{eq:coherences_steady_state})), yielding:} \added{Therefore, solving the populations' equation simplified with the SMA, one obtains:}
\begin{equation}
\medmuskip=0mu
\thinmuskip=-1mu
\thickmuskip=-1mu
  \begin{split}
    &\hspace*{-5mm}{\tenscomp{n}}^{(1),\alpha}_{\rm SMA}({\bm{q}})_{s}=-\frac{\hbar}{k_{B} \bar{T}^2}{\tenscomp{v}^\alpha}(\bm{q})_{s,s}\frac{\omega(\bm{q})_{s}\overline{\tenscomp{N}}({\bm{q}})_{s}[\overline{\tenscomp{N}}({\bm{q}})_{s}+1]}{\Gamma(\bm{q})_{s}}.
  \end{split}
  \raisetag{5mm}
  \label{eq:SMA_pop}
\end{equation}
When the approximated SMA solution for the populations~(\ref{eq:SMA_pop}) 
is inserted into the expression for the heat flux~(\ref{eq:def_flux}), the populations' conductivity (term $\kappa^{\alpha\beta}_{P}$ in Eq.~(\ref{eq:thermal_conductivity_final_sum})) becomes exactly equal to the SMA expression for the Peierls-Boltzmann conductivity \cite{callaway1991quantum}: 
\begin{equation}
\begin{split}
    &\kappa^{\alpha \beta}_{\scriptscriptstyle{\rm P,SMA}} {=}\frac{1}{(2\pi)^3}\!
  \int_{\mathfrak{B}}\! \sum\limits_{s}C(\bm{q})_s{\tenscomp{v}^\alpha\hspace*{-0.5mm}(\bm{q})}_{\hspace*{-0.5mm}s,s}\hspace*{-0.4mm}{\tenscomp{v}}^\beta\hspace*{-0.5mm}(\bm{q})_{\hspace*{-0.5mm}s,s}
  \frac{1}{\Gamma(\bm{q})_{s}}d^3\!q.\hspace*{7mm}
\end{split}
\raisetag{8mm}
  \label{eq:SMA_k}
\end{equation}
Eq.~(\ref{eq:SMA_k}) shows that the populations' conductivity can be interpreted in terms of particle-like phonon wavepackets labeled by $(\bm{q})_s$, which carry the specific heat $C(\bm{q})_s$ and propagate between collisions over a length ${{\Lambda}^\alpha(\bm{q})_{s}}{=}{\tenscomp{v}}^\alpha(\bm{q})_{s,s}[\Gamma(\bm{q})_s]^{-1}$ (as mentioned before, ${\tenscomp{v}}^\alpha(\bm{q})_{s,s}$ is the propagation (group) velocity in direction $\alpha$, and $[\Gamma(\bm{q})_s]^{-1}$ is the typical inter-collision time).  
Importantly, we note that $\kappa^{\alpha\beta}_{P}$ can be interpreted in terms of microscopic carriers propagating particle-like without relying on the SMA approximation, since it has been shown that the conductivity deriving from the exact solution of the population's equation~(\ref{eq:populations_steady_state}) can be determined exactly and in a closed form as a sum over relaxons   \cite{cepellotti2016thermal,PhysRevX.10.011019}, \textit{i.e.} particle-like collective phonon excitations that are the eigenvectors of (a symmetrized version of) the full LBTE scattering matrix $\mathsf{\Omega}^{\rm tot}(\bm{q},{\bm{q'\!'}})_{s,{s'\!'}}$ discussed after Eq.~(\ref{eq:populations_steady_state}).

The microscopic conduction mechanisms that determine the term $\kappa_{_{\rm C}}$ in the conductivity expression~(\ref{eq:thermal_conductivity_final_sum}) is that of phonons tunnelling between two different bands at the same wavevector $\bm{q}$.
Such an interband tunneling mechanism originates from the wave-like nature of phonons, and in general becomes stronger as the coherence between two phonons $(\bm{q})_s$ and $(\bm{q})_{s'}$ becomes stronger (\textit{i.e.} their frequency difference $\omega(\bm{q})_s{-}\omega(\bm{q})_{s'}$ becomes smaller).
We note in passing that this mechanism has some analogies with the Zener interband tunnelling of electrons discussed e.g. in Refs.~\cite{PhysRevB.35.9644,Hubner1996,PhysRevB.86.155433,gebauer2004current,gebauer2004kinetic} relying on the density-matrix formalism.

It has been shown in Ref. \cite{PhysRevX.10.011019}  that in ``simple'' crystals, \textit{i.e.} those characterized by phonon interband spacings much larger that the linewidths (e.g. silicon, diamond), particle-like mechanisms dominate and thus $\kappa^{\alpha \beta}_{\rm P}{\gg}\kappa^{\alpha \beta}_{\rm C}$.
We will discuss in Sec.~\ref{sec:numerical_results} how in the ``complex'' crystal regime, characterized by phonon interband spacings smaller that the linewidths, particle-like and wave-like mechanisms coexist and are both relevant, implying $\kappa^{\alpha \beta}_{\rm P}{\lesssim}\kappa^{\alpha \beta}_{\rm C}$. 
\replaced{
Finally, we note that the Wigner formulation can be applied also to amorphous solids, describing them as limiting cases of disordered but periodic crystals in the limit of infinitely large primitive cells ($N_{at}{\to}\infty$ and consequently $\mathcal{V}{\to}\infty$). Importantly, it is possible to show analytically that in the ordered limit describing a harmonic glass (\textit{i.e.} first letting the primitive cell volume go to infinity $\mathcal{V}{\to}\infty$, and then letting each linewidth go to the same infinitesimal broadening $\hbar\eta$, $\hbar\Gamma(\bm{q})_s{\to}\hbar\eta{\to} 0$, $\forall \;\bm{q},s$ \cite{allen1989thermal,allen1993thermal}) the normalized trace of the LWTE conductivity~(\ref{eq:thermal_conductivity_final_sum}) becomes equivalent to the Allen-Feldman formula for the conductivity of glasses~\cite{simoncelli2019unified}. 
%\deleted{Therefore, i} In this regime wave-like conduction mechanisms dominate. 
More generally, the Wigner formulation extends Allen-Feldman theory accounting also for the effects of anharmonicity on thermal transport in disordered solids, since Eq.~(\ref{eq:thermal_conductivity_final_sum}) accounts for anharmonicity thorough the linewidths.
In practice, evaluating the LWTE conductivity~(\ref{eq:thermal_conductivity_final_sum}) in disordered solids accounting for anharmonicity is a challenging task, since it requires computing all the quantities appearing in Eq.~(\ref{eq:thermal_conductivity_final_sum}) from atomistic models that are sufficiently large to realistically describe disorder and thus computationally expensive.
Moreover, we recall that, analogously to Allen-Feldman theory, the Wigner formulation relies on the hypothesis that atoms vibrate around equilibrium positions; thus, it can be applied only to disordered solids and glasses with a structural stability for which this hypothesis is realistic (see e.g. Refs.~\cite{Egami2019,moon2021examining,Ruta2014,Ross2014,Buchenau1986,Song2019,yu2013beta,PhysRevB.105.014110} for a discussion of structural stability and related phenomena in various glasses). 
Because of the multiple challenges highlighted above, the application of the Wigner formulation to glasses will be subject of future work.
%significantly with the chemical composition and network structure \cite{Egami2019,moon2021examining,Ruta2014,Ross2014,Buchenau1986,Song2019}; therefore
%Moon2018,PhysRevMaterials.3.065601,Kim2021,Liu2009,Braun2016,Martin2022,wingert2016thermal,kwon2017unusually,PhysRevB.101.144203,yu2013beta,PhysRevB.105.014110,
%Therefore applying the Wigner formulation  
%several works have discussed out how the structural stability of glasses is affected  \cite{Moon2018,PhysRevMaterials.3.065601,Kim2021,Liu2009,Braun2016,Martin2022,wingert2016thermal,kwon2017unusually,PhysRevB.101.144203,yu2013beta,PhysRevB.105.014110,Egami2019,moon2021examining}
%depending on the chemical composition and network structure, glasses might hav
%atomic diffusion \cite{moon2021examining} and structural rearrangements 
%of disordered solids with finite-size atomistic models is a challenge that is subject of ongoing research, since disordered solids 
%in disordered solids requires employing finite-size atomistic models to evaluate all the quantities appearing in 
%Describing realistically the vibrational and thermal properties of disordered solids with finite-size atomistic models is a challenge that is subject of ongoing research, since disordered solids \cite{Moon2018,PhysRevMaterials.3.065601,Kim2021,Liu2009,Braun2016,Martin2022,wingert2016thermal,kwon2017unusually,PhysRevB.101.144203,yu2013beta,PhysRevB.105.014110,Egami2019,}
}
{
Finally, we note that for a harmonic glass the coherences conductivity dominates ($\kappa^{\alpha \beta}_{\rm C}{\gg}\kappa^{\alpha \beta}_{\rm P}$) and the normalized trace of Eq.~(\ref{eq:thermal_conductivity_final_sum}) becomes equivalent to the Allen-Feldman formula for the conductivity of glasses.
This can be easily verified evaluating Eq.~(\ref{eq:thermal_conductivity_final_sum}) in the ordered limit describing a harmonic glass, \textit{i.e.} considering: (i) a disordered but periodic crystal with an increasingly larger primitive cell ($N_{at}{\to}\infty$, \textit{i.e.} $\mathcal{V}{\to}\infty$ and thus with the Brillouin zone reducing to the point $\bm{q}{=}\bm{0}$ only); (ii) vanishing anharmonicity, \textit{i.e.} each linewidth reducing to the same infinitesimal broadening $\hbar\eta$  ($\hbar\Gamma(\bm{q})_s{\to}\hbar\eta{\to} 0$, $\forall \;\bm{q},s$), see Ref.~\cite{SimoncelliPhD} for details.}  
A summary of the results obtained here relying on the Wigner framework, and their comparison with the standard Peierls-Boltzmann framework is reported in Table~\ref{tab:compare}. 
%\added{More details on how the Wigner formulation generalizes the Peierls-Boltzmann framework are reported in Appendix~\ref{sec:details_on_generalization}.}

\replaced{It is worth mentioning}{We also note} that Eq.~(\ref{eq:thermal_conductivity_final_sum}) --- with the populations conductivity in the SMA approximation~(\ref{eq:SMA_k}) --- can be obtained from the Green-Kubo formalism considering the Lorentzian spectral-function limit \cite{caldarelli2022}. 
\replaced{We conclude by noting that the thermal conductivity formula~(\ref{eq:thermal_conductivity_final_sum}) has been derived employing two approximations that might be improved: \\
(i) We considered a time-independent and spatially uniform temperature gradient;\\
(ii) We employed the standard perturbative treatment of anharmonic effects \cite{ziman1960electrons,peierls1955quantum,rossi2011theory}, which considers only the lowest third-order anharmonic terms and treats them as a perturbation to the harmonic terms (we recall that in this standard perturbative treatment temperature effects are accounted for through the Bose-Einstein distributions entering in the scattering operator~(\ref{eq:formula_scattering})).\\  
Approximation (i) can potentially be improved following the procedure discussed in Refs.~\cite{PhysRevB.98.085427,allen2021phonon} to account for time- and space-dependent effects on the conductivity. 
Improving approximation (ii) requires accounting for anharmonic terms in the heat flux expression (\textit{i.e.} considering the term $\sigma$ in Sec.~\ref{sub:vibrational_energy_field_and_flux} and thus refining Eq.~(\ref{eq:def_flux})) as well as for higher-than-third-order phonon scattering processes \cite{PhysRevB.93.045202,PhysRevB.96.161201} in the collision operator~(\ref{eq:formula_scattering}).
Another related possible improvement concerns accounting for temperature-renormalization effects on the phonon frequencies and collision operator with accuracy beyond perturbation theory, a challenge that can potentially be tackled relying on the stochastic self-consistent harmonic approximation (SSCHA) \cite{PhysRevB.96.014111,PhysRevLett.122.075901,PhysRevB.100.214307,PhysRevB.103.224307}.
Rigorously accounting for all these anharmonic effects in the Wigner formulation requires additional work, to ensure a consistent treatment of the approximations that have to be performed.
More details on the capability of the approximated treatment of anharmonicity employed here to accurately describe real solids will be provided later in Sec.~\ref{sec:numerical_results} and Appendix~\ref{sec:validation_of_the_perturbative_treatment_of_anharmonicity}. 
%Nevertheless, we will show later that the  is accurate enough for the scope of the present work. In fact, the frequencies and linewidths we computed agree reasonably well with those measured in Raman experiments at various temperatures, and the frequencies we computed yield a specific heat in good agree with that measured in experiments over a broad temperature range.
}{Finally, we note that the thermal conductivity formula~(\ref{eq:thermal_conductivity_final_sum}) has been derived considering a time- independent and spatially uniform temperature gradient; this derivation might be generalized to account for time- and space-dependent effects following the procedure discussed in Refs.~[82,90].}

\begin{table*}[!ht]
	\caption{Comparison between the Peierls-Boltzmann and Wigner frameworks.}
	\label{tab:compare}
\hspace*{-4mm}
	\begin{tabular}{c|c|c}
	\hline

	\hline
	 & \begin{minipage}{0.3\textwidth}    \textbf{Wigner framework}    \end{minipage} & \textbf{Peierls-Boltzmann framework} \\
	 \hline
	\begin{minipage}{0.12\textwidth}
	\vspace*{2mm}		
		Description of vibrational eigenstates 
		\vspace*{2mm}
		\end{minipage}& 
						\multicolumn{2}{c}{\begin{minipage}{0.88\textwidth} 
							\vspace*{2mm}	
Dynamical matrix (\ref{eq:dynamical}),
  $\tenscomp{D}(\bm{q})_{b\alpha,b'\!\alpha'}{=}\sum_{\bm{R}}\tenscomp{G}_{\bm{R}b\alpha,\bm{0}b'\!\alpha'}e^{-i\bm{q}\cdot(\bm{R}{+}\bm{\tau}_b{-}\bm{\tau}_{b'})}$, where  $\bm{\tau}_b$ is the position of the atom $b$ in the primitive cell, $\bm{R}$ is a Bravais-lattice vector, and $\tenscomp{G}_{\bm{R}b\alpha,\bm{0}b'\!\alpha'}$ is the mass-rescaled harmonic force constant tensor~(\ref{eq:matrix_G}). 
  The square vibrational frequencies $\omega^2(\bm{q})_s$ and the phonon eigenvectors $\mathcal{E}(\bm{q})_{s,b\alpha}$ (which describe atoms' displacements) are given by Eq.~(\ref{eq:diagonalization_dynamical_matrix}), 
$\sum_{b'\alpha'}\tenscomp{D}(\bm{q})_{b\alpha,b'\!\alpha'}\mathcal{E}(\bm{q})_{s,b'\!\alpha'}{=}\omega^2(\bm{q})_s\mathcal{E}(\bm{q})_{s,b\alpha}$.\vspace*{2mm}		
\end{minipage}} \\
  \hline
  \begin{minipage}{0.12\textwidth}    
    \vspace*{2mm} Central quantity\vspace*{2mm}
    \end{minipage}
   & Wigner distribution~(\ref{eq:Population_Peierls_Generalized}), $\tenscomp{n}(\bm{R},\bm{q},t)_{s,s'}$
&   Phonon population, $\tenscomp{n}(\bm{R},\bm{q},t)_{s}=\tenscomp{n}(\bm{R},\bm{q},t)_{s,s}$
   \\
		\hline
	\begin{minipage}{0.12\textwidth}		
		\vspace*{2mm}	
		Description of velocity
			\vspace*{2mm}	
		\end{minipage}& 
	\begin{minipage}{0.42\textwidth}		
		\vspace*{2mm}	
		Velocity operator~(\ref{eq:vel_op}),\\[2mm]
		$\tenscomp{v}^\beta(\bm{q})_{s,s'}{=}\!\!\!\!\sum\limits_{{b\alpha,b'\!\alpha'}}\!\!\!\!\!\mathcal{E}^\star(\bm{q})_{{s},b\alpha}{{\nabla^{\!\beta}_{\!\bm{q}}\!\sqrt{\tenscomp{D}(\bm{q})}} }_{b\alpha,b'\hspace*{-0.5mm}\alpha'} \mathcal{E}(\bm{q})_{s'\hspace*{-0.5mm},b'\hspace*{-0.5mm}\alpha'}$
			\vspace*{2mm}	
		\end{minipage}&
      \begin{minipage}{0.42\textwidth}    
    \vspace*{2mm} 
    Group velocity, 
    $\tenscomp{v}^\beta\!(\bm{q})_{s}=\tenscomp{v}^\beta\!(\bm{q})_{s,s}$
      \vspace*{2mm} 
    \end{minipage}
		\\
	\hline
\begin{minipage}{0.12\textwidth}		
Evolution equation 
\end{minipage}
& 
\begin{minipage}{0.42\textwidth}	
\vspace*{2mm}	
Wigner transport equation~(\ref{eq:Wigner_evolution_equation_N})
\includegraphics[width=7.5cm]{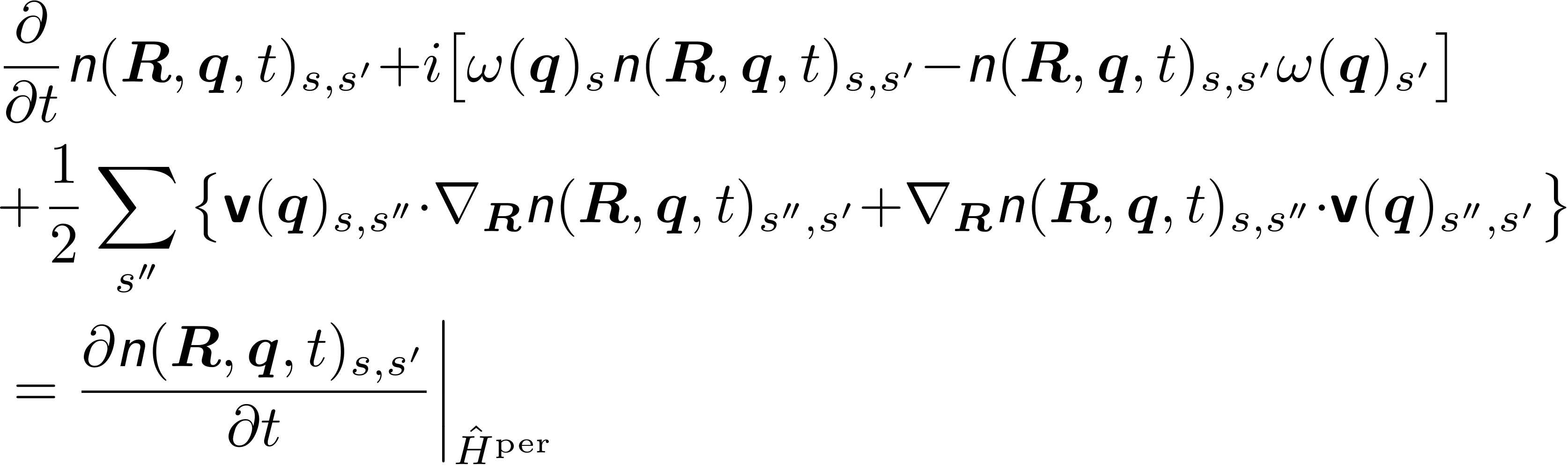}
\vspace*{-3mm}	
\end{minipage}
&
\begin{minipage}{0.42\textwidth}  
      \vspace*{1mm} 
Peierls-Boltzmann transport equation~(\ref{eq:BTE_std}) 
(equal to the diagonal elements of Eq.~(\ref{eq:Wigner_evolution_equation_N}) \cite{FootnoteTable})
\begin{eqnarray}
  \tfrac{\partial  }{\partial  t} \tenscomp{n}({\bm{R}},{\bm{q}},t)_{s}
{+}{\tens{v}}(\bm{q})_{s}{\cdot}{\nabla}_{\bm{R}} \tenscomp{n}(\bm{R},\bm{q},t)_{s}{=}\tfrac{\partial \tenscomp{n}({\bm{R}},{\bm{q}},t)_{s} }{\partial  t} \!\big|_{\!\hat{H}^{\rm per}}
\nonumber
\end{eqnarray}
      \vspace*{1mm} 
\end{minipage}
\\
		\hline
\multirow{2}{*}{	
	\begin{minipage}{0.12\textwidth}
  \vspace*{1.3cm}
  Linearized
		description of vibrations' scattering
		\end{minipage} }& 
						\multicolumn{2}{c}{\begin{minipage}{0.82\textwidth}\raggedright 
							\vspace*{2mm}		
 Phonon linewidths $\Gamma(\bm{q})_s$ and repumping scattering matrix $\mathsf{\Omega}^{\rm in}(\bm{q},\bm{q'\!'})_{s,s'\!'}$ (related by Eq.~(\ref{eq:relation_scatt_in_out})). 
\vspace*{2mm}		
\end{minipage}} \\
\cline{2-3}
& 	
\begin{minipage}{0.42\textwidth}
\vspace*{1mm}   
Wigner scattering superoperator, Eq.~(\ref{eq:formula_scattering})\\[1mm]
\includegraphics[width=7cm]{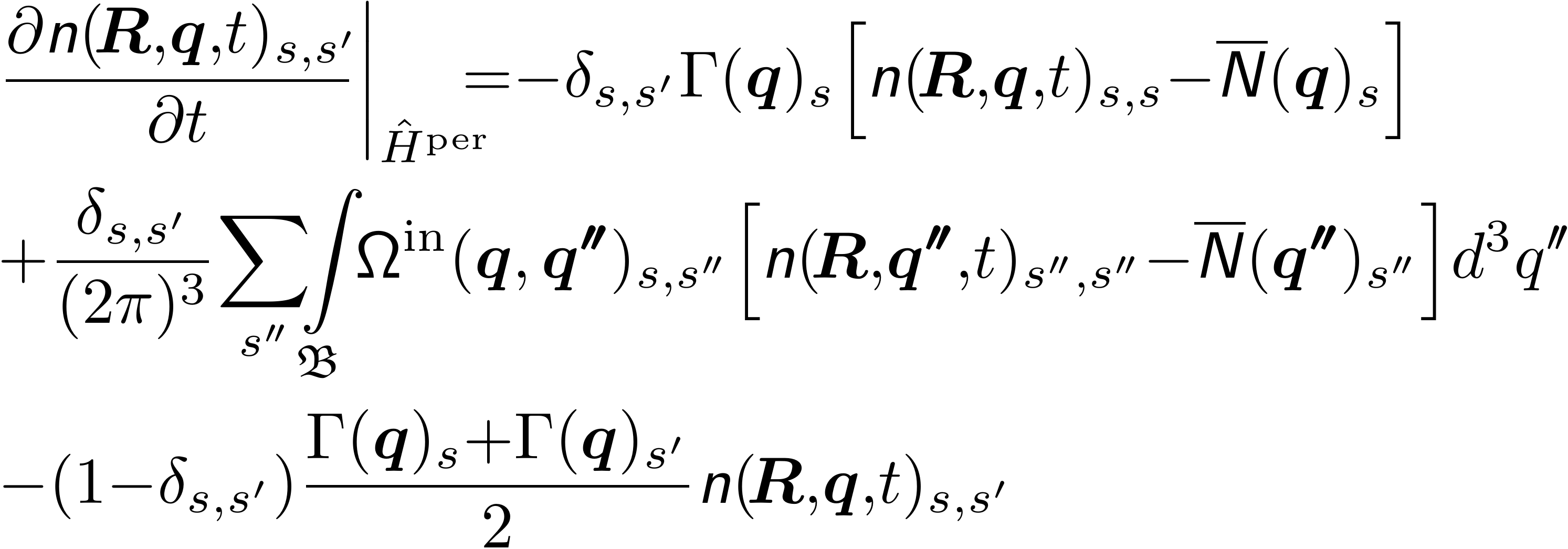}
\vspace*{2mm}   
\end{minipage} 
& \begin{minipage}{0.42\textwidth}
\vspace*{1mm}	
LBTE scattering superoperator, \textit{i.e.} first two lines in Eq.~(\ref{eq:formula_scattering}).\\[1mm]
\includegraphics[width=6.5cm]{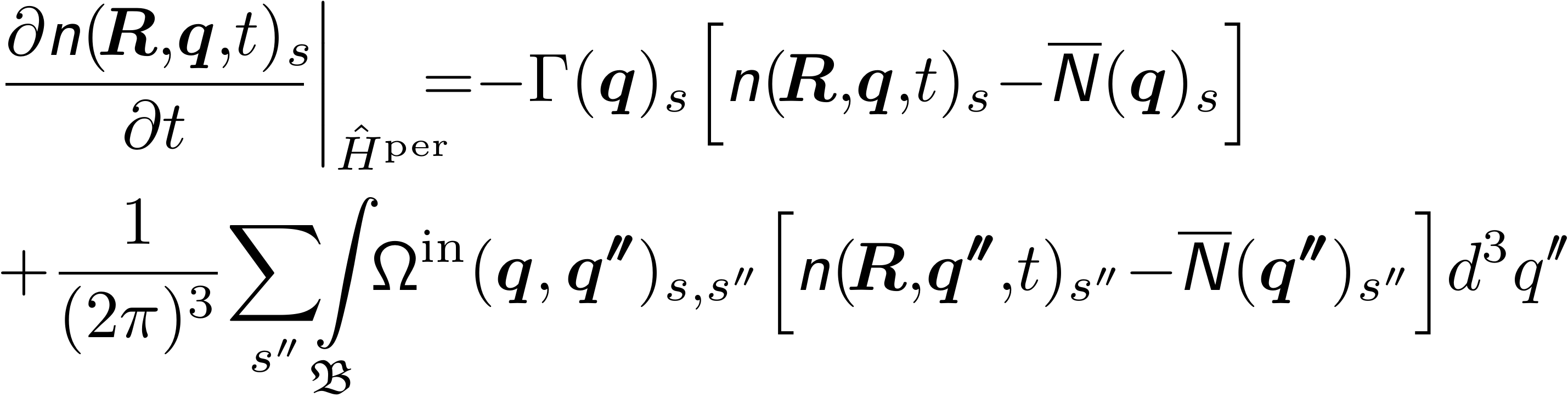}\vspace*{1mm} 
		\end{minipage} \\
        \hline
  \begin{minipage}{0.12\textwidth}    
    \vspace*{2mm} 
Heat flux
      \vspace*{2mm} 
    \end{minipage}& 
  \begin{minipage}{0.45\textwidth}    
    \vspace*{2mm} 
    Wigner's heat flux~(\ref{eq:def_flux}),\\[2mm]
    \includegraphics[width=7.5cm]{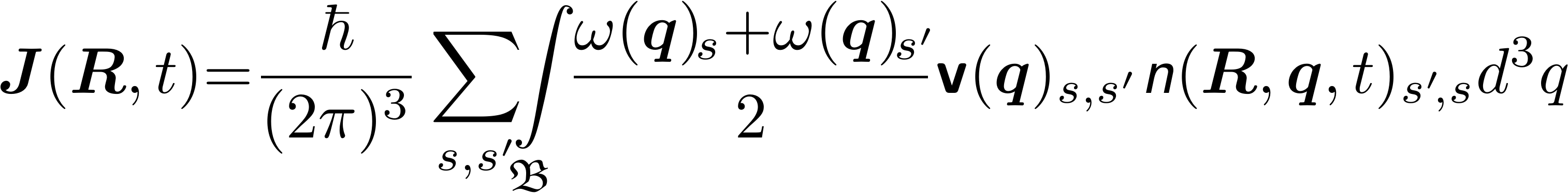}\vspace*{1mm}
    \end{minipage}&
      \begin{minipage}{0.42\textwidth}    
    \vspace*{2mm} 
Obtained from Wigner's heat flux (\ref{eq:def_flux}) considering only diagonal elements ($s{=}s'$), \\[2mm]
    \includegraphics[width=6.2cm]{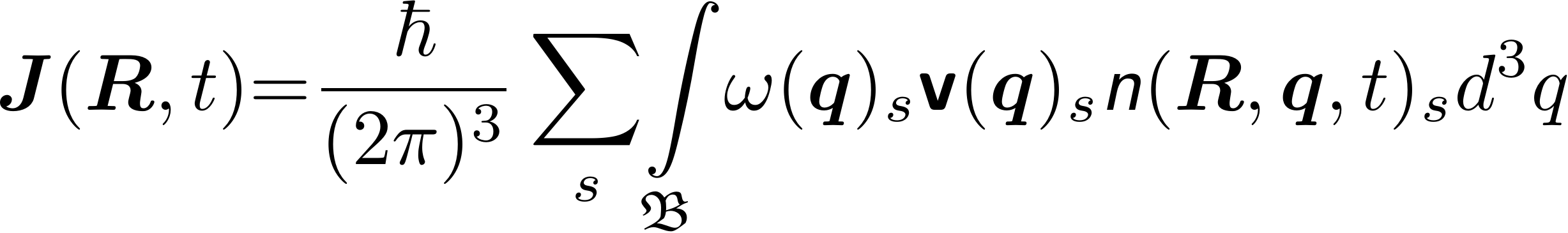}\vspace*{1mm}
    \end{minipage}
    \\
	\hline
\begin{minipage}{0.12\textwidth}		
Thermal conductivity
\end{minipage}
 &
\begin{minipage}{0.42\textwidth}  
\vspace*{2mm} 
$\kappa^{\alpha\beta}=\kappa_{\rm P}^{\alpha\beta}+\kappa_{\rm C}^{\alpha\beta}$ (Eq.~(\ref{eq:thermal_conductivity_final_sum})),\\
 $\kappa_{\rm P}^{\alpha\beta}$ from the solution of Eq.~(\ref{eq:populations_steady_state}),\\ and 
$\kappa_{\rm C}^{\alpha\beta}$ from the solution of  Eq.~(\ref{eq:coherences_steady_state}):
\begin{eqnarray}
&\kappa^{\alpha\beta}_{\rm C}{=}\frac{1}{(2\pi)^3}\!\int\limits_{\mathfrak{B}}\!\!\sum\limits_{s\neq s'}\!\!\!\frac{\omega(\bm{q})_{s}{+}\omega(\bm{q})_{s'}}{4}\!
\left[\frac{C(\bm{q})_s\!}{\omega(\bm{q})_s}{+}\frac{C(\bm{q})_{s'}\!}{\omega(\bm{q})_{s'}}\right]\!\!{\tenscomp{v}^\alpha}(\bm{q})_{s,s'\!}\!
\nonumber\\
&
\hspace*{7mm}{\times}{\tenscomp{v}}^\beta(\bm{q})_{s'\!,s}\frac{\tfrac{1}{2}\big[\Gamma(\bm{q})_{s}{+}\Gamma(\bm{q})_{s'}\big]}{[\omega(\bm{q})_{s'}{-}\omega(\bm{q})_{s}]^2+\tfrac{1}{4}[\Gamma(\bm{q})_{s}{+}\Gamma(\bm{q})_{s'}]^2}d^3q.\nonumber
\end{eqnarray}\vspace*{-2mm} 
\end{minipage}
& 
\begin{minipage}{0.42\textwidth}	
$\kappa_{\rm P}^{\alpha\beta}$, from the solution of Eq.~(\ref{eq:populations_steady_state}).
\end{minipage}
\\
\hline
\multirow{2}{*}{  
  \begin{minipage}{0.12\textwidth}
  \vspace*{1.3cm}
 Regime of validity
    \end{minipage} }& 
            \multicolumn{2}{c}{
            \begin{minipage}{0.85\textwidth}  
              \vspace*{2mm}   
  \added{Vibrations are well defined quasiparticles excitations, $\hbar\omega(\bm{q})_s{>}\hbar\Gamma(\bm{q})_s\;\forall \;\bm{q},s$ (when this condition is not verified, vibrations are overdamped, neither the Wigner nor the Peierls-Boltzmann approach can be applied and spectral-function approaches are needed \cite{caldarelli2022}).}
              \vspace*{2mm}   
              \end{minipage}
            } \\
\cline{2-3}
&   
\begin{minipage}{0.42\textwidth}  
\vspace*{1mm} 
\begin{itemize} 
  \setlength\itemsep{-0.1em}
\item Simple  crystals, where $\kappa^{\alpha\beta}_{\rm P}\gg \kappa^{\alpha\beta}_{\rm C}$;
\item complex crystals, where $\Delta\omega_{\rm avg}
{\sim} \Gamma(\bm{q})_s$ for most phonons $(\bm{q})_s$, and  $\kappa^{\alpha\beta}_{\rm P}\sim \kappa^{\alpha\beta}_{\rm C}$;
\item glasses, where $ \kappa^{\alpha\beta}_{\rm P}{\ll}\kappa^{\alpha\beta}_{\rm C}$.
\end{itemize}
\vspace*{0.5mm} 
\end{minipage}
& \begin{minipage}{0.42\textwidth}  
\vspace*{2mm} 
Simple crystals characterized by %the average phonon interband spacing $\Delta\omega_{\rm avg}$ is much larger than the linewidths
%the average phonon interband spacing $\Delta\omega_{\rm avg}{=}\frac{\max_{\bm{q}s}\omega(\bm{q})_{s}}{3\;N_{\rm at}}$ is much larger than the linewidths: 
\begin{equation}
  \Delta\omega_{\rm avg}
=\frac{\omega_{\rm max }}{3N_{\rm at}}
\gg \Gamma(\bm{q})_s\;\forall \bm{q},s
\nonumber
\end{equation}
where $\omega_{\rm max }$ is the maximum phonon frequency, and $3N_{\rm at}$ is the number of phonon bands.
\vspace*{2mm} 
 \end{minipage} \\

  \hline

  \hline
	\end{tabular}
  \vspace*{-5mm}
\end{table*}

\begin{comment}
\begin{minipage}{0.12\textwidth}  Regime of validity\end{minipage}
&
\begin{minipage}{0.42\textwidth}  
\vspace*{1mm} 
\begin{itemize} 
  \setlength\itemsep{-0.1em}
\item Simple  crystals, where $\kappa^{\alpha\beta}_{\rm P}\gg \kappa^{\alpha\beta}_{\rm C}$;
\item complex crystals, where $\Delta\omega_{\rm avg}
{\sim} \Gamma(\bm{q})_s$ for most phonons $(\bm{q})_s$, and  $\kappa^{\alpha\beta}_{\rm P}\sim \kappa^{\alpha\beta}_{\rm C}$;
\item glasses, where $ \kappa^{\alpha\beta}_{\rm P}{\ll}\kappa^{\alpha\beta}_{\rm C}$.
\end{itemize}
\vspace*{0.5mm} 
\end{minipage}
& 
\begin{minipage}{0.42\textwidth}  
\vspace*{2mm} 
Simple crystals characterized by %the average phonon interband spacing $\Delta\omega_{\rm avg}$ is much larger than the linewidths
%the average phonon interband spacing $\Delta\omega_{\rm avg}{=}\frac{\max_{\bm{q}s}\omega(\bm{q})_{s}}{3\;N_{\rm at}}$ is much larger than the linewidths: 
\begin{equation}
  \Delta\omega_{\rm avg}
=\frac{\omega_{\rm max }}{3N_{\rm at}}
\gg \Gamma(\bm{q})_s\;\forall \bm{q},s
\nonumber
\end{equation}
where $\omega_{\rm max }$ is the maximum phonon frequency, and $3N_{\rm at}$ is the number of phonon bands.
\vspace*{2mm} 
 \end{minipage}\\
\end{comment}

\section{Phase convention and size consistency}
\label{sec:size_consistency_in_silicon_supercell}

As anticipated in Sec.~\ref{sec:preliminaries}, different phase conventions, smooth or step-like, are employed in the literature for the Fourier transform.
{Up to now all the derivations have been performed using the smooth phase convention, 
since in Sec.~\ref{sec:preliminaries} we discussed how quantities computed employing such a  convention are invariant under different possible choices of the crystal's primitive cell, while quantities computed using the step-like phase convention can be ill-defined (\textit{i.e.} they can depend on the non-univocal possible choice for the primitive cell). 
\deleted{However, we note that the primitive cell of a crystal can be univocally chosen using a standardization criterion (see e.g. Ref.~\cite{hinuma2017band}); thus, this could be used to remove the ambiguities above and claim that additional arguments are needed to adopt a specific phase convention.}
From an intuitive point of view, the choice of using the smooth phase convention in the derivation (via the Taylor expansion~(\ref{eq:Taylor})) of the \RedNmTh{}~(\ref{eq:Wigner_evolution_equation_N}), might be motivated noting that the derivative of the square root of the smooth dynamical matrix~(\ref{eq:dynamical}) has variations over its elements that are smoother than those of the derivative of the square root of the step-like dynamical matrix~(\ref{eq:dynamical_Ziman}). Therefore, the use of the smooth phase convention intuitively yields to a more appropriate description of the smooth variations typical of the close-to-equilibrium regime in focus here.
In the following we report quantitative evidence that confirms this intuitive argument. 

We start by discussing how the \RedNmTh{}~(\ref{eq:Wigner_evolution_equation_N}) and related thermal conductivity expression~(\ref{eq:thermal_conductivity_final_sum}) depend on the phase convention adopted in the derivation.
As shown by Eqs.~(\ref{eq:relation_dynamical_matrices},\ref{eq:eigenvectors_conventions_relation}), quantities in Fourier space obtained using the step-like phase convention are related by a unitary transformation to those obtained using the smooth phase convention.
In practice, using the step-like phase convention implies removing \deleted{all} the atomic positions $\bm{\tau}_b,\bm{\tau}_{b'}$ from the complex exponentials in the derivations of Secs.~\ref{sec:density_matrix_formalism_for_atomic_vibrations},\ref{sec:Wigner_thermal_transport_equation}; from Eq.~(\ref{eq:bosonic_reciprocal}) to Eq.~(\ref{eq:evol_density_matrix}) this produces no differences in the physical description.
However, the phase convention adopted affects the \RedNmTh{}~(\ref{eq:Wigner_evolution_equation_N}), since the velocity operator appearing in such an equation depends on the phase convention; it is \replaced{possible}{straightforward} to demonstrate instead that all other coefficients in Eq.~(\ref{eq:Wigner_evolution_equation_N}), which are not obtained from a differentiation procedure, are unaffected by the phase convention.
Specifically, using the relation~(\ref{eq:relation_dynamical_matrices})
between the dynamical matrices in the smooth (${{\tenscomp{D}(\bm{q})}}_{b\alpha,b'\hspace*{-0.5mm}\alpha'}$, Eq.~(\ref{eq:dynamical})) or step-like ($\underline{{\tenscomp{D}(\bm{q})}}_{b\alpha,b'\hspace*{-0.5mm}\alpha'}$, Eq.~(\ref{eq:dynamical_Ziman})) phase conventions, it is \replaced{possible}{straightforward} to show that  the velocity operator in the step-like phase convention $\underline{\tenscomp{v}}^\beta\!(\bm{q})_{s,s'}{=} \sum_{{b,b'\!,\alpha,\alpha'\!}}\underline{\mathcal{E}^\star(\bm{q})}_{{s},b\alpha}{{\nabla^{\beta}_{\bm{q}} \underline{\sqrt{\tenscomp{D}(\bm{q})}}}}_{b\alpha,b'\hspace*{-0.5mm}\alpha'} \underline{\mathcal{E}(\bm{q})}_{s'\hspace*{-0.5mm},b'\hspace*{-0.5mm}\alpha'}$ (where $\underline{\mathcal{E}(\bm{q})}_{s'\hspace*{-0.5mm},b'\hspace*{-0.5mm}\alpha'}$ are the eigenvectors of the step-like dynamical matrix defined in Eq.~(\ref{eq:eigenvectors_conventions_relation})) is related to
that obtained using the smooth phase convention (${\tenscomp{v}}^\beta\!(\bm{q})_{s,s'}$, defined in Eq.~(\ref{eq:vel_op})) by:
\begin{equation}
\begin{split}
  \underline{\tenscomp{v}}^\beta\!(\bm{q})_{\!s,s'\!}{=}&{\tenscomp{v}}^\beta\!(\bm{q})_{\!s,s'\!}
  {+}
  i\big[\omega(\bm{q})_{\!s'\!}{-}\omega(\bm{q})_{\!s}\big]
  \!\!\sum_{b\alpha}\!\mathcal{E}^\star\!(\bm{q})_{\!s,b\alpha}
  \tau_{b}^{\beta}\!
  \mathcal{E}(\bm{q})_{\!s'\!,b\alpha}.
\end{split}
\raisetag{12mm}
\label{eq:relation_velocity_operator}
\end{equation}
\begin{figure}[t]
  \centering
  \includegraphics[width=\WidthFigure]{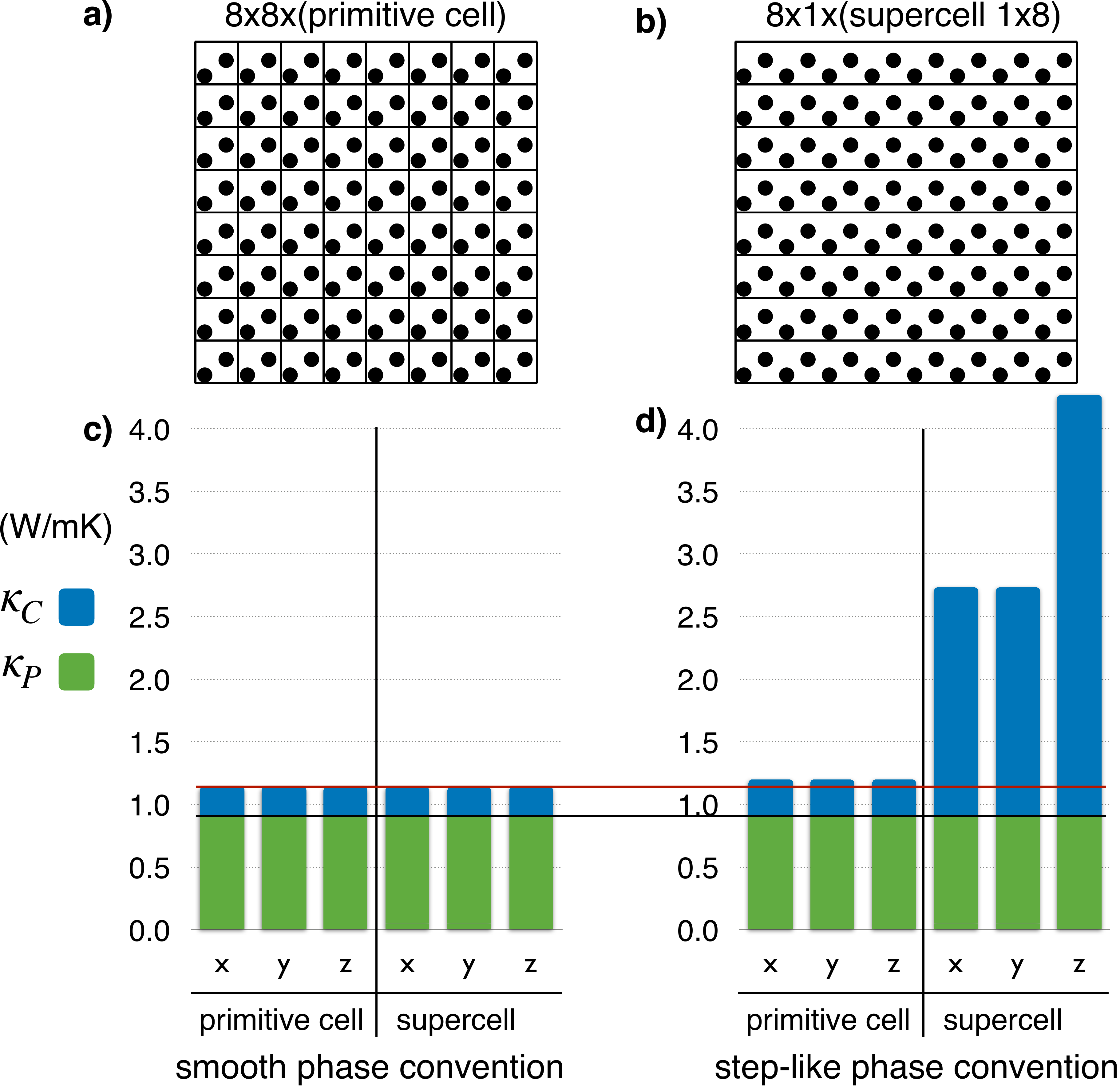}
  \vspace*{-5mm}
  \caption{
  \textbf{Size consistency of the \RedNmTh{} conductivity in the smooth phase convention.}
   Size consistency is tested computing the conductivity~(\ref{eq:thermal_conductivity_final_sum}) of a silicon crystal at $T{=}300\;K$ and using a fictitious linewidth of $\hbar\Gamma(\bm{q})_s{=}30\;{\rm cm}^{-1}\;\forall\;\bm{q},s$ (see text). 
   We sketch in panels \textbf{a,b)} two equivalent crystal representations: \textbf{a)} repeating a 2-atom primitive cell the same number of times along each Cartesian direction; \textbf{b)} repeating a 16-atom unit cell that is  larger than the primitive.
   Panel \textbf{c)} shows that the smooth phase convention (used in Eq.~(\ref{eq:dynamical})  yields size-consistent numerical results, since the conductivity is unchanged when the silicon crystal with fictitious linewidths is modeled as:
   (i) a $16{\times}16{\times}16$ repetition of the 2-atom primitive cell;
   (ii) a (physically equivalent) $2{\times}2{\times}16$ repetition of a unit cell that is a supercell $8{\times}8{\times}1$ of the primitive.
  Panel \textbf{d)} shows instead that the step-like phase convention (used in Eq.~(\ref{eq:dynamical_Ziman})) is not size-consistent, since it yields different conductivities for the two physically equivalent representations of the crystal.
   The black and red lines serve as a guide to the eye, to show that the Peierls-Boltzmann (populations) conductivity does not depend on the phase convention adopted (black), while the coherences conductivity depends on the phase convention adopted (red).
   }
  \label{fig:silicon}
\end{figure}
From Eq.~(\ref{eq:relation_velocity_operator}) it is evident that the elements ${\tenscomp{v}}^\beta\!(\bm{q})_{s,s'}$ of the velocity operator, either on the diagonal ($s{=}s'$) or coupling degenerate vibrational modes (\textit{i.e.} $s{\neq} s'$ with $\omega(\bm{q})_s{=}\omega(\bm{q})_{s'}$), do not depend on the phase convention.
In contrast, off-diagonal elements of the velocity operator that couple non-degenerate vibrational modes depend on the phase convention.
We note that in any subspace spanned by degenerate vibrational modes with wavevector $\bm{q}$ and with frequency $\omega_d$ \footnote{\textit{i.e.}  all the vibrational modes satisfying $\omega(\bm{q})_{s_d}=\omega_d\;\forall \;s_d\in\{i,\dots,j \}$, with $1\leq i<j\leq3\cdot N_{at}$}, one can exploit the freedom of rotating the degenerate eigenstates of the dynamical matrix (Eq.~(\ref{eq:diagonalization_dynamical_matrix})) to render the representation of the velocity operator in direction $\beta$ diagonal in the mode indexes \cite{fugallo2013ab}.
As mentioned in Sec.~\ref{sub:evolution_of_the_wigner_density}, the different Cartesian components of the velocity operator in general do not commute, thus they can not be all diagonalized simultaneously. % (see e.g. Ref.~\cite{PhysRevB.68.033105} for a general discussion).
Nevertheless, since one has the freedom to diagonalize in the degenerate subspace at least one Cartesian component of the velocity operator --- thus to recover along such direction a populations' conductivity exactly equivalent to the Peierls-Boltzmann conductivity discussed in Ref.~\cite{fugallo2013ab} --- we consider the diagonal or degenerate velocity-operator elements as contributing exclusively to the Peierls-Boltzmann conductivity, implying that coherences emerge exclusively from off-diagonal and non-degenerate velocity-operator elements.}

We investigate the differences between the smooth and step-like phase conventions on the conductivity, 
focusing on the general property that the conductivity has to respect --- namely, size consistency. 
To this aim, we compute the conductivity~(\ref{eq:thermal_conductivity_final_sum}) of a silicon crystal described in two mathematically different but physically equivalent ways: (i) repeating $N{\times}N{\times}N$ times a primitive cell; 
(ii) repeating $\tfrac{N}{8}{\times}\tfrac{N}{8}{\times}{N}$ times a unit cell that is a supercell $8{\times}8{\times}1$ of the primitive cell used at point (i). 
A schematic representation of these two cases is shown in Fig.~\ref{fig:silicon}\textbf{a,b} for $N =8$, with panel \textbf{a)} representing case (i) and panel \textbf{b)} representing case (ii).
Since the differences between the conductivities in smooth and step-like phase conventions can emerge only from the off-diagonal elements of the velocity operator (Eq.~(\ref{eq:relation_velocity_operator})), we perform tests where we use fictitious large linewidths $\hbar\Gamma(\bm{q})_s{=}30\;{\rm cm}^{-1}\;\forall\;\bm{q},s$ to enhance the importance the off-diagonal elements of the velocity operator in the conductivity formula~(\ref{eq:thermal_conductivity_final_sum}); therefore, the conductivities obtained in these tests are informative about the size-consistency of Eq.~(\ref{eq:thermal_conductivity_final_sum}) in the smooth or step-like phase convention, but their values have no physical meaning. 
Results show that the primitive-cell and supercell descriptions, which are equivalent from the physical point of view, result having the same thermal conductivity when the smooth phase convention is adopted (Fig.~\ref{fig:silicon}\textbf{c}), while they lead to different conductivities when the step-like phase convention is adopted (Fig.~\ref{fig:silicon}\textbf{d}).
This reiterates that the smooth phase convention~(\ref{eq:dynamical}) has to be used in the derivation of the \RedNmTh{}~(\ref{eq:Wigner_evolution_equation_N}) (as opposed to the step-like one used in Ref.~\cite{simoncelli2019unified}, albeit with negligible numerical differences for that case study), in agreement the intuitive arguments made earlier.

\section{Case studies: \texorpdfstring{L\MakeLowercase{a}{$_2$}Z\MakeLowercase{r}$_2$O$_7$ and C\MakeLowercase{s}P\MakeLowercase{b}B\MakeLowercase{r}$_3$}{La2Zr2O7 and CsPbBr3} }
%\section{Wave-particle duality in heat conduction}
\label{sec:numerical_results}
In this section we consider complex crystals, whose thermal conductivity is very low and not correctly described by the LBTE, as case studies for this more general \nametheory{} framework (Eq.~(\ref{eq:thermal_conductivity_final_sum})).
In particular, we analyze the La$_2$Zr$_2$O$_7$ lanthanum zirconate  and the CsPbBr$_3$ perovskite, as materials used for thermal barrier coatings and thermoelectric devices, respectively. Experimental measurements in these materials \cite{Suresh1997,Vassen2000,Chen2009,Wan2010,Yang2016,wang2018cation} have highlighted a high-temperature trend for the temperature-thermal conductivity relation that has a decay much slower than the $T^{-1}$ predicted by Peierls-Boltzmann (we recall that the LBTE predicts this $T^{-1}$ trend to be universal for all crystals \cite{ziman1960electrons}). The LWTE conductivity~(\ref{eq:thermal_conductivity_final_sum}) encompasses the $T^{-1}$-decaying Peierls-Boltzmann conductivity $\kappa_{\rm P}$, but also adds the coherences conductivity $\kappa_{\rm C}$, which can increase with $T$ (e.g. when it reduces to the Allen-Feldman conductivity in the limiting case of a harmonic glass \cite{allen1989thermal,McGaughey2009predicting}). These analytical considerations suggest that the LWTE conductivity allows for the emergence of much milder decays or glass-like trends in the thermal conductivity, and further motivates the study of these materials with such more general \nametheory{} framework.

\subsection{Lanthanum zirconate} 
\label{sub:zirconate_start}
La$_2$Zr$_2$O$_7$ is a material characterized by wide thermal stability and ultralow thermal conductivity; thus, it is widely used for thermal barrier coatings \cite{zhang2017lanthanum}.
This is a good test case for the Wigner formulation discussed here, 
since at high temperatures La$_2$Zr$_2$O$_7$ falls in the category we defined of complex crystals, with many overlapping phonon bands and a temperature-conductivity curve that is not correctly described by the LBTE.
Here, we calculate the thermal conductivity of La$_2$Zr$_2$O$_7$ as a function of temperature, computing from first principles all the quantities needed to evaluate Eq.~(\ref{eq:thermal_conductivity_final_sum}) (see Appendix~\ref{sub:la__2_zr__2_o__7_} for details). 

La$_2$Zr$_2$O$_7$ is characterized by a cubic structure (spacegroup: Fd-3m, number 227) and by an isotropic thermal conductivity tensor; in the following only the single independent component of the conductivity tensor  will be reported ($\kappa{=}\kappa^{xx}{=}\kappa^{yy}{=}\kappa^{zz}$). 
Moreover, we rely on the SMA approximation to reduce the computational cost for solving the populations' part of the \RedNmTh{} (\textit{i.e.} the LBTE, see Sec.~\ref{sub:steady_state_solution_of_the_wbte}), 
since it is known that for materials with ultralow conductivity the solution of the LBTE determined within the SMA approximation is practically indistinguishable from the exact solution of the LBTE \cite{lindsay_first_2016,fugallo2013ab,simoncelli2019unified,luo2020vibrational}.
Therefore, in this work the conductivity is computed from Eq.~(\ref{eq:thermal_conductivity_final_sum}) with the population term evaluated in the SMA~(\ref{eq:SMA_k}).

\subsubsection{Thermal conductivity }
\label{ssub:thermal_conductivity}
The conductivity as a function of temperature for La$_2$Zr$_2$O$_7$ is shown in Fig.~\ref{fig:k_vs_T_Zirconate}.
In La$_2$Zr$_2$O$_7$ at low temperature, the populations conductivity $\kappa_{\rm P}$ dominates over the coherences conductivity $\kappa_{\rm C}$.
\begin{figure}[b]%67 words
  \centering
  \includegraphics[width=\WidthFigure]{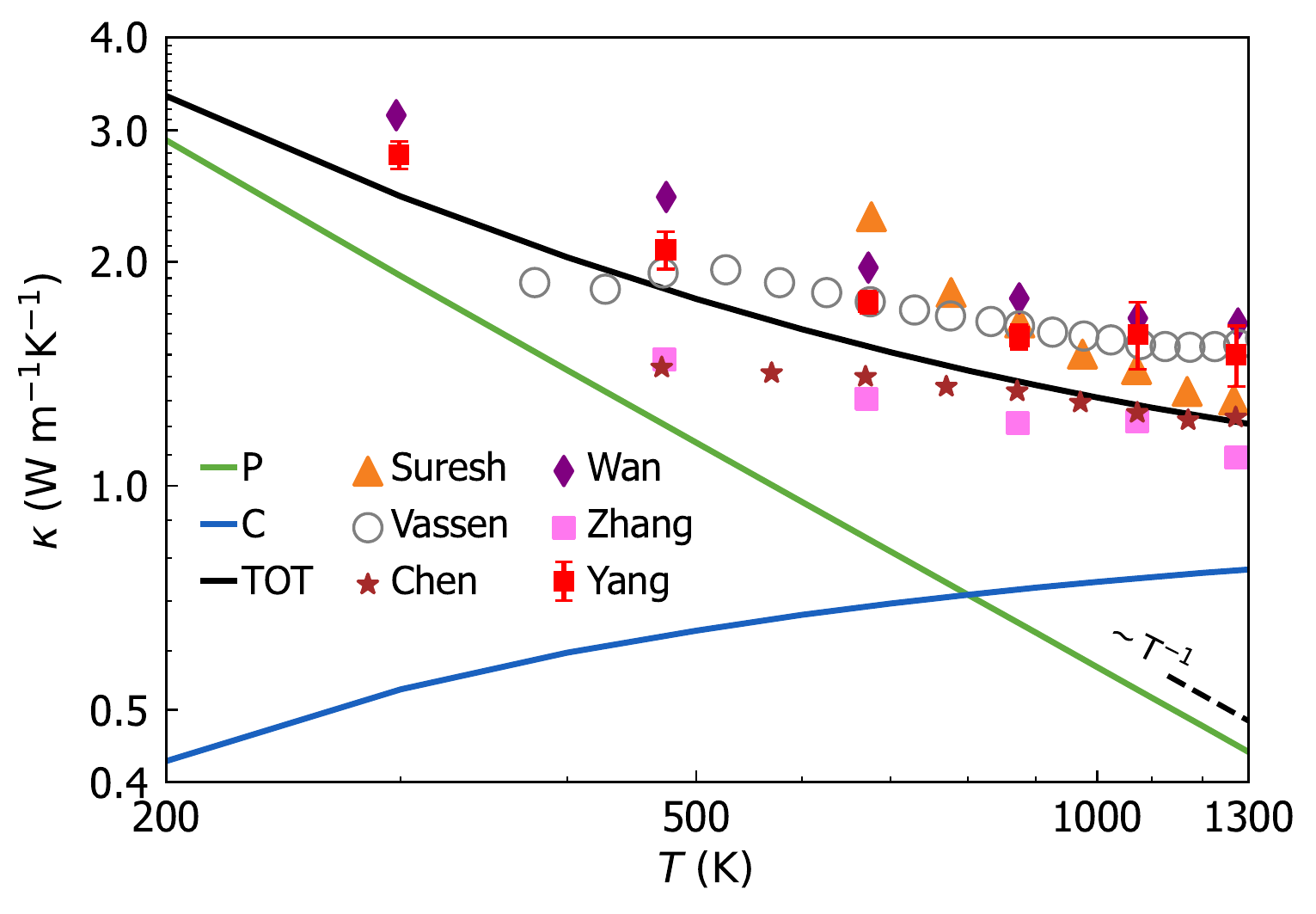}\vspace*{-3mm}
  \caption{\textbf{Bulk thermal conductivity of La$_2$Zr$_2$O$_7$.} 
    Scatter points represent measurements from Suresh \textit{et al.}   \cite{Suresh1997}, Vassen \textit{et al.}   \cite{Vassen2000}, Chen \textit{et al.}   \cite{Chen2009}, Wan \textit{et al.}   \cite{Wan2010}, \added{Zhang \textit{et al.}} \cite{zhang2020microstructure},   Yang \textit{et al.} \cite{Yang2016} {(the experiment which best represents the bulk crystal studied here is that from Yang \textit{et al.}, see text)}. 
Green, Peierls' LBTE conductivity ($\kappa_{\rm P}$), which displays the universal $T^{-1}$ asymptotics (dashed line).     
    Blue, coherences' conductivity ($\kappa_{\rm C}$). 
     Black, total conductivity from equation~(\ref{eq:thermal_conductivity_final_sum}): {$\kappa_{\rm TOT}{=}\kappa_{\rm P}{+}\kappa_{\rm C}$} (we report the single independent component of the diagonal and isotropic conductivity tensor).
  }
  \label{fig:k_vs_T_Zirconate}
\end{figure}

\begin{figure*}
\vspace*{-2mm}
  \centering
  \includegraphics[width=\WidthFigure]{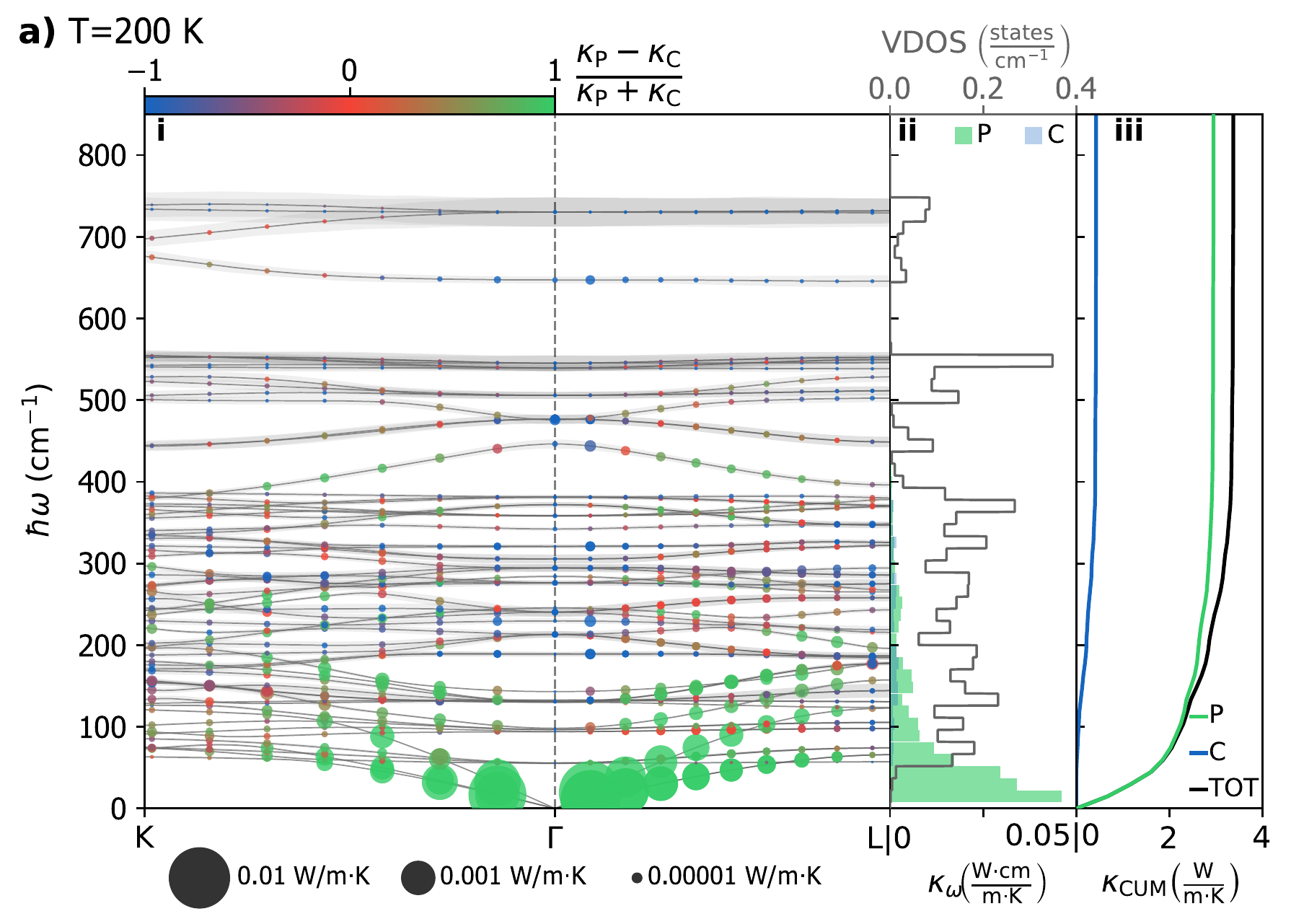}
    \includegraphics[width=\WidthFigure]{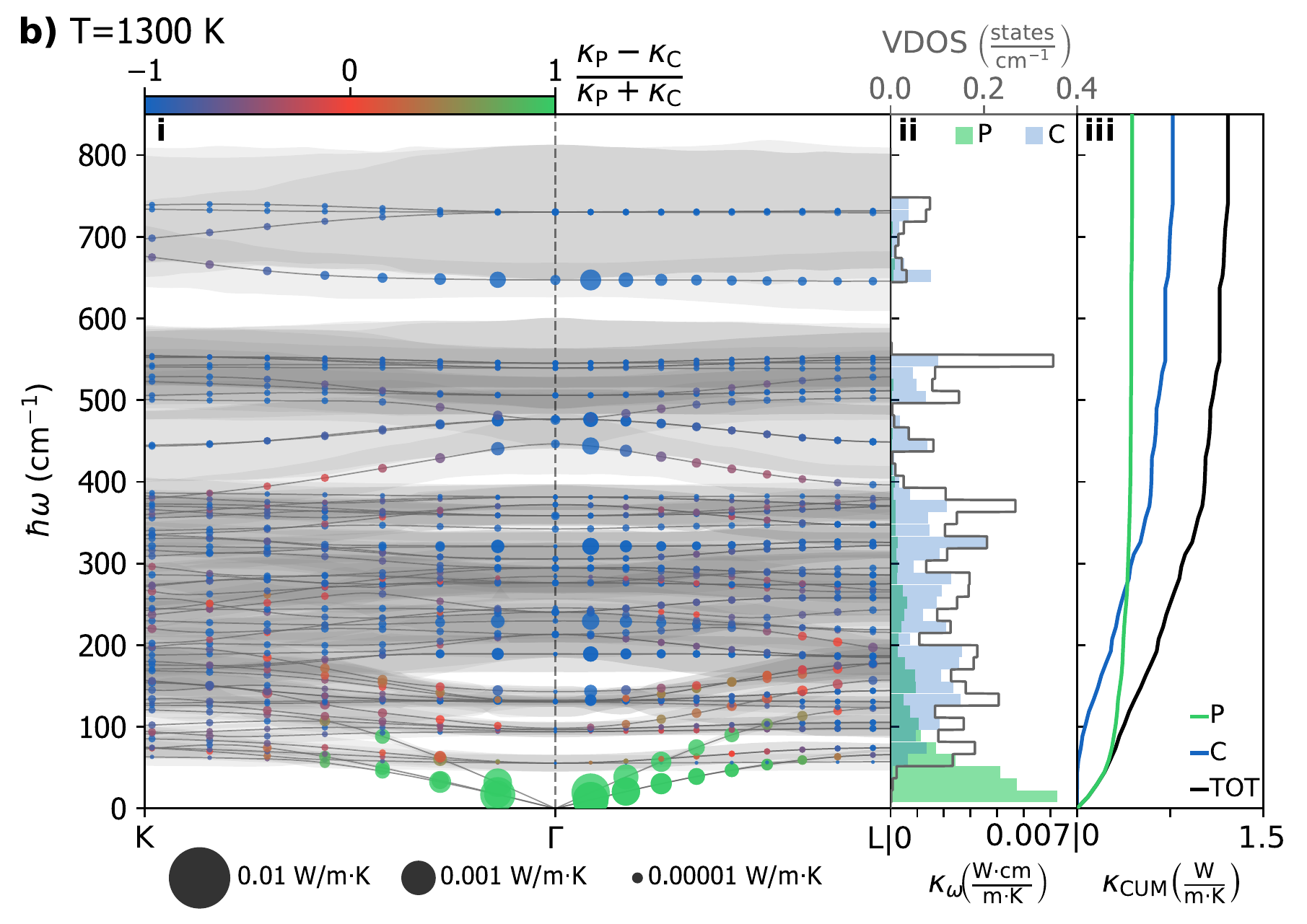}
   \vspace*{-1mm}
  \caption{\textbf{Vibrational properties and heat conduction mechanisms in La$_2$Zr$_2$O$_7$.} Panels \textbf{a}(i), \textbf{b}(i) show the phonon spectrum of La$_2$Zr$_2$O$_7$ (gray lines) with the phonon linewidths (the full amplitude of the shaded gray areas correspond to the full phonon linewidth $\Gamma(\bm{q})_s$) at 200 K and 1300 K, respectively. Colored circles represent the  phonon eigenstates $(\bm{q})_s$ sampled in the  calculation; the area of each circle is related to its contribution to $\kappa$ (area $\propto\sqrt{\text{contribution to}\;\kappa}$   \cite{footnote})
  and the color identifies the origin of the contribution: green for populations' propagation and blue for coherences' anharmonic couplings (red corresponds to $50\%$ of each, see Eq.~(\ref{eq:coloring}) and related discussion for details). In the coherences' couplings between two modes $(\bm{q})_s$ and $(\bm{q})_{s'}$ the contribution of the single mode $s$ is determined as ${C(\bm{q})_s}/[{C(\bm{q})_s{+}C(\bm{q})_{s'}}]$, where $C(\bm{q})_s$ 
  %{=}\tfrac{\hbar^2}{k_{B} {T}^2}\omega^2\hspace*{-0.5mm}(\bm{q})\hspace*{-0.4mm}_{s}\bar{\tenscomp{n}}^{T}\hspace*{-0.6mm}({\bm{q}})_{\hspace*{-0.4mm}s}[\bar{\tenscomp{n}}^{T}\hspace*{-0.5mm}({\bm{q}})_{\hspace*{-0.4mm}s}{+}1]
  is the specific heat of a given phonon (see Appendix~\ref{sec:ioffe_regel_limit_in_space_and_center_of_the_non_sharp_particle_wave_crossover} for the details). Panels \textbf{a}(ii), \textbf{b}(ii) show the thermal conductivity density of states $\kappa_{\omega}$ of populations (``P'', green) and coherences (``C'', blue) at 200 K and 1300 K, respectively. The gray line shows the vibrational density of states (VDOS). Panels \textbf{a}(iii), \textbf{b}(iii) show the cumulative total thermal conductivity (in black) as a sum of the populations' contribution (``P'') in green and coherences' one (``C'') in blue, at 200 K and 1300 K, respectively. At 200 K (panel \textbf{a}) La$_2$Zr$_2$O$_7$ is a simple crystal with interband spacings larger than the linewidths and $\kappa_P{\gg} k_C$, while at 1300 K it is a complex crystal with interband spacings smaller than the linewidths and $\kappa_P{\lesssim} k_C$.
  }
    \label{fig:cond_mech_Zirc}
\end{figure*}
Increasing the temperature, $\kappa_{\rm P}$ decreases following the universal $1/T$ decay of the Peierls-Boltzmann equation \cite{sun2010lattice,lory2017direct,li2015ultralow,ziman1960electrons} --- a trend that is  in broad disagreement with experiments  \cite{Suresh1997,Vassen2000,Chen2009,Wan2010,Yang2016}.
The contribution of the coherences to the conductivity instead increases with temperature, and becomes dominant at high temperature in the complex-crystal regime, where it offsets the incorrect $1/T$ decay of Peierls-Boltzmann conductivity, leading to a total conductivity ($\kappa_{\rm TOT}{=}\kappa_{\rm P}{+}\kappa_{\rm C}$) that is in much closer agreement with experiments.
The experimental values of the thermal conductivity of La$_2$Zr$_2$O$_7$ reported in Fig.~\ref{fig:k_vs_T_Zirconate} can be considered as mainly determined by the lattice thermal conductivity, with radiative effects having negligible impact on these values.
In particular, the data from Yang \textit{et al.}    \cite{Yang2016} refer to experimental measurements of the lattice (phonon) contribution to the conductivity of bulk La$_2$Zr$_2$O$_7$.
Importantly, Yang \textit{et al.}    \cite{Yang2016} discuss how radiative effects are significantly reduced in sintered composites, and all the other experimental data reported in Fig.~\ref{fig:k_vs_T_Zirconate} (Refs. \cite{Suresh1997,Vassen2000,Chen2009,Wan2010}) refer to sintered samples for which radiative effects are suppressed.

Finally, given that the smooth phase convention is the only one yielding a size-consistent conductivity,
the predictions reported in Fig.~\ref{fig:k_vs_T_Zirconate} have been computed using the smooth phase convention.
For completeness, we discuss in Appendix~\ref{sec:effects_of_the_phase_convention_on_the_conductivity} how the conductivity would change if the step-like phase convention was adopted, showing that the coherences' conductivity at high temperature becomes overestimated, while  the populations' conductivity is not affected (this is clear from Eq.~(\ref{eq:relation_velocity_operator}), which shows that the diagonal elements of the velocity operator that enter in the populations' equation~(\ref{eq:populations_steady_state}) are the same in both the smooth and step-like phase conventions).

\subsubsection{Transport mechanisms: from simple to complex crystal}
\label{sub:heat_conduction_mechanisms_in_the_}
In Fig.~\ref{fig:cond_mech_Zirc}\textbf{a}(i) 
we show that in La$_2$Zr$_2$O$_7$ at low temperature (200 K) the predominance of the populations conductivity over the coherences conductivity is related to linewidths smaller than the phonon interband spacings,  demonstrating that at 200 K La$_2$Zr$_2$O$_7$ behaves as a simple crystal. 
More precisely, we show in Figs.~\ref{fig:cond_mech_Zirc}\textbf{a}(ii,iii) that such a dominant populations conductivity is mainly determined by low-frequency phonon bands with large group velocities and weak anharmonicity.
Fig.~\ref{fig:cond_mech_Zirc}\textbf{b}(i) investigates instead the behavior of La$_2$Zr$_2$O$_7$ at high temperature,
showing that at 1300 K the predominance of the coherences conductivity over the populations conductivity is related to interband spacings smaller than the linewidths, \textit{i.e.} to a complex-crystal behavior. Figs.~\ref{fig:cond_mech_Zirc}\textbf{a}(ii,iii) show that the dominant coherences conductivity originates from highly anharmonic flat bands.

\begin{figure}[t]
  \centering
  \includegraphics[width=\WidthFigureParticular]{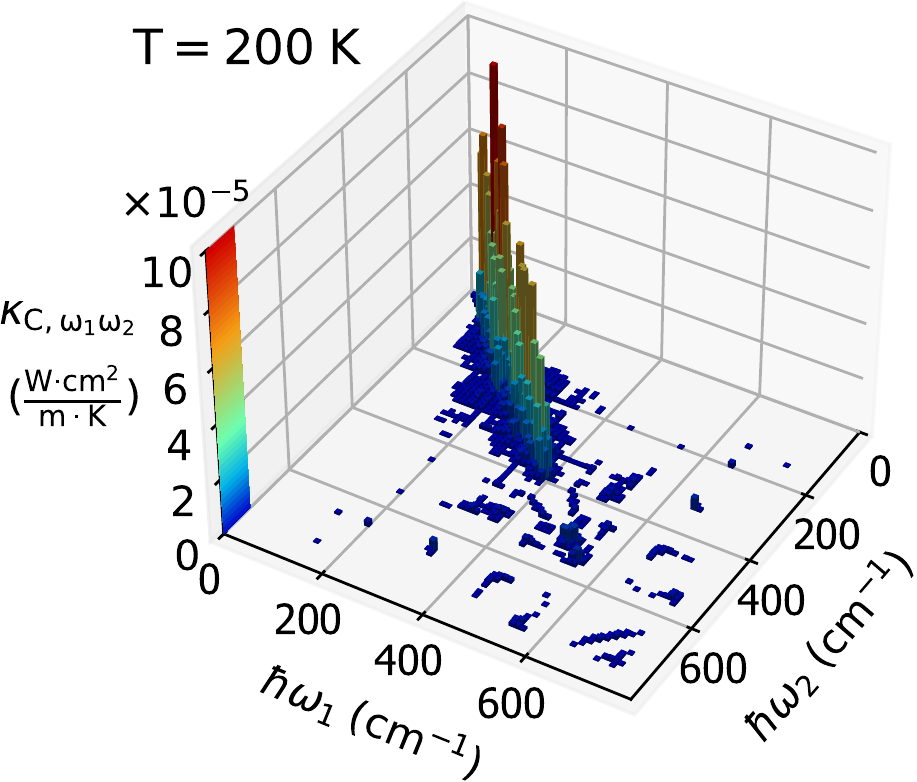}
  \includegraphics[width=\WidthFigureParticular]{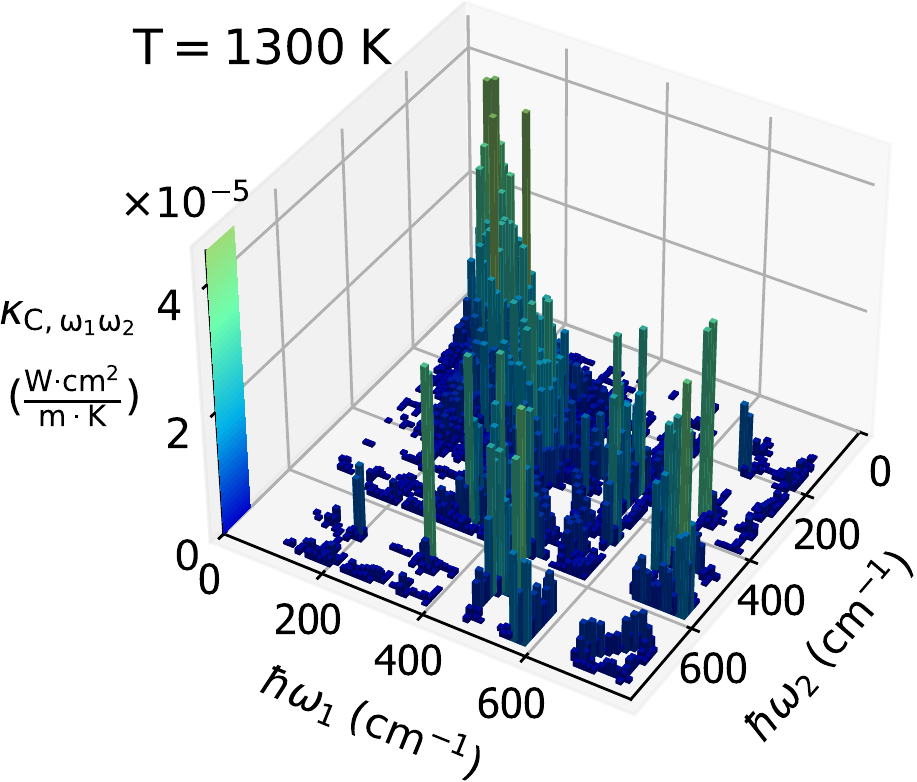}
   %\vspace*{-5mm}
  \caption{\textbf{2-dimensional density of states for the thermal conductivity,} $\kappa_{\rm{C},\omega_1\omega_2}$, which resolves 
how much a Zener-like coupling between two vibrational modes having frequencies $\omega_1,\;\omega_2$ contributes to the coherences' conductivity.
We show $\kappa_{\rm{C},\omega_1\omega_2}$ at 200~K (left panel) and at 1300~K (right panel). The points close to the diagonal $\omega_1{\sim}\omega_2$ correspond to quasi-degenerate vibrational eigenstates; the Allen-Feldman framework considers couplings only between these.   \vspace*{-2mm}
}
  \label{fig:coe_DOS_2D}
\end{figure}
The  density of states of the coherences' conductivity (blue histogram in Fig.~\ref{fig:cond_mech_Zirc}\textbf{a,b}(ii)) can be also resolved in terms of the  frequencies $\omega_1,\;\omega_2$ of the two modes coupled, as shown in Fig.~\ref{fig:coe_DOS_2D}.
At 200~K, such a 2-dimensional  density of states for the coherences' thermal conductivity ($\kappa_{\rm{C},\omega_1\omega_2}$, Fig.~\ref{fig:coe_DOS_2D}, left panel)  shows couplings between quasi-degenerate vibrational modes, similarly to the case of harmonic  glasses ($\Gamma(\bm{q})_s{\to}\eta{\to}0\;\forall \;\bm{q},s$). 
At 1300~K (Fig.~\ref{fig:coe_DOS_2D}, right panel) $\kappa_{\rm{C},\omega_1\omega_2}$ instead includes couplings between phonon modes having very different frequencies, driven by the large anharmonicity ---  therefore the corresponding heat conduction mechanism is intrinsically different from the one of harmonic glasses. 
We conclude by noting that the results presented in this section were obtained employing the approximated treatment of anharmonicity discussed in Sec.~\ref{sub:steady_state_solution_of_the_wbte}. We show in Appendix~\ref{sec:validation_of_the_perturbative_treatment_of_anharmonicity} that  the frequencies computed with this approximated scheme yield predictions for the specific heat in agreement with experiments in the temperature range analyzed. Moreover, we also show that simulating the Raman spectra using the frequencies and linewidths obtained from this approximated scheme yields predictions in agreement with experiments over a temperature range in which coherences are not negligible. 
This shows that the standard perturbative treatment for anharmonicity adopted is accurate enough for the illustrative scope of this analysis.

\subsection{Halide perovskite CsPbBr$_3$ }
\label{sec:tc_CsPbBr3}
\begin{figure}[t]%67 words
  \centering
  \includegraphics[width=\WidthFigure]{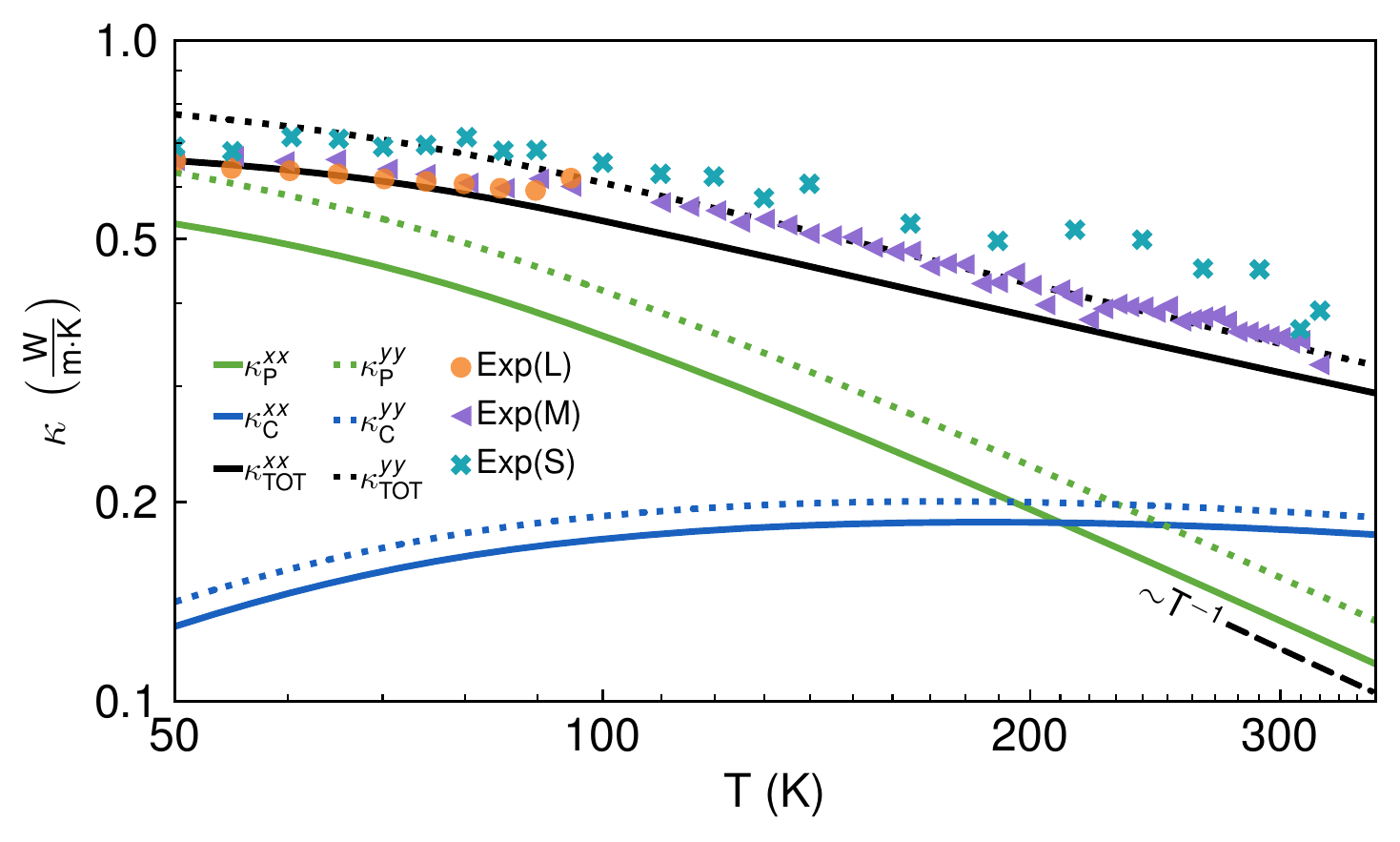}
  \vspace*{-7mm}
  \caption{\textbf{Bulk thermal conductivity of CsPbBr$_3$ in the smooth phase convention.} 
    ``Exp(L)'', ``Exp(M)'' and ``Exp(S)''  refer to measurements   \cite{wang2018cation} of $\kappa^{xx}$ in nanowires having respectively sections of $800{\times} 380\;\rm{nm}^2$, $320{\times} 390\;\rm{nm}^2$ and $300{\times} 160\;\rm{nm}^2$: their broad agreement supports the hypothesis of negligible finite-size boundary scattering \cite{wang2018cation}. 
    Lines are theoretical predictions of the conductivity from equation~(\ref{eq:thermal_conductivity_final_sum}) using the smooth phase convention, with solid lines referring to the tensor component $\kappa^{xx}$ (to be compared with experiments), and dotted lines referring to the component  $\kappa^{yy}$ (this last is reported here for completeness and  is practically indistinguishable from the not-reported component $\kappa^{zz}$). 
Green, Peierls' LBTE conductivity ($\kappa_{\rm P}$), which displays the universal $T^{-1}$ asymptotics (dashed line)   \cite{sun2010lattice,lory2017direct,li2015ultralow,ziman1960electrons}.     
    Blue, coherences' thermal conductivity {($\kappa_{\rm C}$)}. 
     Black, total conductivity from equation~(\ref{eq:thermal_conductivity_final_sum}): {$\kappa_{\rm TOT}=\kappa_{\rm P}+\kappa_{\rm C}$}.  
  }
  \label{fig:k_vs_T}
\end{figure}
The perovskite CsPbBr$_3$ belongs to a family of ultralow-thermal-conductivity materials that are promising for thermoelectric energy conversion   \cite{wang2018cation,lee2017ultralow}, and it has been used in Ref.   \cite{simoncelli2019unified} to showcase the predictive capabilities of the \nametheory{} formulation.
The temperature-conductivity curve for CsPbBr$_3$  reported in our earlier work   \cite{simoncelli2019unified} was computed using the step-like phase convention, which has been shown in Sec.~\ref{sec:size_consistency_in_silicon_supercell} to yield size-inconsistent results.  
Therefore, in this section we update the \RedNmTh{} predictions for the temperature-conductivity curve of CsPbBr$_3$ using the correct size-consistent smooth phase convention discussed before. 
We show in Fig.~\ref{fig:k_vs_T} a comparison between the conductivity predicted with the \RedNmTh{} conductivity formula~(\ref{eq:thermal_conductivity_final_sum}) in the smooth phase convention and the experimental measurements \cite{wang2018cation}, finding results which are not significantly different from those reported in Ref.~\cite{simoncelli2019unified} (see Appendix~\ref{sec:effects_of_the_phase_convention_on_the_conductivity} for a detailed analysis on how the $x$ component of the conductivity tensor of CsPbBr$_3$ changes between the size-consistent smooth phase convention and the size-inconsistent step-like phase convention).
Specifically, Fig.~\ref{fig:k_vs_T} shows that
at 300~K  the populations conductivity contributes just $\sim\tfrac{1}{3}$ of the total conductivity, while the coherences' term provides an additional $\sim\tfrac{2}{3}$, leading to a much closer agreement with experiments that becomes even more relevant in the high-temperature asymptotics.
Conversely, at low temperature the populations'  conductivity becomes dominant, and at 50~K it provides $\sim \tfrac{3}{4}$ of the total conductivity. 
It can be seen that in the high-temperature limit $\kappa^{xx}_{\rm P}{\propto}\;T^{-1}$, as  predicted by Peierls'  theory   \cite{sun2010lattice,lory2017direct,li2015ultralow,ziman1960electrons}; this in disagreement with experiments. Instead, the \RedNmTh{} formulation predicts a decay of $\kappa^{xx}$ much milder than $T^{-1}$, as shown here for CsPbBr$_3$, and as present in many other complex crystals   \cite{li2015ultralow,lory2017direct,PhysRevB.46.6131,Weathers_PRB_2017,chen2015twisting,voneshen2013suppression,Mukhopadhyay1455}. 
Finally, in Appendix~\ref{sec:validation_of_the_perturbative_treatment_of_anharmonicity} we check the accuracy of the standard approximated treatment of anharmonicity we employed, showing that also for CsPbBr$_3$ the frequencies and linewidths we computed yield predictions for the specific heat and Raman spectra in agreement with experiments. 

%\added{We conclude by noting that in the high-temperature limit the populations conductivity decays proportionally to $T^{-1}$ in both La$_2$Zr$_2$O$_7$ and CsPbBr$_3$, while the high-temperature trend of the coherences conductivity is different in La$_2$Zr$_2$O$_7$ and CsPbBr$_3$. An analysis of the quantities that determine the trend of the temperature-coherences conductivity curve is reported in Appendix~\ref{sec:trend_of_the_coherences_conductivity_as_a_function_of_temperature}.}

\section{Particle-wave duality for heat transport}
\label{sec:Particle_wave_crossover_phonons}
We recall that the central object of the Wigner formulation is a matrix distribution $\tenscomp{n}(\bm{R},\bm{q},t)_{s,s'}$, where position $(\bm{R})$, wavevector $(\bm{q})$ and time $(t)$ are the arguments, and the two matrix indexes are the phonon band labels $s,s'$.
Importantly, in the out-of-equilibrium steady-state regime considered, the Wigner matrix distribution~(\ref{eq:Population_Peierls_Generalized}) is not diagonal in the basis which diagonalizes the leading phonon Hamiltonian~(\ref{eq:diagonalization_dynamical_matrix}) (from Eq.~(\ref{eq:Wigner_evolution_equation_N}) it is clear that at fixed $\bm{R},\bm{q},t$, the matrices $\tenscomp{n}(\bm{R},\bm{q},t)_{s,s'}$, ${\omega}(\bm{q})_s\delta_{s,s'}$, and $\tens{v}(\bm{q})_{s,s'}$, do not commute in general), and the evolution of the diagonal elements ($s{=}s'$) of the Wigner distribution is decoupled from the evolution of the off-diagonal elements ($s{\neq}s'$), see Eq.~(\ref{eq:populations_steady_state}) and Eq.~(\ref{eq:coherences_steady_state}).
As discussed by  \citet{rossi2011theory}, the semiclassical limit corresponds to neglecting the off-diagonal Wigner distribution elements ($s{\neq}s'$), thus considering only the diagonal Wigner-distribution elements $\tenscomp{n}(\bm{R},\bm{q},t)_{s,s}$ and interpreting these as number of phonon wavepackets that behave as if they were particles of a classical gas, since these have well defined energies $\hbar\omega(\bm{q})_s$, group velocities $\tens{v}(\bm{q})_{s,s}{=}\tfrac{\partial \omega(\bm{q})_s}{\partial \bm{q}}$, and lifetimes $\tau(\bm{q})_s{=}1/\Gamma(\bm{q})_{s}$  \footnote{We stress that this definition of phonon lifetime is general and always valid, but its implications on the thermal conductivity depend on the regime of thermal transport considered. 
In the kinetic regime of thermal transport Umklapp processes dominate, thus the SMA approximation is valid  \cite{fugallo2013ab,lindsay_first_2016,cepellotti2015phonon}, and the phonon lifetime enters directly in the populations conductivity (Eq.~(\ref{eq:SMA_k})). Instead, in the hydrodynamic regime of thermal transport where normal processes dominate, the phonon lifetime can be still defined but it no longer determines directly the populations conductivity, since this latter is determined by the lifetime of collective excitations of phonon wavepackets (relaxons  \cite{cepellotti2016thermal,PhysRevX.10.011019}) and the relaxon lifetime is related in a non-trivial way to the phonon lifetime  \cite{cepellotti2016thermal}.}.
As mentioned before, such a semiclassical approximation turns out to be accurate in simple crystals characterized by well-separated phonon bands (\textit{i.e.} interbranch spacings much larger than the linewidths), where the exact solution of the LBTE yields a populations' thermal conductivity that is
order of magnitude larger than the coherences conductivity \cite{PhysRevX.10.011019} and 
 in good agreement with that measured in experiments \cite{broido2007intrinsic,PhysRevLett.106.045901,Esfarjani_PRB_11,Chen_Science_12,Nanoscale_Thermal_14}. 
In contrast, when off-diagonal coherences $s{\neq}s'$ cannot be neglected, such a particle-like (semiclassical) picture is no longer valid, since off-diagonal elements are related to the wave-like coherences between quantum vibrational eigenstates (phonons) \cite{rossi2011theory}.
Specifically,  the presence of the difference between frequencies $\omega(\bm{q})_s{-}\omega(\bm{q})_{s'}$ in the coherences evolution equation~(\ref{eq:coherences_steady_state}) demonstrates that: 
(i) coherences between phonons are possible thanks to their wave-like nature and capability to interfere (such interference is possible also in the presence of damping); 
(ii) coherences do not have an absolute energy (frequency) akin to that of a particle-like phonon wave-packet.
In addition, when there is wave-like coherence between two phonon eigenstates, heat is transfered between these via an interband (Zener-like) tunneling mechanism (see Eq.~(\ref{eq:thermal_conductivity_final_sum})); such a wave-like tunneling between phonon bands is intrinsically different from the particle-like propagation of a phonon wavepacket, since in the latter the ``identity'' of a phonon (defined by the wavevector and index of the band to which a phonon belongs) is preserved while in the former is not. 
In the following we derive a quantitative criterion that allows to assess when the  particle-like (semiclassical) picture breaks down and wave-like coherences have to be considered.

\subsection{Wigner and Ioffe-Regel limits in time} % (fold)
\label{sub:winger_limit_in_time}
We start by analyzing the strength of particle-like and wave-like conduction mechanisms, focusing on the conditions under which they become comparable.
We start by resolving the contributions of a phonon $(\bm{q})_s$ to the particle-like and to the wave-like conductivities, recasting the normalized trace of the conductivity tensor~(\ref{eq:thermal_conductivity_final_sum}) in the SMA approximation as \footnote{We consider the normalized trace of the conductivity tensor to simplify, later in this work, the discussion of CsPbBr$_3$, whose conductivity tensor is not isotropic. We employ the SMA approximation because it is accurate for the complex crystals with ultralow thermal conductivity in focus here, as well as for simple crystal that are not in the hydrodynamic regime of thermal transport, e.g. silicon~\cite{PhysRevX.10.011019}. Therefore, within the SMA approximation, one can distinguish the two regimes mentioned above.}
\begin{equation}
  \kappa^{\rm avg}{=}
\frac{1}{(2\pi)^3}
\sum_{s} \!\int_{\mathfrak{B}}\!\Big(\mathcal{K}^{\rm avg}_P(\bm{q})_s+ 
\mathcal{K}^{\rm avg}_C(\bm{q})_s\Big)d^3 q,
\label{eq:contr_single}
\end{equation}
where $\mathcal{K}^{\rm avg}_P(\bm{q})_s$ and  $\mathcal{K}^{\rm avg}_C(\bm{q})_s$ quantify the contributions of the phonon $(\bm{q})_s$ to the particle-like and to the wave-like (coherences) conductivities of Eq.~(\ref{eq:thermal_conductivity_final_sum}), respectively (see Appendix~\ref{sec:ioffe_regel_limit_in_space_and_center_of_the_non_sharp_particle_wave_crossover} for details).
The ratio between these  wave-like and particle-like conductivity contributions 
quantifies the relative strength of the corresponding microscopic conduction mechanisms.
To discuss 
the relative strength of  particle-like and wave-like conduction mechanisms, it is useful to introduce the average phonon interband spacing,
\begin{equation}
\Delta{\omega_{\rm avg}}=\frac{\omega_{\rm max}}{3 N_{\rm at }},
\label{eq:Delta_omega_avg}
\end{equation}
which is defined as the ratio between the maximum phonon frequency ($\omega_{\rm max}$) and the number of phonon bands ($3 N_{\rm at }$).
In fact, 
the ratio between the wave-like and particle-like contributions is approximately equivalent to the ratio between the phonon linewidth and  the average phonon interband spacing,
\begin{equation}
 \frac{\mathcal{K}^{\rm avg\!}_C(\bm{q})_s}{\mathcal{K}^{\rm avg\!}_P(\bm{q})_s}\simeq
 \frac{\Gamma(\bm{q})_s}{\Delta\omega_{\rm avg}}=\frac{[\Delta\omega_{\rm avg}]^{-1}}{\tau(\bm{q})_s}.
 %\simeq\frac{a}{\Lambda(\bm{q})_s},
  \label{eq:Ioffe_Regel_th}
\end{equation}
The approximated equality in Eq.~(\ref{eq:Ioffe_Regel_th}) 
is demonstrated with order-of-magnitude accuracy 
in Appendix~\ref{sec:ioffe_regel_limit_in_space_and_center_of_the_non_sharp_particle_wave_crossover}, relying on the assumption that at fixed $s$ and for variable $s'$, the ratio between the off-diagonal and diagonal velocity-operator elements has an order of magnitude equal to one, $\left[\frac{\!\sum_\alpha|{\tenscomp{v}^\alpha}(\bm{q})_{s,s'\!}|^2\!}{
  \sum_\alpha |\tenscomp{v}^\alpha(\bm{q})_{s,s}|^2
}\right]\simeq1$.
Such an assumption can be understood and justified 
%in La$_2$Zr$_2$O$_7$ and CsPbBr$_3$  
by noting that in  Eq.~(\ref{eq:thermal_conductivity_final_sum})
the trend of the velocity-operator elements $\sum_{\alpha}|{\tenscomp{v}^\alpha}(\bm{q})_{s,s'\!}|^2$ as a function of the frequency difference $\omega(\bm{q})_{s'}{-}\omega(\bm{q})_{s}$ determines how the coherences conductivity behaves at high temperature. %\added{(see also Appendix~\ref{sec:trend_of_the_coherences_conductivity_as_a_function_of_temperature} for a more in-depth discussion)}.
In particular, an approximatively saturating trend for the temperature-coherences conductivity curve  --- which appears in complex crystals with glass-like conductivity such as La$_2$Zr$_2$O$_7$ (Fig.~\ref{fig:k_vs_T_Zirconate}) and CsPbBr$_3$   (Fig.~\ref{fig:k_vs_T}) --- is related to off-diagonal velocity operator elements $|{\tenscomp{v}^\alpha}(\bm{q})_{s,s'\!}|^2$ that for fixed $s$ and variable $s'$ are approximately independent from $s'$ (\textit{i.e.} they are independent from the frequency difference $\omega(\bm{q})_{s'}{-}\omega(\bm{q})_{s}$).

We highlight how Eq.~(\ref{eq:Ioffe_Regel_th}) predicts the crossover from particle-like to wave-like conduction for phonons to be not sharp, with the possibility to have phonons that contribute simultaneously and comparably to populations' particle-like conductivity and to coherences' wave-like conductivity. 
Importantly, phonons with a lifetime $\tau(\bm{q})_s$ equal to the inverse of the average interband spacing $[\Delta\omega_{\rm avg}]^{-1}$ are predicted to be at the center of such non-sharp crossover. 
Since such prediction is obtained via the Wigner framework, hereafter we will refer to the time scale $[\Delta\omega_{\rm avg}]^{-1}$ as  ``Wigner limit in time''.

\begin{figure}[b]
  \centering
  \includegraphics[width=\WidthFigure]{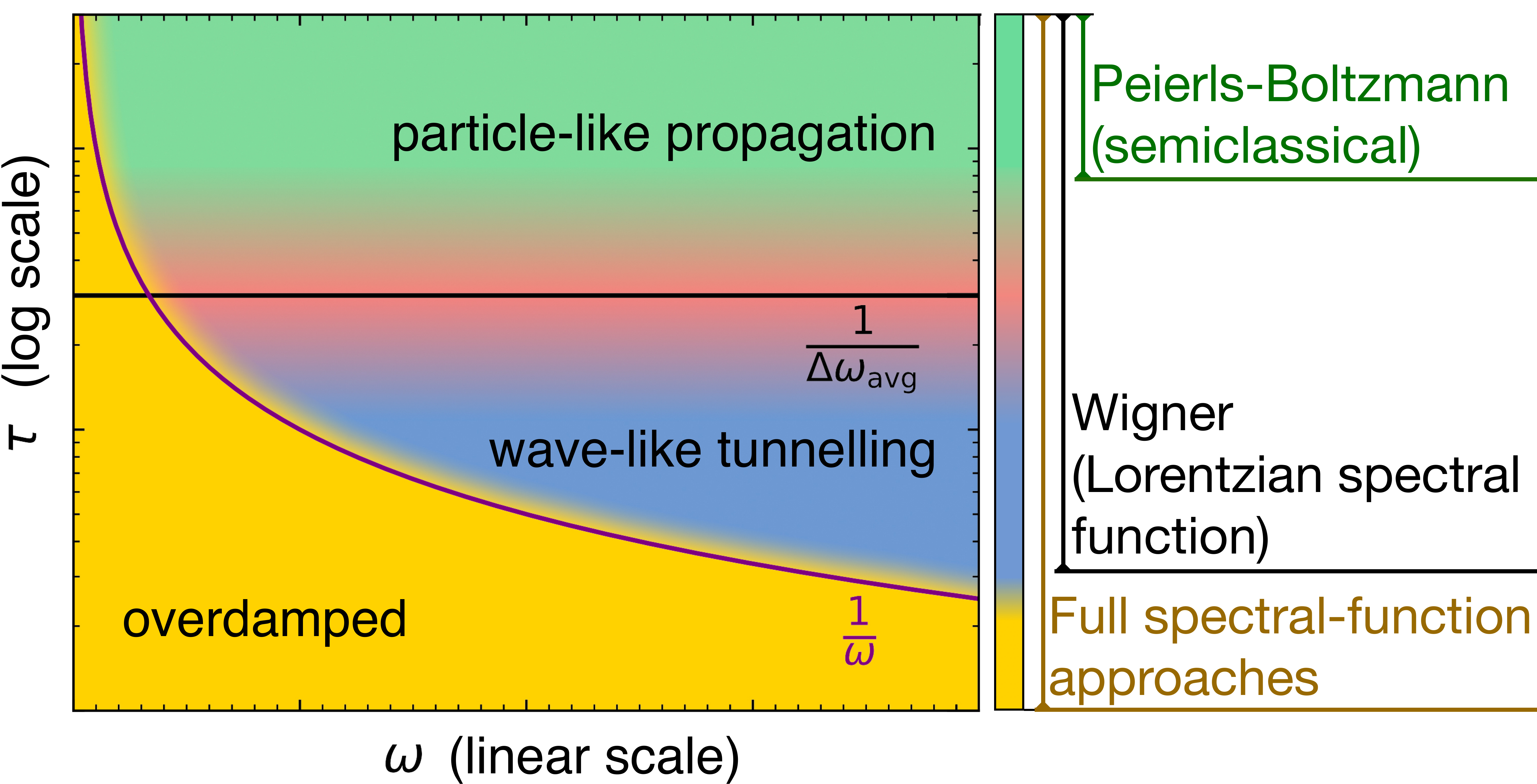}
  \caption{
  \textbf{Regime diagram for thermal transport.} Three regimes can be distinguished in a diagram reporting the phonon lifetime ($\tau$) as a function of the phonon frequency ($\omega$). 
  The horizontal black like is the Wigner limit in time, $\tau=1/\Delta\omega_{\rm avg}$ (\textit{i.e.} the inverse average interband spacing, see Eq.~(\ref{eq:Delta_omega_avg}), and denotes the center of the non-sharp particle-wave crossover for phonons. The green region well above the line represents phonons that propagate particle-like and mainly contribute to the populations conductivity;
  the blue region below the line represents phonons that tunnel wave-like and mainly contribute to the coherences conductivity; the red points around the line are phonons that contribute simultaneously and comparably to the particle-like and wave-like conductivities. 
  The purple line $\tau{=}1/\omega$ is the center of the non-sharp Ioffe-Regel limit in time~\cite{PhysRevLett.72.234,iofferegel1960}. 
  The region above it represents well defined phonons; the Wigner transport equation~(\ref{eq:Wigner_evolution_equation_N}) 
  describes all these phonons (blue, red, green), while the semiclassical Peierls-Boltzmann equation accounts only for phonons that propagate particle-like (green). 
  The yellow region below the Ioffe-Regel limit represents overdamped phonons, which require full spectral-function approaches~\cite{caldarelli2022,dangic2020origin} to be described correctly. 
  }
  \label{fig:transport_regimes}
\end{figure}

\begin{figure*}
  \vspace*{-2mm}
  \centering
  \includegraphics[width=\textwidth]{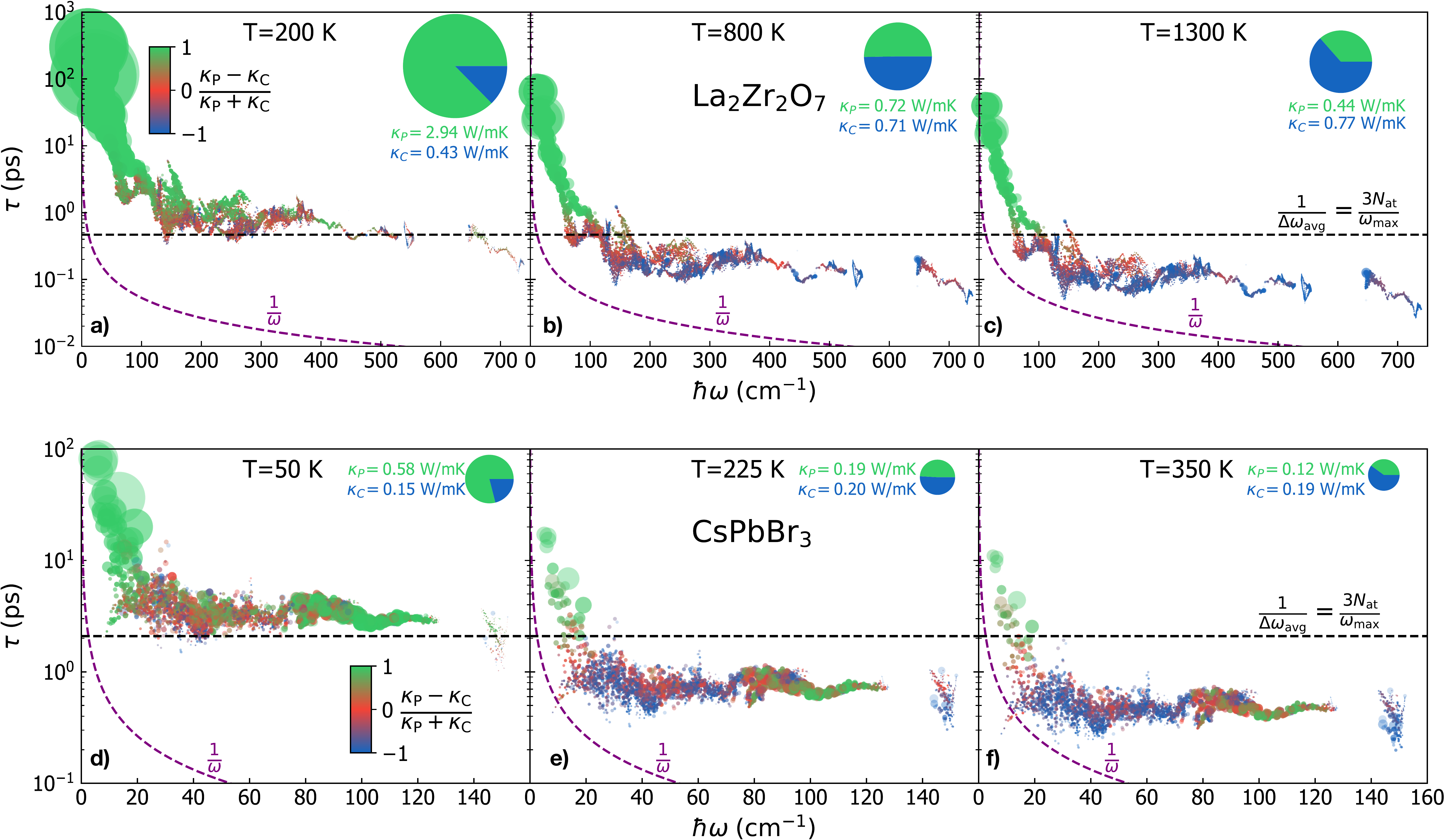}\vspace*{-2mm}

  \caption{  
  \textbf{Phonon lifetimes in La$_2$Zr$_2$O$_7$ and CsPbBr$_3$.}
  We show the phonon lifetimes $\tau(\bm{q}){=}[\Gamma(\bm{q})_s]^{-1}$ 
  as a function of the energy $\hbar\omega({\bm{q}})_s$ 
  for La$_2$Zr$_2$O$_7$ (at a temperature of 200 K (\textbf{a}), 800 K (\textbf{b}), and 1300 K (\textbf{c})), and for CsPbBr$_3$ (at a temperature of 50 K (\textbf{d}), 225 K (\textbf{e}), and 350 K (\textbf{f})). 
  The area of each circle is proportional to the contribution to the total conductivity and colored according to the origin of the contribution: green for particle-like propagation of populations; blue for wave-like  tunneling of coherences; intermediate colors represents phonons contributing to both mechanisms, with red corresponding to $50\%$ of each (see Eq.~(\ref{eq:coloring})).  The dashed-black line is the Wigner limit in time,  corresponding to a phonon lifetime equal to the inverse of the average interband spacing ($\tau=[\Delta\omega_{\rm avg}]^{-1}$ see Eq.~(\ref{eq:Delta_omega_avg})).
  We highlight how the non-sharp crossover from dominant particle-like conduction to dominant wave-like conduction occurs around this value, in agreement with the theoretical predictions from the Wigner framework (Eq.~(\ref{eq:Ioffe_Regel_th})). 
  The dashed-purple hyperbola is the Ioffe-Regel limit in time, $\tau=1/\omega$, the \nametheory{} formulation is applicable when phonons have lifetime longer than this value, 
  a condition which is always verified here. 
  The pie charts have an area proportional to the total conductivity, and the slices resolve the populations conductivity (particle-like, green) and the coherences conductivity (wave-like, blue).
  }
  \label{fig:lifetimes}
\end{figure*}

It is worth mentioning that another condition on the phonon lifetime is discussed in the literature, in relation to the validity of theoretical frameworks that describe transport relying on the concept of phonon as quasiparticle excitation. Specifically, such condition requires phonons to be above the ``Ioffe-Regel limit in time'', \textit{i.e.} to have a lifetime longer than their reciprocal angular frequency, $\tau(\bm{q})_s{=}[\Gamma(\bm{q})_s]^{-1} {>}\tfrac{1}{\omega(\bm{q})_s}$ \cite{PhysRevLett.72.234,iofferegel1960}, implying that the atomic vibrations are not overdamped \cite{knudsen2002elements}. We note that also this limit is not sharp, and definitions of this limit differing by a factor of $2\pi$ can be found in the literature \cite{Taraskin_Elliott_2000}.
The definition employed here follows the book of \citet{landau1980statistical}, where the comparison between the quasiparticle energy ($\hbar\omega(\bm{q})_s$) and its ``degree of broadening'' ($\hbar\Gamma(\bm{q})_s{=}{\hbar}{[\tau(\bm{q})_s]^{-1}}$) is used to assess if the concept of quasiparticle excitations is well defined.
Therefore, having phonons with a lifetime longer the Ioffe-Regel limit in time is a necessary condition for the validity of the \nametheory{} formulation discussed here, since the  particle-like and wave-like conduction mechanisms discussed here are derived relying on the assumption that phonons are well defined quasiparticle excitations (\textit{i.e.} one can label them with a wavevector $\bm{q}$ and with a mode index $s$).
We also note that having phonons with a lifetime longer than the Ioffe-Regel limit in time is a necessary but not sufficient condition for the validity of the semiclassical particle-like description, since having well defined phonons does not imply that  wave-like conduction mechanisms are negligible.
When, instead, the phonon lifetimes become shorter than the Ioffe-Regel limit in time ($\tau(\bm{q})_s{<}\tfrac{1}{\omega(\bm{q})_s}$), \textit{i.e.} in the so-called overdamped regime, phonons are no longer well defined quasiparticle excitations.
This implies that it is no longer possible to describe scattering in terms of the phonon wavevector $\bm{q}$ and mode $s$, thus Eq.~(\ref{eq:formula_scattering}) is no longer valid and one has to address this regime considering phonon spectral functions, as discussed in Refs.~\cite{caldarelli2022,dangic2020origin}.
All these regimes are summarized in Fig.~\ref{fig:transport_regimes}.

With the goal of validating the analytical considerations above (Eq.~(\ref{eq:Ioffe_Regel_th})),
we plot in Fig.~\ref{fig:lifetimes} the  phonon lifetime  $\tau(\bm{q})_s$ as a function of the phonon energy  $\hbar\omega(\bm{q})_s$ and at different temperatures  for La$_2$Zr$_2$O$_7$ (panels \textbf{a}, \textbf{b}, \textbf{c}) and CsPbBr$_3$ (panels \textbf{d}, \textbf{e}, \textbf{f}).
The particle-like and wave-like conductivity contributions of Eq.~(\ref{eq:contr_single}) ($\mathcal{K}^{\rm avg\!}_P(\bm{q})_s$ and $\mathcal{K}^{\rm avg\!}_C(\bm{q})_s$, respectively)  are used in Fig.~\ref{fig:lifetimes} to quantify the magnitude and  type of the contribution to the conductivity: the area of each circle is proportional to the total contribution to the conductivity, $\mathcal{K}^{\rm avg\!}_P(\bm{q})_s{+}\mathcal{K}^{\rm avg\!}_C(\bm{q})_s$, and the color of each circle is set according to the value of 
\begin{equation}
  c{=}\frac{\mathcal{K}^{\rm avg\!}_P(\bm{q})_s-\mathcal{K}^{\rm avg\!}_C(\bm{q})_s}{\mathcal{K}^{\rm avg\!}_P(\bm{q})_s+\mathcal{K}^{\rm avg\!}_C(\bm{q})_s}.
  \label{eq:coloring}
\end{equation}
Specifically, green corresponds to $c{=}1$ (dominant particle-like conductivity contribution), blue corresponds to $c{=}{-}1$ (dominant wave-like conductivity contribution), and intermediate colors represent coexisting particle- and wave-like conduction (e.g. red corresponds to 50\% of each, $c{=}0$). 
Results in Fig.~\ref{fig:lifetimes} are in broad agreement with the predictions of Eq.~(\ref{eq:Ioffe_Regel_th}), with phonons above the Wigner limit in time (\textit{i.e.} with long lifetime) contributing mainly to the populations' particle-like conductivity, and phonons below the Wigner limit in time (\textit{i.e.} with extremely short lifetime) contributing mainly to the coherences' wave-like conductivity.
These results can be rationalized intuitively:
phonons with an extremely short lifetime give a negligible contribution to the particle-like propagation mechanisms because they are suppressed too quickly (\textit{i.e.} before having the time to propagate significantly), but they can give a significant contribution to the wave-like conductivity since they are still allowed to interfere and tunnel (here ``interference'' is used in the sense of the frequency difference appearing in the coherences equation~(\ref{eq:coherences_steady_state})).
Finally, we note that in both La$_2$Zr$_2$O$_7$ and CsPbBr$_3$ all the phonons are above the Ioffe-Regel limit in time (\textit{i.e.} not overdamped, for orthorhombic CsPbBr$_3$ below 350 K this is in broad agreement with recent experiments \cite{lanigan2021two}). This confirms that phonons in these materials are well defined quasiparticle excitations, validating a key hypothesis employed in the derivation of the \RedNmTh{}~(\ref{eq:Wigner_evolution_equation_N}).

\begin{figure*}
  \vspace*{-2mm}
  \centering
  \includegraphics[width=\textwidth]{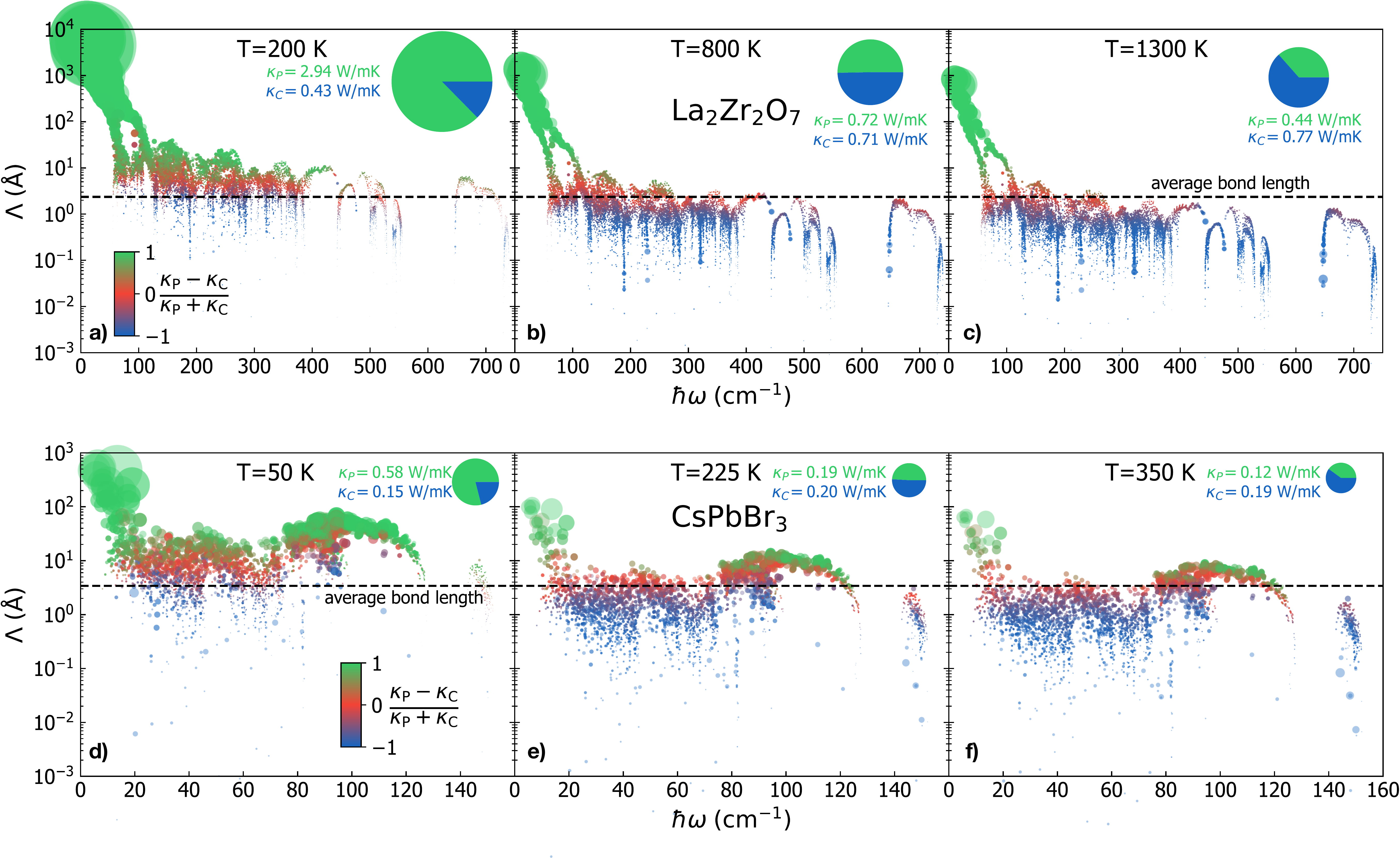}\vspace*{-2mm}

  \caption{  
  \textbf{Phonon mean free paths in La$_2$Zr$_2$O$_7$ and CsPbBr$_3$.}
  We plot the phonon mean free paths (MFP) $\Lambda(\bm{q})_s{=}\replaced{\tenscomp{v}^{\rm avg}(\bm{q})_{s,s}}{\tfrac{|\!|{\tens{v}}\!(\bm{q})_{s,s}|\!|}{\sqrt{3}}}\tau(\bm{q})_s$ 
  as a function of the energy $\hbar\omega({\bm{q}})_s$ 
  for La$_2$Zr$_2$O$_7$ (at a temperature of 200 K (\textbf{a}), 800 K (\textbf{b}), and 1300 K (\textbf{c})), and for CsPbBr$_3$ (at a temperature of 50 K (\textbf{d}), 225 K (\textbf{e}), and 350 K (\textbf{f})). 
  The area of each circle is proportional to the contribution to the total conductivity and colored according to the origin of the contribution: green for particle-like propagation of populations; blue for wave-like  tunneling of coherences; intermediate colors represents phonons contributing to both mechanisms, with red corresponding to $50\%$ of each (see Eq.~(\ref{eq:coloring})).    
  The dashed-black line in panels \textbf{a},\textbf{b},\textbf{c} (\textbf{d},\textbf{e},\textbf{f}) is the average bond length in La$_2$Zr$_2$O$_7$ (CsPbBr$_3$).
  We highlight how phonons at the center of the non-sharp crossover from dominant particle-like conduction to dominant wave-like conduction have a mean free path approximately equal to the average bond length (\textit{i.e.} they are at the Ioffe-Regel limit in space~\cite{allen1999diffusons,iofferegel1960}), a behavior which is explained by Eq.~(\ref{eq:Ioffe_Regel_space}).
  The pie charts have an area proportional to the total conductivity, and the slices resolve the populations conductivity (particle-like, green) and the coherences conductivity (wave-like, blue).
  }
  \label{fig:MFP}
\end{figure*}
\subsection{Ioffe-Regel limit in space} % (fold)
\label{sub:ioffe_regel_limit_in_space}
The analysis of the previous section has been performed in the frequency/lifetime domain because frequencies and lifetimes appear explicitly in the \RedNmTh{}~(\ref{eq:Wigner_evolution_equation_N}) and related conductivity formula~(\ref{eq:thermal_conductivity_final_sum}).
Such an analysis can be recast also in the the frequency/mean free path domain, a representation that is commonly employed in the literature \cite{Kittel_thermal_cond_glasses,allen1993thermal,allen1999diffusons,agne2018minimum,Mukhopadhyay1455,luo2020vibrational} and thus worth to be discussed also here.  

We plot in Fig.~\ref{fig:MFP} the phonon mean free path (MFP, $\Lambda(\bm{q})_s{=}\tenscomp{v}^{\rm avg}(\bm{q})_{s,s}\tau(\bm{q})_s$, \added{where $\tenscomp{v}^{\rm avg}(\bm{q})_{s,s}{=}\sqrt{\frac{1}{3}\sum_{\alpha=1}^3 |\tenscomp{v}^{\alpha}(\bm{q})_{s,s}|^2}$ is the spatially-averaged modulus of the group velocity, more details are provided in Eq.~(\ref{eq:populations_density}))} as a function of the phonon energy $\hbar\omega(\bm{q})_s$ at different temperatures for La$_2$Zr$_2$O$_7$ (panels \textbf{a}, \textbf{b}, \textbf{c}) and CsPbBr$_3$ (panels \textbf{d}, \textbf{e}, \textbf{f}). 
The non-sharp crossover from dominant particle-like propagation mechanisms to dominant wave-like tunneling mechanisms is evident also in the frequency/MFP domain, and its center is approximatively identified by having MFPs equal to the typical interatomic spacing $a$, \textit{i.e.} by phonons which are at the so-called ``Ioffe-Regel limit in space'' \cite{allen1993thermal,allen1999diffusons}.
More precisely, phonons with $\Lambda(\bm{q})_s> a$ (above the Ioffe-Regel limit in space) contribute mainly to the populations' particle-like conductivity, while  
phonons with $\Lambda(\bm{q})_s< a$ (below the Ioffe-Regel limit in space) contribute mainly to the coherences wave-like conductivity.

The appearance of the interatomic spacing as length scale determining the center of the particle-wave crossover can be rationalized with an order-of-magnitude analysis (similar to that leading to Eq.~(\ref{eq:Ioffe_Regel_th})), which yields
\begin{equation}
 \frac{\mathcal{K}^{\rm avg\!}_C(\bm{q})_s}{\mathcal{K}^{\rm avg\!}_P(\bm{q})_s}\simeq
\frac{[\Delta\omega_{\rm avg}]^{-1}}{\tau(\bm{q})_s}
 \simeq\frac{a}{\Lambda(\bm{q})_s}.
  \label{eq:Ioffe_Regel_space}
\end{equation}
Eq.~(\ref{eq:Ioffe_Regel_space}) follows \deleted{straightforwardly} from Eq.~(\ref{eq:Ioffe_Regel_th}), recasting the lifetime in terms of the mean free path, $\tau(\bm{q})_s{=}\Lambda(\bm{q})_s/\replaced{\tenscomp{v}^{\rm avg}(\bm{q})_{s,s}}{\big(\tfrac{|\!|{\tens{v}}\!(\bm{q})_{s,s}|\!|}{\sqrt{3}}\big)}$, 
and using the following estimate for the group velocity:
\begin{equation}
  \replaced{\tenscomp{v}^{\rm avg}(\bm{q})_{s,s}}{\frac{ |\!|\tens{v}(\bm{q})_{s,s}|\!|}{\sqrt{3}}}{=}\frac{|\!|\nabla_{\bm{q}}\omega(\bm{q})_s|\!|}{\sqrt{3}}\simeq 
  \Delta\omega_{\rm avg}\frac{a {\sqrt[3]{N_{\rm at}}}}{\pi}\simeq a \Delta\omega_{\rm avg}.
  \label{eq:group_velocity_est}
\end{equation}
This first approximated equality in Eq.~(\ref{eq:group_velocity_est}) derives from the following  considerations: (i) each phonon band spans throughout half the Brillouin zone a frequency interval that is approximatively equal to the average interband spacing $\Delta\omega_{\rm avg}$ (we consider half the Brillouin zone because of the time reversal symmetry $\omega(\bm{q})_s=\omega(-\bm{q})_s$; for example, in a representative one-dimensional case, all the variations taking place in the  positive half of the Brillouin zone are mirrored in the negative part); 
(ii) the typical magnitude of a reciprocal lattice vector can be estimated as $\frac{2\pi}{a {\sqrt[3]{N_{\rm at}}}}$ ($a$ is the typical interatomic spacing, $N_{\rm at}$ the number of atoms in the primitive cell, and thus
$a^3 {N_{\rm at}}$ is an estimate of the primitive-cell volume and  $a {\sqrt[3]{N_{\rm at}}}$ is an estimate of a direct lattice vectors' length). 
These considerations allow to estimate the typical group velocity as ratio between the typical frequency interval spanned by a phonon band over half the Brillouin zone and half the average length of the reciprocal lattice vector.
We note that in the last approximated equality in Eq.~(\ref{eq:group_velocity_est}) we have approximated the quantity $a\frac{\pi}{{\sqrt[3]{N_{\rm at}}}}\simeq a$.
Within the order-of-magnitude analysis performed here, this means that multiplying the typical bond length $a$ by the factor $\frac{\pi}{{\sqrt[3]{N_{\rm at}}}}$ yields a quantity that can still be considered as typical bond length. Such an approximation is realistic for typical complex crystals, since these usually have a primitive cell containing tens of atoms and for $N_{\rm at}$ ranging from 10 to 100 the ratio $\frac{\pi}{ {\sqrt[3]{N_{\rm at}}}}$ varies from 1.5 to 0.7 \footnote{For La$_2$Zr$_2$O$_7$ $N_{\rm at}= 22$, and for orthorhombic CsPbBr$_3$ $N_{\rm at}= 20$, thus in these cases $\frac{\pi}{ {\sqrt[3]{N_{\rm at}}}}\simeq 1.1$.}.
In general, complex crystals contain different chemical elements and bonds of different lengths, thus for the scope of this order-of-magnitude analysis one has some freedom --- compatible with a factor of $\frac{\pi}{{\sqrt[3]{N_{\rm at}}}}$  --- in the definition of the typical interatomic spacing. This freedom justifies the last approximation performed in Eq.~(\ref{eq:group_velocity_est}).

In summary, we have \replaced{shown}{demonstrated} the existence of a non-sharp crossover from the regime where phonons mainly propagate particle-like to the regime where phonons mainly tunnel wave-like. We have shown that the Wigner limit in time (corresponding to phonons with lifetime equal to the inverse average interband spacing, $\tau(\bm{q})_s=[\Delta\omega_{\rm avg}]^{-1}$), or equivalently the Ioffe-Regel limit in space (corresponding to phonons with MFP equal to the typical interatomic spacing, $\Lambda(\bm{q})_s=a$), can be used to identify the center of the non-sharp particle-wave crossover.
The analysis and data reported in this section 
demonstrate that the Winger formulation allows to describe thermal transport beyond the Ioffe-Regel limit in space, solving rigorously a long-standing problem that has hitherto been tackled with phenomenological models \cite{Mukhopadhyay1455,luo2020vibrational}.
The results reported in Figs.~\ref{fig:lifetimes},\ref{fig:MFP}, and their explanation with Eqs.~(\ref{eq:Ioffe_Regel_th},\ref{eq:Ioffe_Regel_space}), shed light on the strengths and weaknesses of the phenomenological heat-conduction models that conjecture the existence of two heat-conduction mechanisms sharply divided by the Ioffe-Regel limit in space \cite{Mukhopadhyay1455,luo2020vibrational}. Specifically, we have shown that: (i) phonons mainly propagate particle-like (tunnel wave-like) above (below) the Ioffe-Regel limit in space, and this supports where the separation between the particle-like and wave-like channels is sharply performed in the phenomenological models of Refs.~\cite{Mukhopadhyay1455,luo2020vibrational};
(ii) the particle-wave crossover for phonons is not sharp, with phonons contributing simultaneously to both populations' particle-like conductivity and coherences' wave-like  conductivity, and this is in contrast with the phenomenological sharp separation between these two mechanisms done in Refs. \cite{Mukhopadhyay1455,luo2020vibrational}.

{{{
\section{Software implementations} % (fold)
\label{sec:software_implementation}

Here we report the numerical protocols to 
implement the general \RedNmTh{} thermal conductivity expression~(\ref{eq:thermal_conductivity_final_sum}) in commonly used computer programs. Modern LBTE solvers contain almost all the quantities needed to evaluate the general \RedNmTh{} thermal conductivity; the only missing quantity is the off-diagonal part of the velocity operator, and this can be obtained with minimal effort.

The starting point to compute the velocity operator~(\ref{eq:vel_op}) is the mass-renormalized interatomic-force-constant tensor $\tenscomp{G}_{\bm{R}b\alpha,\bm{R'}b'\!\alpha'}$ (Eq.~(\ref{eq:matrix_G})). 
Once the tensor $\tenscomp{G}_{\bm{R}b\alpha,\bm{R'}b'\!\alpha'}$ is known, one has to compute its Fourier transform in the smooth phase convention using Eq.~(\ref{eq:dynamical}),  
to obtain the smooth dynamical matrix in reciprocal Bloch representation ${{\tenscomp{D}}(\bm{q})}_{b\alpha,b'\alpha'}$.
We remark that some LBTE solvers (e.g. \textsc{Phono3py}~\cite{phono3py}) employ the smooth phase convention as default, others instead employ the step-like phase convention (e.g. \textsc{D3Q}~\cite{paulatto2015first,fugallo2013ab}). As discussed before, it is fundamental to use the smooth phase convention to implement correctly the generalized \RedNmTh{} conductivity~(\ref{eq:thermal_conductivity_final_sum}), and we refer the reader to Sec.~\ref{sec:preliminaries} for  details on how to transform quantities from the step-like phase convention to the smooth phase convention needed here. Then, one needs to compute the square root of the smooth dynamical matrix in reciprocal Bloch representation,  $\sqrt{{\tenscomp{D}}(\bm{q})}_{b\alpha,b'\alpha'}$.
To compute this quantity, we rely on the property that the square-root matrix $\sqrt{{\tenscomp{D}}(\bm{q})}_{b\alpha,b'\alpha'}$ has the same eigenvectors $\mathcal{E}(\bm{q})_{s,b\alpha}$ of the matrix ${{\tenscomp{D}}(\bm{q})}_{b\alpha,b'\alpha'}$, and the eigenvalues of $\sqrt{{\tenscomp{D}}(\bm{q})}_{b\alpha,b'\alpha'}$ are the square roots of the eigenvalues of ${{\tenscomp{D}}(\bm{q})}_{b\alpha,b'\alpha'}$. 
Eigenvectors and eigenvalues of ${{\tenscomp{D}}(\bm{q})}_{b\alpha,b'\alpha'}$ have been introduced in Eq.~(\ref{eq:diagonalization_dynamical_matrix}) (and are available in modern LBTE solvers), and allow to compute the square root of the smooth dynamical matrix in reciprocal Bloch representation as
\begin{equation}
\begin{split}
  \sqrt{{\tenscomp{D} }(\bm{q})}_{b\alpha,b'\alpha'}
  %\sum_{s,s'}\mathcal{E}(\bm{q})_{s,b\alpha}\times\\
  %&\times
  %\sum_{c,\beta}\sum_{c',\beta'}\sqrt{\mathcal{E}^\star(\bm{q})_{{s},c\beta} {\tenscomp{D}}(\bm{q})_{c\beta,c'\beta'} \mathcal{E}(\bm{q})_{s',c'\beta'}}\mathcal{E}^\star(\bm{q})_{{s'},b'\alpha'}\\
  &=\sum_{s}\mathcal{E}(\bm{q})_{s,b\alpha}\omega(\bm{q})_{s}\mathcal{E}^\star(\bm{q})_{{s},b'\alpha'}\;.
  \end{split}
  \raisetag{8mm}
  \label{eq:exception_sum}
\end{equation}
Once the square root of the dynamical matrix is known, its derivative --- needed to evaluate the velocity operator~(\ref{eq:vel_op}) --- is computed as:
\begin{equation}
\begin{split}
  &\nabla_{\!\bm{q}}^{\!\beta}\!{\sqrt{\!\tenscomp{D}(\bm{q})}_{b\alpha,b'\!\alpha'\!}}
  {=}
  \!\!\!\lim\limits_{\delta q^\beta{\to 0}}\!\!\!
  {\frac{\!\! \sqrt{\!\tenscomp{D}(\bm{q}{+}\delta q^\beta)}_{b\alpha,b'\!\alpha'}{-}\sqrt{\!\tenscomp{D}(\bm{q}{-}\delta q^\beta)}_{b\alpha,b'\!\alpha'\!\!}}{2\delta q^\beta}\! },
\end{split}
\raisetag{1mm}
\label{eq:velocity_operator_fin_diff}
 \end{equation}
and the limit here can be evaluated numerically employing standard finite-difference methods.
Once Eq.~(\ref{eq:velocity_operator_fin_diff}) is evaluated numerically, inserting its result into Eq.~(\ref{eq:vel_op}) allows to evaluate numerically the velocity operator.

After having computed the velocity operator, one can use the vibrational frequencies, phonon specific heats, and phonon linewidths --- which are available in modern LBTE solvers \cite{phono3py,alamode,li2014shengbte,carrete2017almabte,fugallo2013ab,paulatto2015first,paulatto2013anharmonic} --- to evaluate numerically the coherence conductivity ($\kappa_{\rm C}^{\alpha\beta}$) appearing in generalized \RedNmTh{} conductivity formula~(\ref{eq:thermal_conductivity_final_sum}). The total conductivity is finally obtained adding the Peierls-Boltzmann conductivity ($\kappa_{\rm P}^{\alpha\beta}$, already computed by LBTE solvers) and the coherences conductivity, $\kappa_{\rm TOT}^{\alpha\beta}{=}\kappa_{\rm P}^{\alpha\beta}{+}\kappa_{\rm C}^{\alpha\beta}$.

Finally, we stress that the total \RedNmTh{} conductivity $\kappa_{\rm TOT}^{\alpha \beta}$ is the physically relevant quantity. 
To have a Peierls-Boltzmann particle-like conductivity $\kappa^{\alpha\beta}_P$ equivalent to that discussed in previous work~\cite{fugallo2013ab}, one has to consider the diagonal or degenerate velocity-operator elements (\textit{i.e.} $\tenscomp{v}^\beta(\bm{q})_{s,s'}$ such that $\omega(\bm{q})_s=\omega(\bm{q})_{s'}$) as contributing exclusively to the Peierls-Boltzmann conductivity. 
A possible way to account for this prescription consists in exploiting the freedom of rotating the degenerate eigenstates of the dynamical matrix in reciprocal Bloch representation to obtain a velocity operator that is diagonal in the degenerate subspaces, as done in Ref.~\cite{fugallo2013ab}.

\section{Conclusions }
\label{sec:conclusion}

In summary, we have derived from the Wigner phase space formulation of quantum mechanics a \nametheory{} transport equation~(\ref{eq:Wigner_evolution_equation_N}) able to account rigorously and simultaneously for the interplay between the quantum Bose-Einstein statistics of atomic vibrations, anharmonicity, and disorder in determining the thermal properties of solids. 

We have discussed how the Wigner phase-space framework determines an expression for the microscopic harmonic energy field and the related microscopic harmonic heat flux; such an expression 
allows to compute the thermal conductivity from the solution of the linearized Wigner transport equation (\RedNmTh{}).
At leading order in the temperature gradient, the solution of the \RedNmTh{}~(\ref{eq:Wigner_evolution_equation_N})  yields a generalized \RedNmTh{} thermal conductivity~(\ref{eq:thermal_conductivity_final_sum}) that is the sum of two terms: a populations conductivity that describes the particle-like intraband propagation of phonons exactly as in the Peierls-Boltzmann LBTE, and a coherences conductivity describing the wave-like interband tunnelling of phonons.
We note in passing that the coherences thermal conductivity has some analogies with the Zener interband electrical conductivity \cite{PhysRevB.35.9644,PhysRevB.96.144303,gebauer2004current,gebauer2004kinetic,cepellotti2021interband}; however, thermal transport is fundamentally different from electronic transport since the full phonon spectrum contributes to the heat current, while only the bands close to the Fermi level
contribute to the electronic Zener current.
The \RedNmTh{} conductivity~(\ref{eq:thermal_conductivity_final_sum}) reduces to the LBTE conductivity in the limit of a simple crystal \cite{PhysRevX.10.011019}
(\textit{i.e.} characterized by interband spacings much larger than the linewidths) and to the Allen–Feldman conductivity in the limit of a harmonic glass; most importantly, the \RedNmTh{} conductivity~(\ref{eq:thermal_conductivity_final_sum}) is more general and encompasses all intermediate cases, applying, for example, also to complex crystals \deleted{and anharmonic glasses, which are} characterized by interband spacings smaller than the linewidths.
%\added{Moreover, we have discussed how the Wigner formulation extends the Allen-Feldman theory accounting for the anharmonicity, and future work will aim at understanding if it can be useful to better understand the anomalous thermal properties of glasses \cite{Moon2018,PhysRevMaterials.3.065601,Kim2021,Liu2009,Braun2016,Martin2022,wingert2016thermal,kwon2017unusually,PhysRevB.101.144203}.}
\added{Moreover, we have discussed how the Wigner formulation extends the Allen-Feldman theory accounting for the interplay between disorder, anharmonicity, and quantum Bose-Einstein statistics of atomic vibrations; however, evaluating the Wigner formulation in disordered solids is a computationally challenging task. Fittingly, recent work \cite{PhysRevB.105.134202} has discussed techniques that allow to account for anharmonicity in disordered systems at a reduced computational cost, and future work will aim at understanding if these techniques can be combined with the Wigner formulation to better understand the anomalous thermal properties of glasses \cite{Moon2018,PhysRevMaterials.3.065601,Kim2021,Liu2009,Braun2016,Martin2022,wingert2016thermal,kwon2017unusually,PhysRevB.101.144203}.}

{
We have discussed how the coefficients appearing in the \RedNmTh{}~(\ref{eq:Wigner_evolution_equation_N}), and in the \RedNmTh{} conductivity~(\ref{eq:thermal_conductivity_final_sum}) that follows from it, depend on the phase convention adopted for the Fourier transform.
Specifically, we have discussed the necessity to compute the dynamical matrix in reciprocal Bloch representation and related off-diagonal velocity operator's elements using the smooth phase convention~(\ref{eq:dynamical}), employed e.g. \replaced{by}{in the book of} Wallace~\cite{wallace1972thermodynamics}, in order to obtain a thermal conductivity that does not depend on the non-univocal possible choice of the crystal's unit cell and is size-consistent.}
{Therefore, the formulation presented here \replaced{improves and corrects}{is an improved version of} our previous work~\cite{simoncelli2019unified}, where the \deleted{other} commonly employed~\cite{peierls1955quantum,ziman1960electrons} step-like convention~(\ref{eq:dynamical_Ziman}) was adopted. We have also shown that \deleted{in practice} these improvements/corrections \replaced{become}{are} increasingly more relevant as \added{the} size and anisotropy of the unit cell used to describe a crystal increase. %within the present framework increase. 
}

We have shown that the present approach overcomes the limitations of the Peierls-Boltzmann formulation,
predicting accurately the temperature-conductivity curves of complex crystals such as La$_2$Zr$_2$O$_7$ and CsPbBr$_3$, both displaying a decay much milder than the universal $T^{-1}$ trend predicted by Peierls' theory (Figs.~\ref{fig:k_vs_T_Zirconate}, \ref{fig:k_vs_T}).
We have shown that the \RedNmTh{} formulation predicts a non-sharp crossover from dominant particle-like heat-conduction mechanisms to dominant wave-like heat-conduction mechanisms, which is centered at the Wigner limit in time (corresponding to $\tau(\bm{q})_s{\sim} [\Delta\omega_{\rm avg}]^{-1}$, where $\tau(\bm{q})_s$ is the phonon lifetime and $\Delta\omega_{\rm avg}$ is the average interband spacing). We have discussed the equivalence between the Wigner limit in time and the Ioffe-Regel limit in space, this latter corresponding to a phonon mean free path ($\Lambda(\bm{q})_s$) equal to the the typical interatomic spacing ($a$), $\Lambda(\bm{q})_s{\simeq} a$. 
Specifically we have shown that phonons with $\tau(\bm{q})_s{\gg} [\Delta\omega_{\rm avg}]^{-1}$ (or equivalently with $\Lambda(\bm{q})_s{\gg} a$) mainly propagate  particle-like and contribute to the populations conductivity; phonons with $\tau(\bm{q})_s{\ll} [\Delta\omega_{\rm avg}]^{-1}$ (or equivalently with $\Lambda(\bm{q})_s{\ll} a$) mainly tunnel wave-like and contribute to the coherences conductivity; phonons with $\tau(\bm{q})_s{\simeq} [\Delta\omega_{\rm avg}]^{-1}$ (or equivalently with $\Lambda(\bm{q})_s{\simeq} a$) behave simultaneously particle- and wave-like and contribute comparably to the populations and coherences conductivities.
We have exploited these findings to propose a classification, based on the heat conduction mechanisms, of the various regimes of thermal transport, also identifying the regime of applicability of the semiclassical Peierls-Boltzmann equation (which accounts for particle-like conduction only) and of the Wigner equation~(\ref{eq:Wigner_evolution_equation_N}) (which accounts for both particle- and wave-like conduction).
We have provided numerical recipes to implement with minimal effort the general \RedNmTh{} thermal conductivity 
formula in LBTE solvers \cite{alamode,li2014shengbte,carrete2017almabte,phono3py,fugallo2013ab}.

It is also worth mentioning that other 
thermal conductivity expressions valid 
%^approaches to describe thermal transport
in both crystals and glasses have been proposed relying on the Green-Kubo theory of linear response \cite{dangic2020origin,isaeva2019modeling,Semwal_1972,PhysRevB.103.024204}, which differ under some aspects from Wigner expression~(\ref{eq:thermal_conductivity_final_sum}) discussed here.
On the one hand, the Green-Kubo formulations reported to date are restricted to the so-called “kinetic regime” of thermal transport, where Umklapp scattering processes dominate and the single-mode relaxation-time approximation is accurate \cite{fugallo2013ab,lindsay_first_2016,cepellotti2015phonon}. 
Such a restriction is not present in the Wigner transport equation~(\ref{eq:Wigner_evolution_equation_N}), which can be used also in the hydrodynamic regime of thermal transport (\textit{i.e.} when normal scattering processes dominate and the single-mode relaxation-time approximation breaks down \cite{cepellotti2015phonon,fugallo2013ab,PhysRevX.10.011019}).
On the other hand, the Green-Kubo framework allows to describe transport in the overdamped regime, \textit{i.e.} below the Ioffe-Regel limit in time, where phonons are no longer well defined quasiparticle excitations and they have to be described using spectral functions \cite{caldarelli2022}. We highlight that the complex crystals studied in this work are not in the overdamped regime (\textit{i.e.} they have phonons above the Ioffe-Regel limit in time), but conditions under which the overdamped regime can be reached have been discussed recently~\cite{lanigan2021two}. 
The approaches discussed up to now account for anharmonicity approximatively, truncating at third or higher order the expansion of the interatomic potential around equilibrium (Eq.~(\ref{eq:expansion_pot})). We note that approaches based on molecular dynamics (MD), either in the equilibrium \cite{marcolongo2016microscopic,PhysRevLett.118.175901,baroni2020heat,ercole2017accurate,verdi2021thermal,mcgaughey2014predicting,Lv2016}, non-equilibrium \cite{PhysRevB.103.174306,PhysRevLett.113.185501,li2019influence,McGaughey2009predicting,felix2020suppression}, or approach-to-equilibrium \cite{Lampin2013,Puligheddu2017} flavors, allow to account for anharmonicity exactly, although are computationally more expensive than the Wigner approach and do not account for the quantum Bose-Einstein statistics of atomic vibrations \cite{Puligheddu2019}, which affects significantly the thermal properties around and below the Debye temperature. %, thus can be used only to describe conduction at high temperature (\textit{i.e.} for $T{\gg} T_D$, where $T_D$  is the Debye temperature). %Also, these molecular-dynamics methods have a prohibitively high computational cost, as they require nanoseconds-long simulations on models containing hundreds to thousands of atoms, and the

%thus have computational costs that h their application.
 
We conclude by noting that this novel formulation is relevant for several technological applications, as it will potentially allow to predict the ultralow thermal conductivity of, for example, target materials for thermoelectric energy conversion   \cite{lee2017ultralow,voneshen2013suppression,Minnich_thermoelectric,GChen_nat,2021_thermo,Zheng2021,Hanus2021,Hu2021,PhysRevB.100.220201,PhysRevB.89.125403}, porous materials for gas-storage technologies   \cite{PhysRevLett.116.025902,kapil2019modeling}, materials for thermal barrier coatings \cite{zhang2017lanthanum,Suresh1997,Vassen2000,Chen2009,Wan2010,Hanus2021}, 
complex crystals used in high-temperature piezoelectric transducers \cite{damjanovic1998materials},
and extremely anisotropic van der Waals thermal conductors \cite{kim2021extremely}
(see also Refs. \cite{xia2020particlelike,PhysRevX.10.041029,PhysRevB.102.201201,PhysRevLett.125.085901,tadano2021firstprinciples,godse2021anharmonic,Hanus2021_b} for recent applications of the present framework to materials science).

\section*{Acknowledgements} % (fold)
\label{sec:section_name}
We thank G. Caldarelli, L. Benfatto, \added{and P. B. Allen} for many useful discussions. 
%We acknowledge support from the Swiss National Science Foundation (SNSF) and the MARVEL NCCR. 
This research was supported by the NCCR MARVEL, a National Centre of Competence in Research, funded by the Swiss National Science Foundation (grant number 182892).
M.S. acknowledges support from the SNSF project P500PT\_203178.

\appendix

\section{Details on the localization properties of the bosonic operators in real space.}
\subsection{Mean square momentum on first excited state.}
\label{sub:app_momentum}
{
Here we perform a calculation analogous to that reported in the main text for the mean square displacement.
From Eq.~(\ref{eq:def_op_a_real}) it follows that the momentum operator for an atom at position $\bm{R'}{+}\bm{\tau}_{b'}$ can be written as:
\begin{equation}
\begin{split}
   \hat{p}(\hspace*{-0.3mm}\bm{R'}\hspace*{-0.3mm})_{b'\!\alpha'}{=}\sqrt{{\!\frac{\hbar M_{b'}}{2} }}\!\!\!\!\!\sum_{\bm{R'\!'}\!,b'\!'\!,\alpha'\!'}\!\!\!\!\!\!
    \sqrt[4]{\!\tenscomp{G}}_{\!\bm{R'}\!b'\!\alpha'\!\!,\bm{R'\!'}\!b'\!'\!\alpha'\!'}\!\!\left[\!\hat{a}(\hspace*{-0.3mm}\bm{R'\!'}\hspace*{-0.3mm})_{b'\!'\!\alpha'\!'} {+}\hat{a}^\dagger\!(\hspace*{-0.3mm}\bm{R'\!'}\hspace*{-0.3mm})_{b'\!'\!\alpha'\!'}\! \right]\!.
    \label{eq:momenutm}
 \end{split}
 \raisetag{4mm}   
 \end{equation} 
Then, the expectation value of the squared momentum on the excited state is:
\begin{equation}
\begin{split}
    &\left<0\right|\hat{a}(\bm{R})_{b\alpha}|\hat{p}^2(\bm{R'})_{b'\alpha'}\big|\hat{a}^\dagger(\bm{R})_{b\alpha}\left|0\right>
    %&=\frac{\hbar}{M_{b'}}\left(\sqrt[4]{\tenscomp{G}^{-1}}_{\bm{R'}b'\alpha',\bm{R}b\alpha}\right)^2 +\frac{\hbar}{2M_{b'}}\sqrt{\tenscomp{G}^{-1}}_{\bm{0}b'\alpha',\bm{0}b'\alpha'}\\
    \\
    &={\hbar M_{b'}}\left(\sqrt[4]{\tenscomp{G}}_{\bm{R'}b'\alpha',\bm{R}b\alpha}\right)^2 
    +\left<0\right|\hat{p}^2(\bm{R'})_{b'\alpha'}\left|0\right>,
\end{split}
  \label{eq:ref_excitation_mom}
\end{equation}
where $\left<0\right|\hat{p}^2(\bm{R'})_{b'\alpha'}\left|0\right>{=}\frac{\hbar M_{b'}}{2}\sqrt{\tenscomp{G}}_{\bm{0}b'\!\alpha',\bm{0}b'\!\alpha'}$ is the mean square momentum due to the zero-point motion.
As mentioned in the main text, the force constant matrix $\tenscomp{G}_{\bm{R}b\alpha,\bm{R'}\!b'\!\alpha'}$ decays to zero for $|\bm{R}{+}\bm{\tau}_b{-}\bm{R'}{-}\bm{\tau}_{b'}|{\to}\infty$, and from this it follows that also $\sqrt[4]{\tenscomp{G}}_{\bm{R'}\!b'\!\alpha'\!,\bm{R}b\alpha}$ goes to zero in the same limit \cite{SimoncelliPhD} (as it happens for $\sqrt[4]{\tenscomp{G}^{-1}}_{\bm{R'}\!b'\!\alpha'\!,\bm{R}b\alpha}$, see main text). 
Therefore, Eq.~(\ref{eq:ref_excitation_mom}) shows that the bosonic operator $\hat{a}^\dagger(\bm{R})_{b\alpha}$ 
creates atomic vibrations in which atoms close to $\bm{R}{+}\bm{\tau}_b$ have a larger mean square momentum than atoms far from $\bm{R}{+}\bm{\tau}_b$. 
}

\subsection{Localization problems of the phonon-mode basis }
\label{sub:localization_problems_of_the_eigenmodes_basis}
{
In this Appendix we show that the standard phonon operators $\hat{a}(\bm{q})_{s}$ defined in Eq.~(\ref{eq:ph_op_mode_basis}), have in general non-unique real-space representation and as such are not suitable to track vibrations in real space. 

As anticipated in Sec.~\ref{sec:density_matrix_formalism_for_atomic_vibrations}, the phase of the eigenvector $\mathcal{E}(\bm{q})_{s,b\alpha}$ appearing in Eq.~(\ref{eq:ph_op_mode_basis}) is undetermined at all $\bm{q}$-points; \textit{i.e.}, there exists a gauge freedom in the definition of the eigenvectors $\mathcal{E}(\bm{q})_{s,b\alpha}$ that allows to apply them unitary transformations 
\begin{equation}
{\mathcal{E}}_{_\mathcal{U}\!}(\bm{q})_{s,b\alpha}{=}\sum_{s'}\mathcal{U}(\bm{q})_{s,s'}\mathcal{E}(\bm{q})_{s',b\alpha},
\label{eq:transformation_matrix}  
\end{equation}
where $\mathcal{U}(\bm{q})_{s,s'}$ is a unitary matrix of size $3N_{\rm at}{\times}3N_{\rm at}$ periodic in $\bm{q}$ and with non-zero matrix elements exclusively within a subspace of phonon bands that is invariant with respect to the space group of the crystal \cite{revmodphys.84.1419}.
The freedom inherent in Eq.~(\ref{eq:transformation_matrix}) implies that the real-space (Wannier) representation of the phonon operator~(\ref{eq:ph_op_mode_basis}) is non-unique and its localization properties depend on the particular choice made for the arbitrary unitary matrix $\mathcal{U}(\bm{q})_{s,s'}$.
To illustrate this, we consider the case in which $\mathcal{U}(\bm{q})_{s,s'}{=}e^{-i\bm{q}\cdot\bm{R}_\mathcal{U}}\delta_{s,s'}$, where $\bm{R}_\mathcal{U}$ is an arbitrary Bravais-lattice vector, and we use the notation $\hat{{a}}_{\!_\mathcal{U}}\!(\bm{q})_{s}=\hat{{a}}(\bm{q})_{s}e^{i\bm{q}\cdot\bm{R}_\mathcal{U}}$ to emphasize that the phonon operator~(\ref{eq:ph_op_mode_basis}) depends on the arbitrary unitary transformation $\mathcal{U}(\bm{q})_{s,s'}$ (\textit{i.e.} on the choice of $\bm{R}_\mathcal{U}$ in this example).
The real-space (Wannier) representation of the phonon operator~(\ref{eq:ph_op_mode_basis}) is:
\begin{equation}
\begin{split}
\hat{{a}}_{_\mathcal{U}\!}(\bm{R})_{s}&{=}
\frac{\mathcal{V}}{(2\pi)^3}\!\!
\int_{\mathfrak{B}}\!\!
\hat{{a}}_{_\mathcal{U}\!}(\bm{q})_{s}
e^{+i\bm{q}{\cdot}\bm{R}} d^3q,
%{=}
%\frac{\mathcal{V}}{(2\pi)^3}\!\!
%\int_{\mathfrak{B}}\!\!
%\hat{a}(\bm{q})_{s}
%e^{+i\bm{q}{\cdot}(\bm{R}+\bm{R}_U)},
\end{split}
\raisetag{2mm}
\end{equation}
and clearly $\hat{{a}}_{_\mathcal{U}\!}(\bm{R})_{s}$ inherits from $\hat{{a}}_{_\mathcal{U}\!}(\bm{q})_{s}$ a dependence from the aforementioned arbitrary unitary transformation. 
Now we investigate the localization properties of $\hat{{a}}_{_\mathcal{U}\!}(\bm{R})_{s}$ performing an analysis analogous to that discussed in Sec.~\ref{sec:density_matrix_formalism_for_atomic_vibrations}. We create on the ground state $\left|0\right>$ the excitation 
$\hat{{a}}^\dagger_{_\mathcal{U}\!}(\bm{R})_{s}\left|0\right>$, and we compute the mean-square atomic displacement at a generic position $\bm{R'}{+}\bm{\tau}_{b'}$ in such an excited state, finding
\begin{equation}
\begin{split}
    &\left<0\right|\hat{a}_{_\mathcal{U}\!}(\bm{R})_{s}|\hat{u}^2(\bm{R'})_{b'\alpha'}\big|\hat{a}^\dagger_{_\mathcal{U}\!}(\bm{R})_{s}\left|0\right>-\left<0\right|\hat{u}^2(\bm{R'})_{b'\alpha'}\left|0\right>
    \\
    &{=}\frac{\hbar}{M_{b'}}\!\!\left(\!\!
    \frac{\mathcal{V}}{(2\pi)^3}
    \int_{\mathfrak{B}}\!\!
    \frac{\mathcal{E}(\bm{q})_{s,b'\!\alpha'}e^{i\bm{q}\cdot\bm{\tau}_{b'}}}{\sqrt{\omega(\bm{q})_s}}
    e^{i\bm{q}\cdot(\bm{R'}-\bm{R}-\bm{R}_\mathcal{U})}d^3q\!\right)^2\!\!\!,\hspace*{16mm}
\end{split}
  \label{eq:ref_excitation_ph}
  \raisetag{4mm}
\end{equation}
where $\left<0\right|\hat{u}^2(\bm{R'})_{b'\alpha'}\left|0\right>$ is the mean-square atomic displacement due to the zero-point motion discussed in Sec.~\ref{sec:density_matrix_formalism_for_atomic_vibrations}.
From Eq.~(\ref{eq:ref_excitation_ph}) it is apparent that the excitation created by 
the operator $\hat{a}^\dagger_{_\mathcal{U}\!}(\bm{R})_{s}$ on the ground state $\left|0\right>$ is centered at a Bravais lattice position $\bm{R'}-\bm{R}-\bm{R}_\mathcal{U}$, which can be arbitrarily shifted since the Bravais-lattice vector $\bm{R}_\mathcal{U}$ appearing in the unitary transformation $e^{i\bm{q}\cdot\bm{R}_\mathcal{U}}$ can be chosen arbitrarily.

In conclusion, we have shown that the standard phonon bosonic operators~(\ref{eq:ph_op_mode_basis}) have 
non-unique real-space representations, each of which creates atomic vibrations with different center.
This mirrors the electronic case, where the phase indeterminacy of the Bloch orbitals is reflected in the non-uniqueness of the transformation into Wannier functions \cite{revmodphys.84.1419}.
In contrast, the localized bosonic operators in real space~(\ref{eq:def_op_a_real})
have by construction well-defined localization properties (see Eq.~(\ref{eq:ref_excitation}) and Fig.~\ref{fig:displ_real}), which allow to track vibrations and thus perform the derivation detailed in Sec.~\ref{sec:Wigner_thermal_transport_equation}.}

\section{Properties of the Wigner matrix distribution}
\label{sec:properties_of_the_wigner_transform}
Here we discuss the properties of the Wigner matrix defined in  Eq.~(\ref{eq:Wigner_transform_Rec}).
Specifically, the Wigner representation of a one-body operator $\hat O$ 
is a distribution that depends on a direct lattice vector $\bm{R}$ and on a wavevector $\bm{q}$, it can depend on time $t$, and carries two matrix indexes ${b\alpha,b'\alpha'}$ that denote positions inside the primitive cell of the crystal; in this Appendix, such a distribution will be denoted with 
 $\mathcal{W}_{[O]}(\bm{R},\bm{q})_{b\alpha,b'\alpha'}{=}\tenscomp{O}(\bm{R},\bm{q})_{b\alpha,b'\alpha'}$ (the time dependence will be reported only if needed). 
Despite this phase-space distribution is discussed for the particular case of vibrational excitations (phonons) here, \replaced{in principle it can be employed also to describe other types of quasiparticles in a periodic potential.}{its extension to other quasiparticles in a periodic potential is straightforward.} 
Specifically, the one-body matrix element 
${O}(\bm{q}{+}\tfrac{\bm{q'\!'}}{2},\bm{q}{-}\tfrac{\bm{q'\!'}}{2})_{b\alpha,b'\alpha'}$ appearing in Eq.~(\ref{eq:Wigner_transform_Rec}) is a 
matrix element in the Zak basis   \cite{PhysRevLett.19.1385,zak1968dynamics,PhysRevB.10.1315}, which has as quantum numbers eigenvalues of translation operators in direct and reciprocal spaces 
(that is a position inside a unit cell $\bm{\tau}$, which is
a continuous position for electrons and the discrete atomic position for phonons, and a wavevector $\bm{q}$ belonging to a Brillouin zone).
We also note that, in order to adapt the formulation discussed here to other quasiparticles, the phase convention for the Zak eigenfunctions has to be chosen according to the convention~(\ref{eq:bosonic_reciprocal}), \textit{i.e.} it must yield a transformation that relates the matrix elements in the Zak basis to the matrix elements in the standard position basis analogous to Eq.~(\ref{eq:one_body_T1}). 

We want now to discuss and demonstrate the mathematical properties satisfied by the direct and inverse Wigner transformations (Eq.~(\ref{eq:Wigner_transform_Rec}) and Eq.~(\ref{eq:Wigner_transform_Rec_inv})).\\

\paragraph{Proof of the inverse Wigner transform~(\ref{eq:Wigner_transform_Rec_inv}).} % (fold)
\label{sub:proof_of_eq_wigner_representation_inv_}
To prove that the transformation~(\ref{eq:Wigner_transform_Rec_inv}) is the inverse of the transformation~(\ref{eq:Wigner_transform_Rec}), it is sufficient to show that inserting the first into the second (or vice versa) one obtains the identity:
\begin{equation}
\begin{split}
O(\bm{q}{+}\tfrac{\bm{q'\!'}}{2},\bm{q}{-}\tfrac{\bm{q'\!'}}{2})_{b\alpha,b'\alpha'} & {=}
  \!\!\!\int_{\mathfrak{B}}\!
O(\bm{q}{+}\tfrac{\bm{q'}}{2},\bm{q}{-}\tfrac{\bm{q'}}{2})_{b\alpha,b'\alpha'}\\
 &
    \frac{\mathcal{V}}{(2\pi)^3} 
\sum_{\bm{R}}
   e^{i(\bm{q'}-\bm{q'\!'})\cdot \bm{R}}     d^3q'\\
&\hspace*{-10mm}=\int_{\mathfrak{B}}\!
O(\bm{q}{+}\tfrac{\bm{q'}}{2},\bm{q}{-}\tfrac{\bm{q'}}{2})_{b\alpha,b'\alpha'}
\delta(\bm{q'}{-}\bm{q'\!'}) d^3q'.
\end{split}
\label{eq:proof_inverse}
\raisetag{17mm}
\end{equation}
In Eq.~(\ref{eq:proof_inverse}) we have used the exponential representation of the Dirac delta \cite{marder2010condensed,callaway1991quantum} 
{
\begin{equation}
\begin{split}
\sum_{\bm{R'\!'}} e^{-i(\bm{q'}{-}\bm{q'\!'})\cdot\bm{R'\!'}}&=
\lim_{N_c\to\infty} N_c \sum_{\bm{G}}\delta_{\bm{q'}{-}\bm{q'\!'},\bm{G}}\\
&= \frac{(2\pi)^3\!\!}{\mathcal{V}}\sum_{\bm{G}}\delta(\bm{q'}{-}\bm{q'\!'}{-}\bm{G})\\
&\hspace*{-3mm}\stackrel{\bm{q'\!}\!,\bm{q'\!'}\in \mathfrak{B}}{=} \frac{(2\pi)^3\!\!}{\mathcal{V}}\delta(\bm{q'}{-}\bm{q'\!'})\\
\end{split}
\label{eq:Dirac_Delta_representation}
\end{equation}
where $N_c\to\infty$ is the number of Bravais lattice vectors over which the sum $\sum_{\bm{R'\!'}}$ runs, $\bm{G}$ is a reciprocal lattice vector, $\delta_{\bm{q'}{-}\bm{q'\!'},\bm{G}}$ is the Kronecker delta, and $\delta(\bm{q'}{-}\bm{q'\!'}{-}\bm{G})$ is the Dirac delta.
In the last line of Eq.~(\ref{eq:Dirac_Delta_representation}) we have exploited the property that only the reciprocal lattice vector $\bm{G}=\bm{0}$ has to be considered when both $\bm{q'}$ and $\bm{q'\!'}$ are restricted to the first Brillouin zone $\mathfrak{B}$ \cite{callaway1991quantum,kittel_book} (we stress that this is the case in Eq.~(\ref{eq:proof_inverse})).}\\

\paragraph{The Wigner transform of a translation-invariant operator does not depend on space.} % (fold)
\label{ssub:wigner_transform_of_a_translation_invariant_operator}
{
Eq.~(\ref{eq:Wigner_transform_Rec}) implies that the Wigner transform of a translation-invariant operator, \textit{i.e.} diagonal in reciprocal space,
is independent from the spatial position $\bm{r}$. 
To show this, we start by considering a translation-invariant one-body operator $O(\bm{R},\bm{R'})_{b\alpha,b'\!\alpha'}{=}O(\bm{R}{-}\bm{R'})_{b\alpha,b'\!\alpha'}$.
Performing the Fourier transform~(\ref{eq:one_body_T1}), and defining $\bm{R''}{=}\bm{R}{-}\bm{R'}$, yields
\begin{equation}
\begin{split}
&O\big(\bm{q}{+}\tfrac{\bm{q'\!'}}{2},\bm{q}{-}\tfrac{\bm{q'\!'}}{2},t\big)_{b\alpha,b'\hspace*{-0.5mm}\alpha'}\\
  &{=}\!\sum_{\bm{R'\!'}}\!  O(\bm{R'\!'}\!\!,t)_{b\alpha,b'\hspace*{-0.5mm}\alpha'} e^{- i \bm{q}{\cdot} (\bm{R'\!'}\!{+}\bm{\tau}_b\!{-}\bm{\tau}_{b'\!})}
  \!\!\sum_{\bm{R}}\! e^{-i\bm{q'\!'}\!\cdot\bm{R}  }
e^{-i\bm{q'\!'}\!{\cdot}\! \frac{{\bm{\tau}_b}\!{-}\!\bm{R'\!'}\!{+}\bm{\tau}_{b'}\!\! }{2} }
  \\
  %\frac{(2\pi)^3}{\mathcal{V}} \delta(\bm{q'\!'})\\
  &=O(\bm{q},t)_{b\alpha,b'\hspace*{-0.5mm}\alpha'} \frac{(2\pi)^3}{\mathcal{V}} \delta(\bm{q'\!'}),
\end{split}
\raisetag{6mm}
\label{eq:diagonal_example}
\end{equation}
where we have used the exponential representation of the Dirac delta~(\ref{eq:Dirac_Delta_representation}) in the case where the wavevector $\bm{q''}\in \mathfrak{B}$.
It is now evident that inserting the translation-invariant operator~(\ref{eq:diagonal_example}) in the Wigner transform~(\ref{eq:Wigner_transform_Rec}) yields a distribution that does not depend on space.
This property has been exploited to transform Eq.~(\ref{eq:trace_Fourier}) into Eq.~(\ref{eq:local_energy_field}).
Specifically, following the procedure above \replaced{one can}{it is straightforward to} verify that the Wigner representation of the square root force-constant matrix $\sqrt{\tenscomp{G}}_{\bm{R}b\alpha,\bm{R'}\!b'\!\alpha'}$---which is translation-invariant, as a consequence of Eq.~(\ref{eq:translation_inv})---does not depend on space and is equal to the square root of the smooth dynamical matrix in reciprocal Bloch representation $\sqrt{\tenscomp{D}(\bm{q})}_{b\alpha,b'\!\alpha'}$. 
}\\

\paragraph{The Wigner matrix distribution is Hermitian.}
From the hermiticity of a generic observable, ${O}(\bm{q},\bm{q'})_{b\alpha,b'\alpha'}{=}{O}^*(\bm{q'},\bm{q})_{b'\alpha',b\alpha}$,  it is straightforward to prove that the Wigner matrix distributions defined by Eq.~(\ref{eq:Wigner_transform_Rec}) are Hermitian, \textit{i.e.} they satisfy ${\tenscomp{O}}(\bm{R},\bm{q})_{b\alpha,b'\alpha'}^* ={\tenscomp{O}}(\bm{R},\bm{q})_{b'\alpha',b\alpha}$. \\

\paragraph{Marginal of the Wigner matrix distribution.} % (fold)
\label{par:marginal_in_reciprocal_space_}
Integrating the Wigner matrix distribution over all the Bravais-lattice vectors yields the diagonal elements of the one-body density matrix in reciprocal space:
 \begin{equation}
\begin{split}
    \!\!\sum_{\bm{ R}}\! {\tenscomp{w}}(\bm{{R}},\bm{ q})_{b\alpha,b'\!\alpha'\!}&{=}
\frac{\mathcal{V}}{\!(2\pi)^3\!}\!\!\int\limits_{\mathfrak{B}}\!\!\!\!
\varrho(\bm{q}{+}\tfrac{\bm{q'\!'\!\!}}{2},\bm{q}{-}\tfrac{\bm{q'\!'\!\!}}{2})_{b\alpha,b'\!\alpha'\!}
\sum_{\bm{ R}} \!\! e^{+i\bm{q'\!'}\!\cdot \bm{ R}} d^3\!q'\!'\\
&=\frac{\mathcal{V}}{\!(2\pi)^3\!}\!\!\int\limits_{\mathfrak{B}}\!\!\!\!
\varrho(\bm{q}{+}\tfrac{\bm{q'\!'\!\!}}{2},\bm{q}{-}\tfrac{\bm{q'\!'\!\!}}{2})_{b\alpha,b'\!\alpha'\!}
\frac{\!(2\pi)^3\!}{\mathcal{V}}\delta(\bm{q'\!'}) d^3\!q'\!'\\
&=\varrho(\bm{q},\bm{q})_{b\alpha,b'\alpha'}
\;,
\end{split}
\raisetag{5mm}
\end{equation}
where we have used the exponential representation of the Dirac delta~(\ref{eq:Dirac_Delta_representation}) in the case where the wavevector $\bm{q''}{\in} \mathfrak{B}$.\\

\paragraph{Inner product.} 
As discussed in Eq.~(\ref{eq:trace_property_main}) of the main text, the trace of the product of two one-body operators in the Dirac framework can be rewritten as phase-space integral involving the Wigner representations of these operators.
Here we prove such a property. We start using the properties of the Dirac delta to recast the first line of Eq.~(\ref{eq:trace_property_main}) as follows
\begin{equation}
  \begin{split}
 &{\rm Tr}\big(\!\hat \rho(t)\hat A \!\big)  \\
&{=}\frac{\mathcal{V}^2}{\!(2\pi)^6\!}\!\sum_{\substack{b\alpha\\b'\!\alpha'} }\;
    \int\limits_{\!\!\mathfrak{B}\mathfrak{B}}\!\!\!\!\!\!\!\!\!\! \int
   \!{\varrho}\hspace*{-0.3mm}\big(\!{{\bm{q}{+}\tfrac{\bm{q'\!'}\!}{2}}}\!,\!\bm{q}{-}\tfrac{\bm{q'\!'}\!}{2}\!,t\big)_{\!b\alpha,b'\!\alpha'\!}
   {A}\!\big(\!\bm{q}{-}\tfrac{\bm{q'\!'}\!}{2},\!\bm{q}{+}\tfrac{\bm{q'\!'}\!}{2}\!\big)_{\!b'\!\alpha'\!,b\alpha}\!\!
 d^3\!q d^3\!q'\!'\\
 &{=}\frac{\mathcal{V}^2}{(2\pi)^6}
 \sum_{\substack{b\alpha\\b'\alpha'} }\;
    \int\!\!\!\!\!\!\!\! \int\limits_{\mathfrak{B}\mathfrak{B}\mathfrak{B}}\!\!\!\!\!\!\!\!\int
   \!\!{\varrho}\hspace*{-0.3mm}\big(\!{{\bm{q}{+}\tfrac{\bm{q'}\!}{2}}}\!,\!\bm{q}{-}\tfrac{\bm{q'}\!}{2}\!,t\big)_{\!b\alpha,b'\!\alpha'\!}
   \delta(\bm{q'}-\bm{q'\!'})\\
   &\hspace{2.7cm}{\times}{A}\!\big(\!\bm{q}{-}\tfrac{\bm{q'\!'}\!}{2},\!\bm{q}{+}\tfrac{\bm{q'\!'}\!}{2}\!\big)_{\!b'\!\alpha'\!,b\alpha}
 d^3\!q d^3\!q'd^3\!q'\!'  . 
\label{eq:trace_property_main_proof}
\raisetag{17mm}
  \end{split}
\end{equation}
Since in Eq.~(\ref{eq:trace_property_main_proof}) both $\bm{q'}$ and $\bm{q''}$ belong to the first Brillouin zone  $\mathfrak{B}$, we can rewrite the Dirac delta using the exponential representation~(\ref{eq:Dirac_Delta_representation}), obtaining:
\begin{equation}
  \begin{split}
 &{\rm Tr}\big(\!\hat \rho(t)\hat A \!\big) =\frac{\mathcal{V}^3}{(2\pi)^9}
 \sum_{\substack{b\alpha\\b'\alpha'} }\;
    \int\!\!\!\!\!\!\!\! \int\limits_{\mathfrak{B}\mathfrak{B}\mathfrak{B}}\!\!\!\!\!\!\!\!\int
   \!\!{\varrho}\hspace*{-0.3mm}\big(\!{{\bm{q}{+}\tfrac{\bm{q'}\!}{2}}}\!,\!\bm{q}{-}\tfrac{\bm{q'}\!}{2}\!,t\big)_{\!b\alpha,b'\!\alpha'\!}\\
     &\hspace{1.6cm} {\times}\sum_{\bm{R}}e^{i(\bm{q'}-\bm{q'\!'})\cdot \bm{R}}
{A}\!\big(\!\bm{q}{-}\tfrac{\bm{q'\!'}\!}{2},\!\bm{q}{+}\tfrac{\bm{q'\!'}\!}{2}\!\big)_{\!b'\!\alpha'\!,b\alpha}
 d^3\!q d^3\!q'd^3\!q'\!'   .
\label{eq:trace_property_main_proof2}
\raisetag{17mm}
  \end{split}
\end{equation}
Now we exchange the order of sum and integrals, \replaced{thus we can}{and it is easy to} recognize that the integrals over $\bm{q'}$ and $\bm{q'\!'}$ yield the phase-space representation of the one-body density matrix and of the one-body operator ${A}(\bm{q},\bm{q'})_{b\alpha,b'\alpha'}$, \textit{i.e.} Eq.~(\ref{eq:trace_property_main}).

\section{Accuracy of the perturbative treatment of anharmonicity in La$_2$Zr$_2$O$_7$ and CsPbBr$_3$} % (fold)
\label{sec:validation_of_the_perturbative_treatment_of_anharmonicity}
\added{As discussed in Sec.~\ref{sub:steady_state_solution_of_the_wbte}, the standard perturbative treatment of anharmonicity employed in this work considers the vibrational frequencies as temperature independent (bare phonon frequencies), and accounts for the effect on temperature via the Bose-Einstein distributions appearing in the scattering operator~(\ref{eq:formula_scattering}).}

\added{To check how accurate it is to consider the bare phonon frequencies (and related atomic displacement patterns defined in Eq.~(\ref{eq:diagonalization_dynamical_matrix})) as a representation of the actual vibrational frequencies (and related atomic displacements) over the entire temperature range analyzed, we
use them to compute the specific heat and compare our calculations with experiments. 
Specifically, the specific heat at constant volume (since thermal expansion is neglected in the standard perturbative treatment we adopted) is computed as
$C_V^{\rm Th}(T){=}\frac{1}{\mathcal{D}(2\pi)^3}\int\tfrac{\hbar^{2}\omega^2(\bm{q})_s}{k_B{T}^2}\overline{\tenscomp{N}}(\bm{q})_s\big[\overline{\tenscomp{N}}(\bm{q})_s{+}1\big]d^3q$, where $\mathcal{D}$ is the volumetric mass density and $\overline{\tenscomp{N}}(\bm{q})_s$ the Bose-Einstein distribution at temperature $T$.
}
\begin{figure}[b]
  \centering
  \includegraphics[width=\columnwidth]{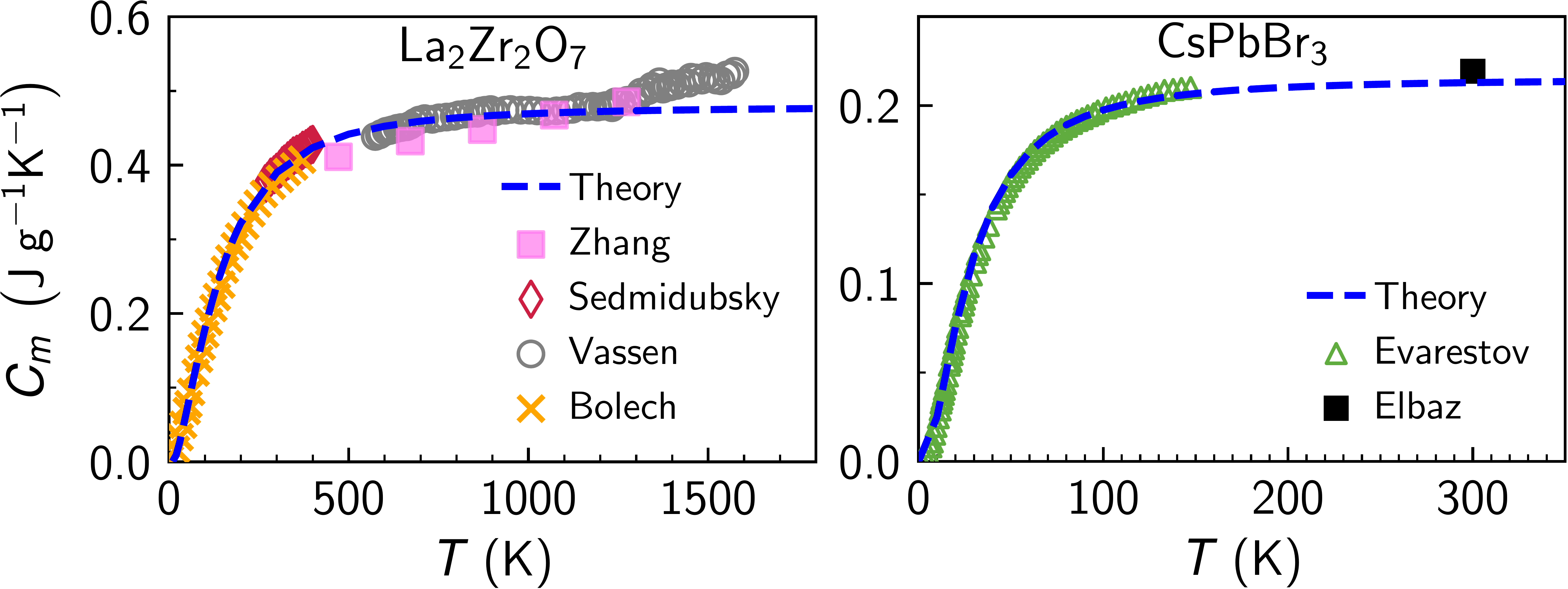}
  \caption{
  \added{\textbf{Specific heat of La$_2$Zr$_2$O$_7$ and CsPbBr$_3$.} 
  Dashed-blue lines are theoretical calculations at constant volume and using the bare phonon frequencies.
  Scatter points are measurements performed at constant pressure. For La$_2$Zr$_2$O$_7$ these are taken from Vassen~\cite{Vassen2000}, Bolech~\cite{bolech1997heat}, Zhang~\cite{zhang2020microstructure} and Sedmidubsk{\`y} \cite{sedmidubsky2005high}.   
   For CsPbBr$_3$ these are taken from Evarestov \cite{evarestov2020first} and Elbaz \cite{elbaz2017phonon}.}}
  \label{fig:spec_heat_P}
\end{figure}
\begin{figure*}
\begin{overpic}[width=0.49\textwidth]{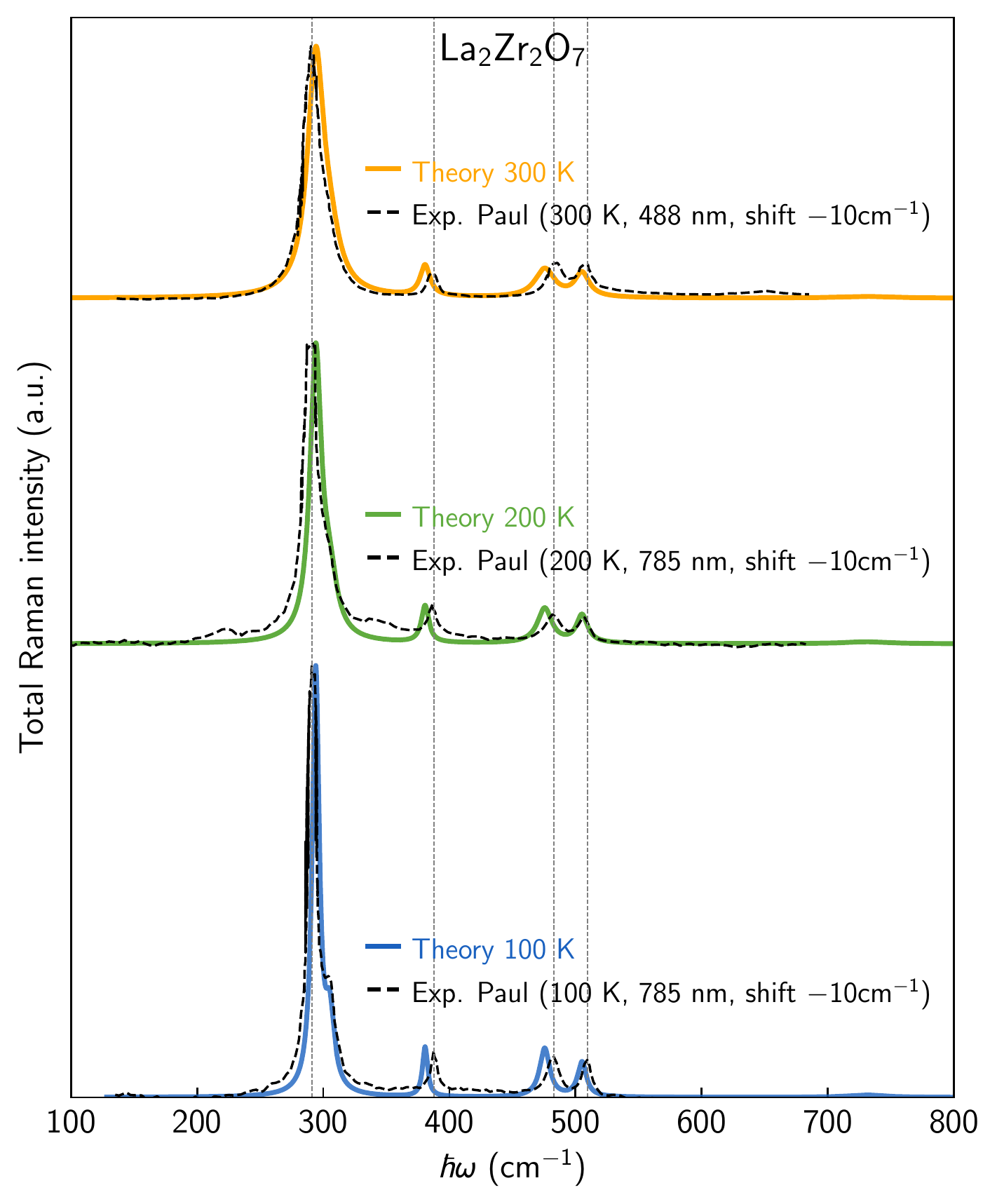}
  \put(1,95){\textbf{a)}}
 \end{overpic}
 \begin{overpic}[width=0.478\textwidth]{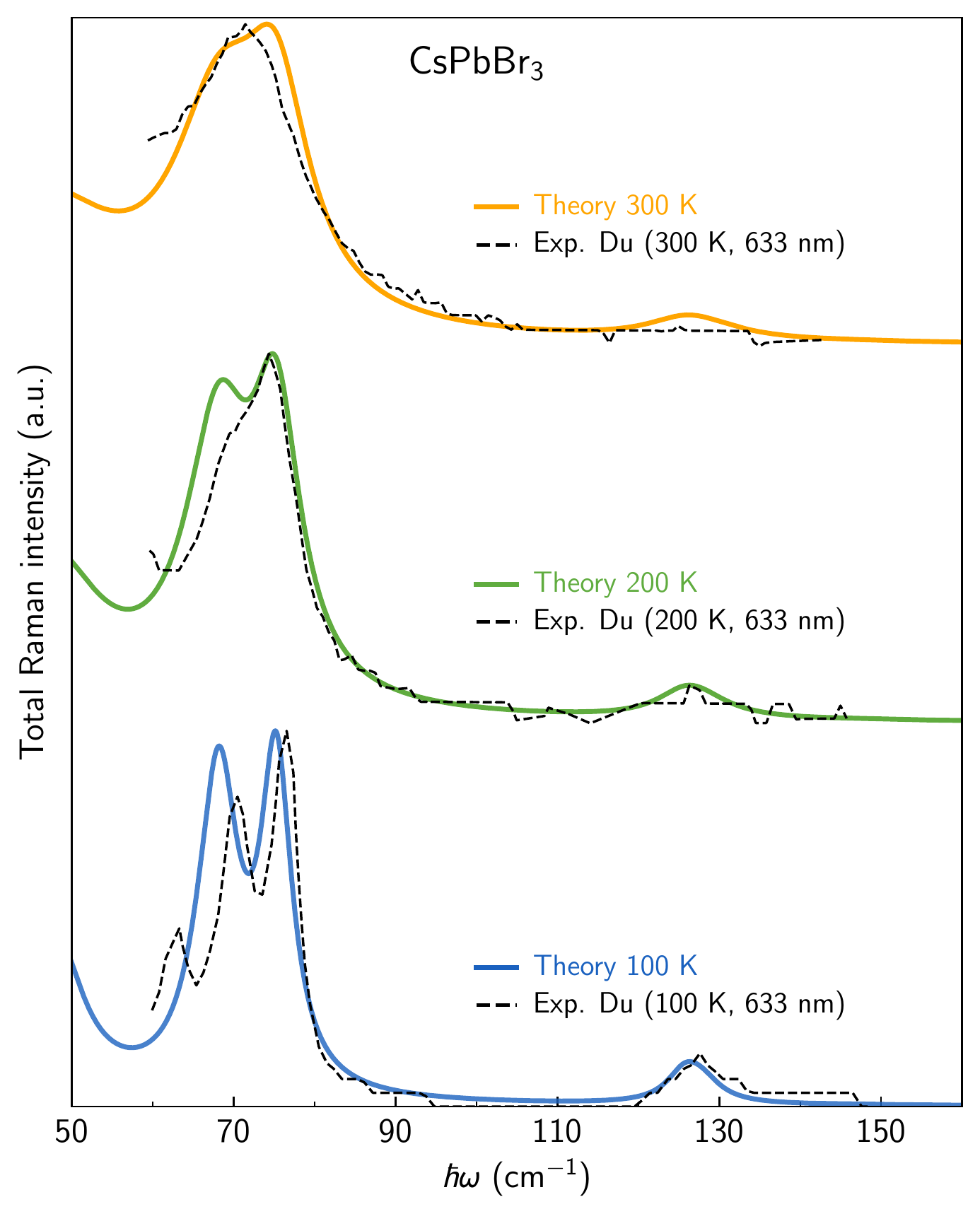}
  \put(1,95){\textbf{b)}}
 \end{overpic}
  \caption{\added{
  \textbf{Raman spectra of La$_2$Zr$_2$O$_7$ and CsPbBr$_3$.}
  Solid lines are temperature-dependent simulations performed relying on the same perturbative treatment of anharmonicity that we employed in out thermal-conductivity calculations (see text for details). Dashed-black lines are experiments, taken from Paul \textit{et al.}~\cite{paul2016structural} for La$_2$Zr$_2$O$_7$, and from Du \textit{et al.}~\cite{du2019unveiling} for CsPbBr$_3$. 
The experimental measurements on La$_2$Zr$_2$O$_7$ in panel \textbf{a)} have been shifted rigidly of 10 cm$^{-1}$ towards left at all temperatures, to ease the comparison with our calculations.
The vertical lines in panel \textbf{a)} highlight that in La$_2$Zr$_2$O$_7$ the experimental Raman peaks  do not have an appreciable temperature-dependent shift, thus the aforementioned rigid shift is not compensating for the temperature-dependent renormalization of the vibrational frequencies not accounted for in our calculations (see text for details). Overall, the temperature-dependent broadening of the experimental Raman peaks is in remarkably good agreement with our calculations, thus the perturbative treatment of anharmonicity employed is accurate enough for our illustrative purposes.}
  }
  \label{fig:raman}
\end{figure*}
Fig.~\ref{fig:spec_heat_P} shows that the vibrational frequencies used in our calculations yield for La$_2$Zr$_2$O$_7$ a specific heat in good agreement with experiments up to 1300 K. Above 1300 K, experiments start deviating from theory, suggesting that at these very high temperatures a more refined treatment of anharmonicity is needed. For this reason, calculations in Sec.~\ref{sec:numerical_results} have been limited to the upper-bound temperature of 1300 K.
For CsPbBr$_3$, the theoretical and experimental specific heats agree reasonably well over the temperature range analyzed (which corresponds to the temperature interval in which the orthorhombic phase is stable).

\added{
To assess the degree of accuracy of our perturbative treatment of anharmonicity in La$_2$Zr$_2$O$_7$ and CsPbBr$_3$, 
we use the theoretical anharmonic linewidths to predict the temperature dependence of the Raman spectrum of these complex crystals, and compare our predictions with experiments \cite{paul2016structural,du2019unveiling}. 
%We adopted this procedure because La$_2$Zr$_2$O$_7$ and CsPbBr$_3$ feature many Raman-active and closely-spaced phonon bands~(see Fig.~\ref{fig:cond_mech_Zirc} and Ref.~\cite{simoncelli2019unified}), thus in these materials the alternative extracting the frequencies and linewidths of individual vibrational modes from the Raman spectrum is a challenging task that requires to known
Specifically, the Raman tensor of La$_2$Zr$_2$O$_7$ and CsPbBr$_3$ is computed following the approach detailed in Ref.~\cite{skelton2017lattice}. The Raman tensor and bare phonon frequencies at $\bm{q}=\bm{0}$ are used to compute the modal Raman intensity $I_s$ for La$_2$Zr$_2$O$_7$ and CsPbBr$_3$ (we use the formula for the powder-averaged intensity discussed in Eq.~(5) of Ref.~\cite{PhysRevB.71.214307}, in such formula we consider laser frequencies equal to those used in the experiments \cite{paul2016structural,du2019unveiling} against which we compare our calculations).
Finally, from the the Raman intensity $I_s$ the Raman spectrum is computed as \cite{PhysRevB.94.155435}:
$I(\omega) \propto \sum_{s} I_s \frac{\frac{1}{2}(\Gamma(\bm{0})_s+\Gamma_{\rm ins})}{(\omega-\omega(\bm{0})_s)^2+\frac{1}{4}(\Gamma(\bm{0})_s{+}\Gamma_{\rm ins})^2}$, where $\omega(\bm{0})_s$ and $\Gamma(\bm{0})_s$ are the bare phonon frequencies and anharmonic linewidths at $\bm{q}{=}\bm{0}$, respectively, and the linewidth $\Gamma_{\rm ins}{=}2$ cm$^{-1}$ accounts for the instrumental broadening (see e.g. Ref.~\cite{PhysRevB.66.115411}). Therefore, the Raman spectrum is simulated under the same approximations employed in our thermal-conductivity calculations, \textit{i.e.} it does not account for the temperature-dependent shift of frequencies, and considers anharmonicity up to the lowest third order.
Fig.~\ref{fig:raman} shows that the simulated Raman spectrum agrees reasonably well with experiments, both for La$_2$Zr$_2$O$_7$ \cite{paul2016structural} and CsPbBr$_3$ \cite{du2019unveiling} over the temperature range from 100 to 300 K. 
We note that in Fig.~\ref{fig:raman}\textbf{a)} the experimental measurements in La$_2$Zr$_2$O$_7$ are shifted rigidly of 10 cm$^{-1}$ at all temperatures, to ease the comparison with our calculations. The vertical lines in Fig.~\ref{fig:raman}\textbf{a)} highlight that experiments in La$_2$Zr$_2$O$_7$ do not show an appreciable temperature-dependent shift of the Raman peaks in the temperature range analyzed, thus the shift we applied is not compensating for the temperature-dependent renormalization of the vibrational frequencies that is not accounted for in our calculations. 
Assessing the origin of such a systematic shift goes beyond the scope of the present study, since Fig.~\ref{fig:spec_heat_P} shows that the bare phonon frequencies calculated for La$_2$Zr$_2$O$_7$ yield a specific heat in satisfying agreement with experiments and are thus accurate enough for our scope.  
%Such a rigid shift might be a systematic error deriving from the approximations involved in the (PBEsol) exchange-correlation functional we employed to simulate La$_2$Zr$_2$O$_7$ goes beyond the scope of the present study.
Although Raman experiments in La$_2$Zr$_2$O$_7$ do not cover the entire temperature range over which   the thermal conductivity is computed, we note that in the range from 100 to 300 K the temperature-dependent broadening of the experimental Raman peaks is in remarkably good agreement with our calculations, and at 300 K the coherences conductivity in La$_2$Zr$_2$O$_7$ is a non-negligible fraction of the total thermal conductivity. These considerations suggest that the perturbative description of anharmonicity employed is accurate enough to describe the main features of the temperature-thermal conductivity curve of La$_2$Zr$_2$O$_7$.
Finally, we note that often in the literature the inverse of the analysis discussed here is performed, \textit{i.e.} the experimental values for the linewidths are extracted, via a fitting procedure, from the measured Raman spectrum and then compared with their theoretical counterparts. 
Such a procedure is error-prone in complex crystals such as La$_2$Zr$_2$O$_7$ and CsPbBr$_3$. In fact, these complex crystals feature many Raman-active vibrational modes that are very close in frequency and have very different intensity, therefore in these cases very often the linewidths of individual vibrational modes cannot be resolved from the Raman measurements. For these reasons such a procedure has not been followed here, and instead the simulated Raman spectra has been compared directly with experiments.}

\added{
In summary, for both La$_2$Zr$_2$O$_7$ and CsPbBr$_3$, the bare phonon frequencies computed yield a specific heat in good agreement with that measured in experiments. Moreover, the anharmonic linewidths  computed yield a temperature-dependent broadening of the simulated Raman peaks in good agreement with experiments over a temperature range in which the coherences conductivity is not negligible. This shows that the perturbative treatment for anharmonicity adopted is accurate enough for the illustrative scope of the present study.
}

\section{Effects of the phase convention on the thermal conductivity} % (fold)
\label{sec:effects_of_the_phase_convention_on_the_conductivity}
In this section we discuss the effects of the phase convention adopted for the Fourier transform on the thermal conductivity, comparing the thermal conductivity obtained using the velocity operator in the smooth phase convention~(\ref{eq:vel_op}) and that obtained using the velocity operator in the step-like phase convention~(\ref{eq:relation_velocity_operator}).
%the operator matrix element appearing in the Wigner transform~(\ref{eq:Wigner_transform_Rec}).
This comparison is reported only for illustrative purposes. We remark that the velocity operator in the smooth phase convention~(\ref{eq:vel_op}) is the only one yielding a size-consistent conductivity --- thus the convention to be used --- and the results reported in Fig. \ref{fig:k_vs_T_Zirconate} and Fig.~\ref{fig:k_vs_T} have been computed using the smooth phases. We show in Fig.~\ref{fig:k_vs_T_phase_zirco}  and Fig.~\ref{fig:k_vs_T_phase} how the predictions of the conductivity of La$_2$Zr$_2$O$_7$ and CsPbBr$_3$, respectively, change if the size-inconsistent step-like phase convention is used in place of the correct (size-consistent) smooth phase convention. We note that in both La$_2$Zr$_2$O$_7$ and CsPbBr$_3$ the populations' terms of the conductivity $\kappa_{\rm P}$ is not affected by the phase convention adopted, since the diagonal elements of the velocity operator do not depend on the phase convention (see Eq.~(\ref{eq:relation_velocity_operator})). 
The differences in the conductivities obtained in the two phase conventions are increasingly more relevant as temperature increases and arise exclusively from the coherences' conductivity $\kappa_{\rm C}$, with the step-like phase convention yielding a larger $\kappa_{\rm C}$.
\begin{figure}[h]
  \centering
  \includegraphics[width=\WidthFigure]{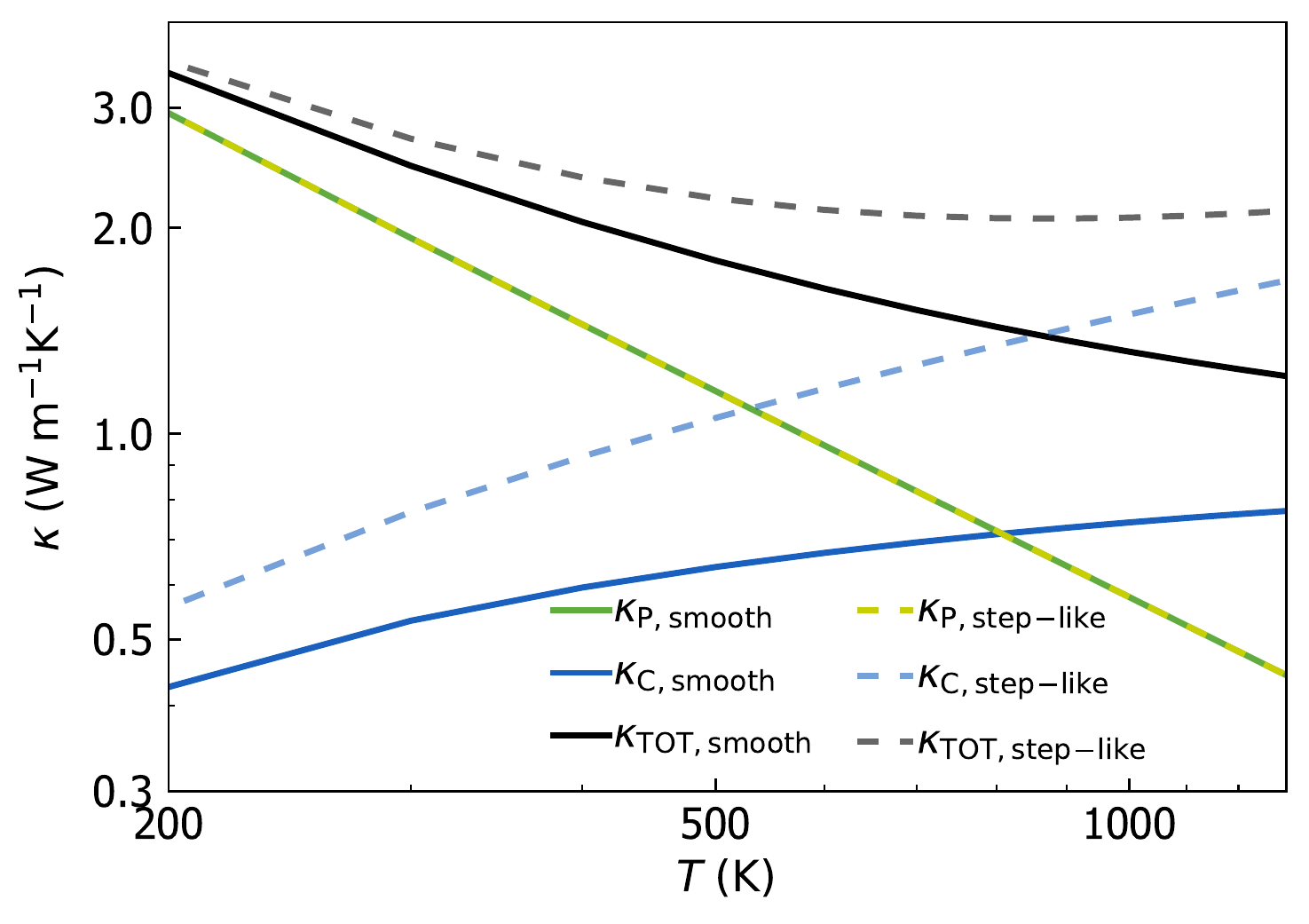}
  \caption{{\textbf{Effect of the phase convention on the thermal conductivity of La$_2$Zr$_2$O$_7$.} 
  Solid lines are the conductivities computed using the correct size-consistent smooth phase convention, dashed lines are the conductivities computed using the size-inconsistent step-like phase convention.
   The populations' conductivities are green ($\kappa_{\rm P}$), and they are equal in the two phase conventions (see text).
    The  coherences'  conductivities ($\kappa_{\rm C}$) are blue.
     The total conductivity (black) is given by the sum {$\kappa_{\rm TOT}=\kappa_{\rm P}+\kappa_{\rm C}$}.
     Using the step-like phase convention yields a total conductivity larger than that obtained using the smooth phase convention. We recall that the populations and coherences conductivity tensors of La$_2$Zr$_2$O$_7$ are proportional to the identity (see Sec~\ref{sub:zirconate_start}), thus populations and coherences conductivities are represented here as scalars. 
      }
  }
  \label{fig:k_vs_T_phase_zirco}
\end{figure}
\begin{figure}[H]%67 words
  \centering
  \includegraphics[width=\WidthFigure]{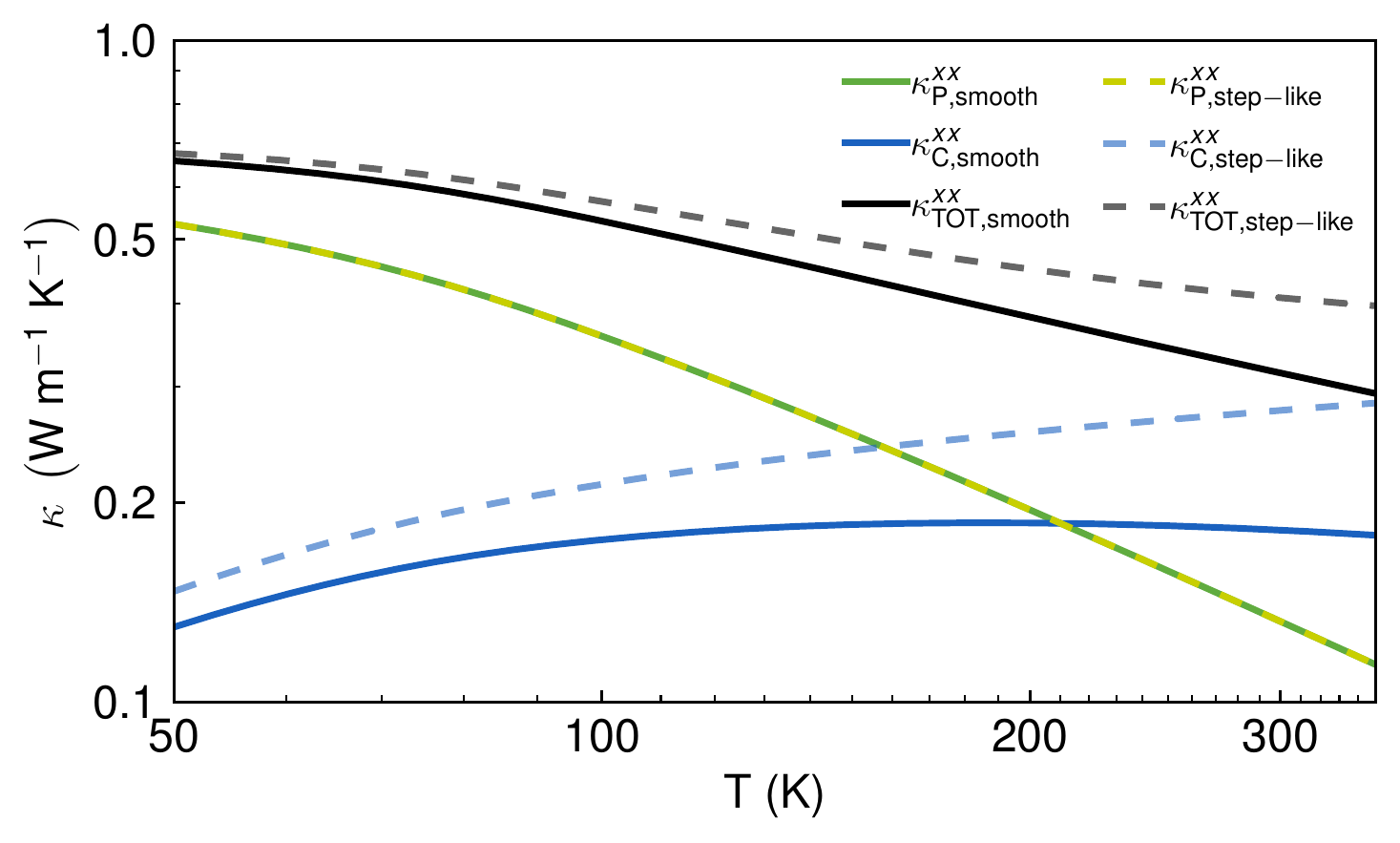}
  \caption{\textbf{Effect of the phase convention on the thermal conductivity of CsPbBr$_3$.} 
  Solid lines are the conductivities computed using the correct size-consistent smooth phase convention, dashed lines are the conductivities computed using the size-inconsistent step-like phase convention (also reported in Fig.~1 of Ref.   \cite{simoncelli2019unified}).
   The populations' conductivities are green ($\kappa^{xx}_{\rm P}$), and they are equal in the two phase conventions (see text).
    The  coherences'  conductivities ($\kappa^{xx}_{\rm C}$) are blue.
     The total conductivity (black) is given by the sum {$\kappa^{xx}_{\rm TOT}=\kappa^{xx}_{\rm P}+\kappa^{xx}_{\rm C}$}.
     Using the step-like phase convention yields a total conductivity larger than that obtained using the smooth phase convention. 
  }
  \label{fig:k_vs_T_phase}
\end{figure}

\section{Ioffe-Regel limit in space and center of the non-sharp particle-wave crossover for phonons} % (fold)
\label{sec:ioffe_regel_limit_in_space_and_center_of_the_non_sharp_particle_wave_crossover}
{
In this Appendix we report the details of the derivation of Eq.~(\ref{eq:Ioffe_Regel_th}). We start by showing how to obtain the single-phonon contribution to the particle-like and wave-like (coherences) conductivities appearing in such an equation.
% ($\mathcal{K}^{\rm avg}_P(\bm{q})_s$ and $\mathcal{K}^{\rm avg}_C(\bm{q})_s$, respectively).
The expression for the particle-like contribution $\mathcal{K}^{\rm avg}_P(\bm{q})_s$ is trivially obtained from the integrand of the populations (Peierls-Boltzmann) conductivity in the SMA~(\ref{eq:SMA_k}) (we recall that the SMA is accurate for the ultralow-thermal-conductivity materials in focus here),
\begin{equation}
\begin{split}
    \mathcal{K}^{\rm avg}_P(\bm{q})_s&{=}C(\bm{q})_s\bigg(\sum_\alpha \frac{|\tenscomp{v}^\alpha(\bm{q})_{s,s}|^2}{3}\bigg)\frac{1}{\Gamma(\bm{q})_s}\\
  &\added{{=}C(\bm{q})_s\sum_\alpha \bar{\tenscomp{v}}^{\rm avg}(\bm{q})_{s,s}\bar{\tenscomp{v}}^{\rm avg}(\bm{q})_{s,s}\Gamma(\bm{q})_s},
\end{split}
  \label{eq:populations_density}
\end{equation}
\added{where we have defined the spatially-averaged group velocity $\bar{\tenscomp{v}}^{\rm avg}(\bm{q})_{s,s}=\sqrt{\frac{1}{3}\sum_{\alpha=1}^3 |\tenscomp{v}^\alpha(\bm{q})_{s,s}|^2}$.}
The expression for the single-phonon contribution to the wave-like conductivity, $\mathcal{K}^{\rm avg\!}_C(\bm{q})_s$, is obtained requiring that in the coherences' coupling between two phonons $(\bm{q})_{s}$ and $(\bm{q})_{s'}$ (which is quantified by the normalized trace of the integrand of the coherence term in Eq.~(\ref{eq:thermal_conductivity_final_sum})), each phonon contributes to the coupling with a weight equal to the relative specific heat (e.g. for phonon $(\bm{q})_{s}$ the weight is $\tfrac{C(\bm{q})_s}{C(\bm{q})_s+C(\bm{q})_{s'}}$, and correspondingly for phonon $(\bm{q})_{s'}$ the weight is $\tfrac{C(\bm{q})_{s'}}{C(\bm{q})_s+C(\bm{q})_{s'}}$.
In practice, denoting with $k(\bm{q})^{\alpha\beta}_{s,s'}$ the integrand of the coherences term in Eq.~(\ref{eq:thermal_conductivity_final_sum}), and noting that for $\alpha{=}\beta$ this term is symmetric in the mode indexes, $k(\bm{q})^{\alpha\alpha}_{s,s'}=k(\bm{q})^{\alpha\alpha}_{s',s}$, we can rewrite the average trace of the coherences conductivity tensor as
\begin{equation}
\begin{split}
    \kappa^{\rm avg}_C &= \frac{1}{3}\sum_{\alpha}\frac{1}{(2\pi)^3}\sum_{s, s'{\neq} s}\int_{\mathfrak{B}} k(\bm{q})^{\alpha\alpha}_{s,s'} d^3q \\
    %&=\frac{1}{3}\sum_{\alpha}\frac{1}{\mathcal{V}N_c}\sum_{\bm{q}}\sum_{s, s'{\neq} s} \frac{C(\bm{q})_s{+}C(\bm{q})_{s'}}{C(\bm{q})_s{+}C(\bm{q})_{s'}}k(\bm{q})^{\alpha\alpha}_{s,s'}\\
%    &=\frac{1}{(2\pi)^3}\!\!\sum_{s}\!\int_{\mathfrak{B}}
%    \!\!\Bigg[\!\sum_{s'{\neq} s}\!\frac{C(\bm{q})_s}{C(\bm{q})_s{+}C(\bm{q})_{s'\!\!}}
%  \!  \bigg(\!\frac{1}{3}\!\sum_{\alpha}\! k(\bm{q})^{\alpha\alpha}_{s,s'}\!\!\bigg)\!\!\Bigg]d^3q\\
%  &+ \frac{1}{(2\pi)^3}\!\!\sum_{s'}\!\int_{\mathfrak{B}}
%    \!\!\Bigg[\!\sum_{s{\neq} s'}\!\frac{C(\bm{q})_{s'}}{C(\bm{q})_s{+}C(\bm{q})_{s'\!\!}}
%  \!  \bigg(\!\frac{1}{3}\!\sum_{\alpha}\! k(\bm{q})^{\alpha\alpha}_{s,s'}\!\!\bigg)\!\!\Bigg]d^3q\\
      &{=} \frac{1}{(2\pi)^3}\!\sum_s\!\int_{\mathfrak{B}}
    \!\!\Bigg[\!\sum_{s'{\neq} s}\!\frac{C(\bm{q})_s}{C(\bm{q})_s{+}C(\bm{q})_{s'\!\!}}
  \!  \bigg(\!\frac{1}{3}\!\sum_{\alpha}\! 2k(\bm{q})^{\alpha\alpha}_{s,s'}\!\!\bigg)\!\!\Bigg]d^3q.%\\
%&= \frac{1}{\mathcal{V}N_c}\sum_{\bm{q}} \sum_{s}
%  \mathcal{K}^{\rm avg}_C(\bm{q})_s,
\end{split}
\raisetag{18mm}
\label{eq:contri_spec_heat}
\end{equation}
The term in square brackets in Eq.~(\ref{eq:contri_spec_heat}) has the following explicit form
  \begin{equation}
  \begin{split}
&\mathcal{K}^{\rm avg\!}_C(\bm{q})_s{=} \!\!
\sum_{s'\neq s}\!\!
  \frac{C(\bm{q})_s}{C(\bm{q})_{s}{+}C(\bm{q})_{s'\!}\!}
\frac{\omega(\bm{q})_{s}{+}\omega(\bm{q})_{s'}\!}{2}\!\!
\left[\!\frac{C(\bm{q})_s}{\omega(\bm{q})_s}{+}\frac{C(\bm{q})_{s'\!}}{\omega(\bm{q})_{s'\!\!}}\!\right]\!\\
&{\times}\!\!
\left[\frac{1}{3}\!\sum_\alpha|{\tenscomp{v}^\alpha}(\bm{q})_{s,s'\!}|^2\!\right]\!\!\frac{\tfrac{1}{2}\big[\Gamma(\bm{q})_{s}{+}\Gamma(\bm{q})_{s'}\big]}{[\omega(\bm{q})_{s'}{-}\omega(\bm{q})_{s}]^2{+}\tfrac{1}{4}[\Gamma(\bm{q})_{s}{+}\Gamma(\bm{q})_{s'}]^2},
  \end{split}
\label{eq:coherence_density}
  \end{equation}
and represents the contribution of the phonon $(\bm{q})_s$ to the wave-like (coherences) conductivity. 

We now report the calculation --- with order-of-magnitude accuracy --- that justifies the 
first relation appearing in Eq.~(\ref{eq:Ioffe_Regel_th}), $\frac{\mathcal{K}^{\rm avg}_C(\bm{q})_s}{\mathcal{K}^{\rm avg}_P(\bm{q})_s}\sim \frac{\Gamma(\bm{q})_s}{\Delta\omega_{\rm avg}}$.
Rewriting such a relation using Eq.~(\ref{eq:populations_density}) and Eq.~(\ref{eq:coherence_density}), and rearranging terms, we obtain
\begin{equation}
  \begin{split}
&\frac{1}{\Delta\omega_{\rm avg}}
\sim 
\sum_{s'\neq s}\!\!
\frac{\omega(\bm{q})_{s}{+}\omega(\bm{q})_{s'}\!}{C(\bm{q})_{s}{+}C(\bm{q})_{s'\!}\!} \frac{1}{2}\!\!
\left[\!\frac{C(\bm{q})_s}{\omega(\bm{q})_s}{+}\frac{C(\bm{q})_{s'\!}}{\omega(\bm{q})_{s'\!\!}}\!\right]\!\\
&{\times}
\left[\frac{\!\sum_\alpha|{\tenscomp{v}^\alpha}(\bm{q})_{s,s'\!}|^2\!}{
  \sum_\alpha |\tenscomp{v}^\alpha(\bm{q})_{s,s}|^2
}\right]\!\!\frac{\tfrac{1}{2}\big[\Gamma(\bm{q})_{s}{+}\Gamma(\bm{q})_{s'}\big]}{[\omega(\bm{q})_{s'}{-}\omega(\bm{q})_{s}]^2{+}\tfrac{1}{4}[\Gamma(\bm{q})_{s}{+}\Gamma(\bm{q})_{s'}]^2}\;.
  \end{split}
  \raisetag{20mm}
\label{eq:expression_ord_magnitude}
  \end{equation}
We now proceed with performing an order-of-magnitude analysis of Eq.~(\ref{eq:expression_ord_magnitude}).
First, we note that the Lorentzian in Eq.~(\ref{eq:expression_ord_magnitude}) is peaked around $\omega(\bm{q})_{s'}{=}\omega(\bm{q})_{s}$, implying the following order-of-magnitude estimate for the term involving frequencies and specific heats: $\frac{\omega(\bm{q})_{s}{+}\omega(\bm{q})_{s'}}{C(\bm{q})_s{+}C(\bm{q})_{s'}}\!
\frac{1}{2}\!\left[\frac{C(\bm{q})_s}{\omega(\bm{q})_s}{+}\frac{C(\bm{q})_{s'}}{\omega(\bm{q})_{s'}}\right]{\sim} 1$.
Second, we rely on the considerations reported in Sec.~\ref{sec:Particle_wave_crossover_phonons} to approximate the ratio between the velocity-operator elements to one, $\frac{\!\sum_\alpha|{\tenscomp{v}^\alpha}(\bm{q})_{s,s'\!}|^2\!}{
  \sum_\alpha |\tenscomp{v}^\alpha(\bm{q})_{s,s}|^2
}\sim 1$.
These considerations imply that the relation reported in Eq.~(\ref{eq:expression_ord_magnitude}), with order-of-magnitude accuracy, can be simplified to:
\begin{equation}
  \begin{split}
&\frac{1}{\Delta\omega_{\rm avg}}
%\frac{2\pi}{{\sqrt[3]{N_{\rm at}}}}
{\sim}\!\! 
\sum_{s'\neq s}\!\!
\frac{\tfrac{1}{2}\big[\Gamma(\bm{q})_{s}{+}\Gamma(\bm{q})_{s'}\big]}{[\omega(\bm{q})_{s'}{-}\omega(\bm{q})_{s}]^2{+}\tfrac{1}{4}[\Gamma(\bm{q})_{s}{+}\Gamma(\bm{q})_{s'}]^2}.
  \end{split}
  \raisetag{1mm}
\label{eq:expression_ord_magnitude_simp}
\end{equation}

Now we distinguish what happens in Eq.~(\ref{eq:expression_ord_magnitude_simp}) in the two regimes of simple and complex crystals.
In the simple-crystal regime,  where the typical phonon interband spacings are much larger than the linewidths ($\Delta\omega_{\rm avg}\gg \Gamma(\bm{q})_s\;\forall \bm{q},s$), Eq.~(\ref{eq:expression_ord_magnitude_simp}) assumes values much smaller than one on both sides and is trivially verified.
In the complex-crystal regime,  where the typical phonon interband spacings are comparable or smaller than the linewidths ($\Delta\omega_{\rm avg}\lesssim \Gamma(\bm{q})_s\;\forall \bm{q},s$),
we need to perform two approximations to proceed with our order-of-magnitude analysis:
(i) we consider the $3N_{\rm at}$ phonon bands as uniformly distributed over the frequency range $(0,\omega_{\rm max}]$, (\textit{i.e.} for a generic mode index $s'$, we have  
$\omega(\bm{q})_{s'}{\simeq}n\cdot \Delta \omega_{\rm avg}$, where $n\in [1,\dots,3 N_{\rm at}]$ \footnote{Strictly speaking, the sum $\Sigma_{n=0}^{3 {N_{\rm at}}{-}1}$ should not contain the term $s= s'$. In the limit of a large number of bands, including in the sum also the term $s{=}s'$ does not alter the order-of-magnitude analysis performed here, and we will see soon that allows to approximate the sum with an integral. For these reasons the term $s{=}s'$ is included approximatively in the sum.});
(ii) we approximate the average linewidth as 
$\frac{\Gamma(\bm{q})_{s}{+}\Gamma(\bm{q})_{s'}}{2}{=}\Gamma(\bm{q})_{s}$.
These approximations allow us to rewrite the sum in Eq.~(\ref{eq:expression_ord_magnitude_simp}) as an integral:
$\frac{1}{\Delta\omega_{\rm avg}}
    {\sim}
  %\sum_{n=1}^{3\cdot {N_{\rm at}}}
%\frac{\Gamma(\bm{q})_{s}}{[n{\cdot} \Delta\omega_{\rm avg}{-}\omega(\bm{q})_{s}]^2+\Gamma^2(\bm{q})_{ s}}
%{\simeq}
\frac{1}{\Delta\omega_{\rm avg}} \int_{0}^{\omega_{\rm max}}
\frac{\Gamma(\bm{q})_{s}}{[\omega{-}\omega(\bm{q})_{s}]^2+\Gamma^2(\bm{q})_{s}}d\omega.$
Now we note that the integral of the Lorentzian distribution yields a number smaller than $\pi$ and of the order of one.
In fact, such an integral would yield $\pi$ if the integration domain was running from $-\infty$ to $+\infty$, in the case considered here the linewidths are a fraction of the finite integration domain (see e.g. Fig.~\ref{fig:cond_mech_Zirc}\textbf{b}) and thus a considerable portion of the Lorentzian is always integrated, yielding a value of the order of unity.
These considerations imply that the two members in Eq.~(\ref{eq:expression_ord_magnitude_simp}) are approximately equal also in the case of a complex crystal, proving with order-of-magnitude accuracy Eq.~(\ref{eq:Ioffe_Regel_th}). 
}

\section{Computational details }
\label{sub:computational_details_and_code_availability}
\subsection{La$_2$Zr$_2$O$_7$ }
\label{sub:la__2_zr__2_o__7_}
The crystal structure of La$_2$Zr$_2$O$_7$ is cubic with spacegroup Fd$\bar{3}$m [227] and has been taken from the materials project database  \cite{Jain2013} (id = mp-4974    \cite{osti_1208525}).
The equilibrium crystal structure has been computed using density functional theory (DFT), specifically performing a ``vc-relax''  calculation with the Quantum \textsc{espresso} \cite{giannozzi2017advanced} software. The following functionals have been tested: LDA (with ultrasoft pseudopotentials from the psilibrary \cite{dal2014pseudopotentials}), PBEsol (with GBRV   \cite{garrity2014pseudopotentials} pseudopotentials), and PBE with Grimme D2 correction    \cite{grimme2006semiempirical} (with pseudopotentials from the SSSP precision library   \cite{lejaeghere2016reproducibility,prandini2018precision}).
A summary of the computational parameters and equilibrium lattice parameter is reported in Table~\ref{tab:cell_param_La2Zr2O7}.
\begin{table}[b]
 \caption{Lattice parameters of La$_2$Zr$_2$O$_7$ (cubic) obtained by first-principles simulations or experiments   \cite{subramanian1983oxide}. A 4{$\times$}4{$\times$}4  mesh of k-points with a (1{$\times$}1{$\times$}1) shift is used.}
  \label{tab:cell_param_La2Zr2O7}
  \centering
  \begin{tabular}{l|c|c}
  \hline

  \hline
  functional & cutoff [Ry] (dual)  & lattice parameter [\AA] \\
  \hline
  LDA    &    80 (8)        &  10.6594462  \\
  PBE+D2   &    50 (8)        &  10.6591719  \\
  PBEsol   &    50 (8)        &  10.7391572  \\
Experiment   \cite{subramanian1983oxide} &            &    10.805       \\
  \hline

  \hline
  \end{tabular}
\end{table}
The PBEsol functional has been selected on the basis of the best agreement between first-principles and experimental lattice parameters   \cite{subramanian1983oxide} (see Table~\ref{tab:cell_param_La2Zr2O7}).
Second-order force constants have been computed on a $4{\times}4{\times}4$ mesh using density-functional perturbation theory   \cite{RevModPhys.73.515} and accounting for the non-analytic term correction due to the dielectric tensor and Born effective charges. 
Third-order force constants have been computed using the finite-difference method implemented in ShengBTE   \cite{li2014shengbte}, on a $2{\times}2{\times}2$ supercell, integrating the Brillouin zone with a $2{\times}2{\times}2$ Monkhorst-Pack mesh, and considering interactions up to 7.8 \AA.
%(which corresponds to the 12$^{\rm th}$ nearest neighbor parameter by Ref. \cite{luo2020vibrational} for this material).
Then, harmonic force constants have been converted from Quantum \textsc{espresso} format to \texttt{hdf5}  format using \textsc{phonopy}   \cite{phonopy}, and third order force constants have been converted to  \texttt{hdf5}  format using \textsc{hiphive}   \cite{hiphive}.
The \RedNmTh{} conductivity formula~(\ref{eq:thermal_conductivity_final_sum}) has been implemented in the \textsc{phono3py} software \cite{phono3py}. 
Results reported in Fig.~\ref{fig:k_vs_T_Zirconate} have been computed 
using a mesh $19{\times}19{\times}19$,  and evaluating the scattering  operator accounting for natural-abundance isotopic scattering   \cite{PhysRevLett.106.045901,tamura1983isotope} and third-order anharmonicity   \cite{fugallo2013ab,paulatto2013anharmonic, carrete2017almabte,phono3py,phonts,alamode}. 
The conductivity has been computed using the single-mode relaxation time approximation (SMA), since for La$_2$Zr$_2$O$_7$   the exact Peierls-Boltzmann thermal conductivity is known to be practically indistinguishable from the SMA value   \cite{luo2020vibrational} (in agreement with expectations for materials with ultralow thermal conductivity   \cite{lindsay_first_2016,fugallo2013ab}).
The tetrahedron method has been used for the computation of the phonon linewidths.
The data shown in Fig.~\ref{fig:k_vs_T_Zirconate} for $\kappa_{\rm P}$ are compatible with the Peierls-Boltzmann conductivity of La$_2$Zr$_2$O$_7$ presented in  Ref.~\cite{luo2020vibrational}.
The data needed to reproduce the calculations reported here (DFT-relaxed primitive cell, second-order  and third-order force constants) are available on the Open Science Platform Materials Cloud \cite{materialsCloud_ref,MaterialsCloud}.

\subsection{CsPbBr$_3$}
\label{sub:cspbbr}
The crystal structure of orthorhombic CsPbBr$_3$ has been taken from Ref.    \cite{ExpCellParam} (Crystallographic Open Database   \cite{COD_database} id 4510745) and converted to \textit{Pnmb} spacegroup using VESTA   \cite{momma2011vesta}.
The equilibrium crystal structure has been obtained performing a ``vc-relax'' calculation with Quantum \textsc{espresso}   \cite{giannozzi2017advanced}, employing the PBEsol exchange-correlation functional and GBRV   \cite{garrity2014pseudopotentials} pseudopotentials.  The PBEsol functional is selected on the basis of its good agreement between  first-principles and experimental lattice parameters   \cite{lee2017ultralow,ExpCellParam} (see Table~\ref{tab:cell_param}) and its capability to accurately describe the vibrational properties of inorganic halide perovskites   \cite{lee2017ultralow}.
Kinetic energy cutoffs of 50 and 400 Ry have been used for the wave functions and the charge density; the Brillouin zone has been integrated with a Monkhorst-Pack mesh of $5{\times}5{\times}4$ points, with a (1,1,1) shift, and 
second-order force constants have been computed on a $4{\times}4{\times}4$ mesh using density-functional perturbation theory   \cite{RevModPhys.73.515} and accounting for the non-analytic term correction due to the dielectric tensor and Born effective charges. 
Third-order force constants have been computed using the finite-difference method implemented in ShengBTE   \cite{li2014shengbte}, on a $2{\times}2{\times}2$ supercell, integrating the Brillouin zone with a $2{\times}2{\times}2$ Monkhorst-Pack mesh, and considering interactions up to $6$ \AA.
Then, anharmonic force constants have been converted from \textsc{ShengBTE} format to \texttt{mat3R} format using the conversion software \texttt{d3\_import\_shengbte.x} provided with the \textsc{D3Q} package   \cite{paulatto2013anharmonic,paulatto2015first}, including interactions up to the third neighbouring cell for the re-centering of the third-order force constants (parameter NFAR=3).
Thermal conductivity calculations have been performed with the software \textsc{D3Q}, using a  mesh $8{\times}7{\times}5$.
The scattering  operator has been computed using a Gaussian smearing of $2\;{\rm cm^{-1}}$, considering third-order anharmonicity \cite{fugallo2013ab,paulatto2013anharmonic}  and scattering due to isotopes at natural abundance \cite{PhysRevLett.106.045901,tamura1983isotope}.
The conductivity is computed using the single-mode relaxation time approximation (SMA), since for this system  the exact thermal conductivity is practically indistinguishable from the SMA value (in agreement with expectations for materials with ultralow thermal conductivity   \cite{lindsay_first_2016,fugallo2013ab}).
The data shown in Fig.~\ref{fig:k_vs_T} for $\kappa^{\alpha\beta}_{\rm P}$ are compatible with the theoretical results presented in  Ref.    \cite{lee2017ultralow} and related Supplementary Material. In particular, the direction $xx$ in this work corresponds to the $[010]$ direction in Ref.    \cite{lee2017ultralow}. Experimental data from Ref.    \cite{lee2017ultralow} do not allow to evaluate the effect of boundary scattering, as they refer to nanowires having roughly the same size, and are smaller than all the nanowires used in Ref.    \cite{wang2018cation}. Therefore, we compared the bulk thermal conductivity predictions of equation~(\ref{eq:thermal_conductivity_final_sum}) with the experimental data from Ref.    \cite{wang2018cation}.
The results reported in Fig.~\ref{fig:k_vs_T} have been reproduced using the software \textsc{phono3py} \cite{phono3py} and the tetrahedron method (the softwares \textsc{phonopy}   \cite{phonopy} and \textsc{hiphive}   \cite{hiphive} have been used to convert force constants from Quantum \textsc{espresso} or ShengBTE format to the \texttt{hdf5} format used by \textsc{phono3py}).
An implementation of Eq.~(\ref{eq:thermal_conductivity_final_sum}) will be made available as an update for the packages \textsc{phono3py}   \cite{phono3py} and \textsc{D3Q}   \cite{paulatto2015first,paulatto2013anharmonic,fugallo2013ab}. The data needed to reproduce the calculations reported here (DFT-relaxed primitive cell, second-order  and third-order force constants) are available on the Open Science Platform Materials Cloud \cite{materialsCloud_ref,MaterialsCloud}.
\begin{table}[H]
 \caption{Lattice parameters of orthorhombic CsPbBr$_3$, obtained by first-principles simulations or experiments   \cite{Miyatae1701217,lee2017ultralow}. The directions $\hat{x},\hat{y}$ and $\hat{z}$ discussed e.g. in figure~(\ref{fig:k_vs_T}) refer respectively to lattice vectors $\bm{a}_1$, $\bm{a}_2$ and $\bm{a}_3$.}
  \label{tab:cell_param}
  \centering
  \begin{tabular}{l|c|c|c}
  \hline

  \hline
  & ${a}_1$ [\AA] & ${a}_2$ [\AA] & ${a}_3$ [\AA] \\
  \hline
    Present (th) & 7.963 & 8.389 & 11.632 \\
    Ref.    \cite{lee2017ultralow} (th) & 7.990 & 8.416 & 11.674 \\
    Ref.    \cite{ExpCellParam} (exp) & 8.198  & 8.244  & 11.735   \\
    Ref.    \cite{Miyatae1701217} (exp) & 8.223 & 8.243 & 11.761   \\
  \hline

  \hline
  \end{tabular}
\end{table}

%\bibliography{biblio}

\begin{thebibliography}{207}
\bibitem [{Foo()}]{FootnoteTable}%
  \BibitemOpen
  \href@noop {} {\bibinfo {title} {{Here the BTE is written replacing in
  Eq.~(\ref{eq:BTE_std}) the continuous position $\bm{r}$ with the
  Bravais-lattice vector $\bm{R}$. This can be justified recalling that the
  phonon population appearing in the BTE is obtained from the diagonal elements
  of the Wigner distribution~(\ref{eq:Population_Peierls_Generalized}) and, as
  discussed in Sec.~\ref{sec:Wigner_thermal_transport_equation}, the knowledge
  of the Wigner distribution at the Bravais-lattice sites only is sufficient to
  fully describe thermal transport.}}}\BibitemShut {Stop}%
\end{thebibliography}

%apsrev4-2.bst 2019-01-14 (MD) hand-edited version of apsrev4-1.bst
%Control: key (0)
%Control: author (8) initials jnrlst
%Control: editor formatted (1) identically to author
%Control: production of article title (0) allowed
%Control: page (0) single
%Control: year (1) truncated
%Control: production of eprint (0) enabled
%

\end{document}